\documentstyle[eqsecnum,rmp,aps,twocolumn,harvard]{revtex}

\citationstyle{agsm}

\input{psfig}

\def\Cgqb{C_{\rm g}^{\rm qb}}
\def\CgSET{C_{\rm g}^{\rm SET}}
\def\CN{C_{\rm T}}
\def\ECqb{E_{\rm C,qb}}
\def\ECSET{E_{\rm C,\rm SET}}
\def\ECint{E_{\rm int}}
\def\ECpar{E_{\rm int}^{\parallel}}
\def\ECper{E_{\rm int}^{\perp}}
\def\Vgqb{V_{\rm g}^{\rm qb}}
\def\VgSET{V_{\rm g}^{\rm SET}}
\def\ngqb{n_{\rm g,qb}}
\def\NgSET{N_{\rm g,SET}}
\def\dHint{\delta {\cal H}_{\rm int}}

\def\Nav{\langle N \rangle}
\def\Gnull{\Gamma^{0}}
\def\Gone{\Gamma^{1}}

\def\Cg{C_{\rm g}}
\def\CJ{C_{\rm J}}
\def\Cqb{C_{\rm qb}}

\def\Ech{\delta E_{\rm ch}}
\def\EC{E_{\rm C}}
\def\Eint{E_{\rm int}}
\def\Epar{E_{\rm int}^{\parallel}}

\def\EJ{E_{\rm J}}
\def\DEav{\Delta E_{\rm av}}

\def\Hqb{{\cal H}_{\rm qb}}

\def\Hcoup{{\cal H}_{\rm coup}}
\def\Hcontr{{\cal H}_{\rm ctrl}}
\def\Hav{{\cal H}_{\rm av}}

\def\i{{\rm i}}

\def\nqb{n_{\rm g}}
\def\ng{n_{\rm g}}
\def\Phix{\Phi_{\rm x}}
\def\Qg{Q_{\rm g}}

\def\Vdeg{V_{\rm deg}}
\def\Vg{V_{\rm g}}
\def\Vqb{V_{\rm g}}

\newcommand{\ket}[1]{\left|#1\right\rangle}
\newcommand{\bra}[1]{\left\langle#1\right|}
\newcommand\G[1]{\Gamma^{#1}}
\newcommand\DM[2]{\hat\rho_{#1}^{#2}}      
\newcommand\HDM[1]{\hat\rho_{#1}}      

\def\picH{%
\mbox{\raisebox{-12pt}{%
\begin{picture}(40,24)
\put(0,12){\line(1,0){8}}
\put(32,12){\line(1,0){8}}
\put(8,0){\framebox(24,24){\sf H}}
\end{picture}
}}
}
\def\picRphi{%
\mbox{\raisebox{-30pt}{%
\begin{picture}(100,60)
\put(0,60){\line(1,0){100}}
\put(0,0){\line(1,0){100}}
\put(38,18){\framebox(24,24){$\phi$}}
\put(50,0){\line(0,1){18}}
\put(50,60){\line(0,-1){18}}
\put(50,60){\circle*{5}}
\put(50,0){\circle*{5}}
\end{picture}
}}
}
\begin{document}

\title{
\null\vskip-11.5mm\hskip0.63\textwidth
{\normalsize\rm Submitted to {\it Reviews of Modern Physics}}\\[1.5mm]
Quantum state engineering with Josephson-junction devices}

\author{Yuriy Makhlin$^{1,2}$, Gerd Sch\"on$^{1,3}$, and Alexander
Shnirman$^{1}$}

\address{
$^1$Institut f\"ur Theoretische Festk\"orperphysik,
Universit\"at Karlsruhe, D-76128 Karlsruhe, Germany\\
$^2$Landau Institute for Theoretical Physics, 
Kosygin st. 2, 117940 Moscow, Russia\\
$^3$Forschungszentrum Karlsruhe, Institut f\"ur Nanotechnologie,
D-76021 Karlsruhe, Germany}

\maketitle

\citationmode{abbr}

\begin{abstract}
Quantum state engineering, i.e., active control over the coherent
dynamics of suitable quantum mechanical systems, has become a fascinating
perspective of modern physics. With concepts
developed in atomic and molecular physics and in the context of NMR, the
field has been  stimulated further by the perspectives
of quantum computation and communication. For this purpose a number of
individual two-state quantum systems (qubits) should be 
addressed and coupled in a controlled way. Several physical
realizations of qubits have been considered, incl.\
trapped ions, NMR, and quantum optical systems. For potential
applications such as logic operations, nano-electronic devices appear
particularly promising because they can be embedded in electronic
circuits and scaled up to large numbers of qubits. 

Here we review the quantum properties of low-capacitance Josephson
junction devices. The relevant quantum degrees of freedom are
either Cooper pair charges on small islands or fluxes in ring
geometries, both in the vicinity of degeneracy points. The coherence
of the superconducting state is exploited to achieve long phase
coherence times. Single- and two-qubit quantum  
manipulations can be controlled by gate voltages or magnetic
fields, by methods established for single-charge systems or the SQUID
technology. Several of the interesting single-qubit properties,
incl.\ coherent oscillations have been demonstrated in recent
experiments, thus displaying in a spectacular way the laws of quantum
mechanics in solid state devices. Further experiments, such as
entanglement of qubit states, which are crucial steps towards a
realization of logic elements, should be within reach. 

In addition to the manipulation of qubits the resulting quantum state
has to  be read out. For a Josephson charge qubit this can be
accomplished by coupling it capacitively to a single-electron
transistor (SET).  To describe the quantum measurement
process we analyze the time evolution of the density matrix of the coupled
system.  As long as the transport voltage is turned off, the transistor
has only a weak influence on the qubit. When the voltage is
switched on, the dissipative current through the SET destroys the phase
coherence of the qubit within a short 
time.  The measurement is accomplished only after a longer time, when the
macroscopic signal, i.e., the current through the SET, resolves
different quantum states. 
At still longer times the measurement process itself destroys the
information about the initial state.  Similar scenarios are found
when  the quantum state of a flux qubit is measured by a dc-SQUID,
coupled to it inductively.

\end{abstract}

\tableofcontents

\section{Introduction}

The interest in `macroscopic' quantum effects in low-capacitance
Josephson junction circuits has persisted for many years. One of the
motivations was to test whether the laws of quantum mechanics apply
in macroscopic systems, in a Hilbert space spanned by
macroscopically distinct states \cite{LeggettMQC}. 
The degrees of freedom studied were the 
phase difference of the superconducting order parameter across a
junction, or the flux in a superconducting ring (SQUID) geometry.
Various quantum phenomena, such as macroscopic quantum tunneling (MQT)
and resonance tunneling were demonstrated
\citeaffixed{Voss-Webb,Clarke_PRB,Lukens}{see e.g.}.  
On the other hand, despite experimental efforts
\citeaffixed{Tesche}{e.g.} coherent oscillations
of the flux between two macroscopically distinct
states (macroscopic quantum coherence, MQC) had not been observed.  

The field received new attention recently, after it was
recognized that suitable Josephson devices may serve as quantum bits
(qubits) in quantum information devices, and that quantum logic
operations \footnote{Since computational applications are 
widely discussed, we employ here and below frequently the terminology
of quantum information theory, referring to a two-state
quantum system as qubit and denoting unitary manipulations of its
quantum state as quantum logic operations or gates.}
can be performed by controlling gate voltages or
magnetic fields 
\citeaffixed{Bouchiat_PhD,Our_PRL,Averin,Our_Nature,%
Nakamura_Nature,Mooij,Blatter}{e.g.}. In this context, as well as for
other conceivable  applications of quantum state 
engineering, the experimental milestones are
the observation of quantum superpositions of macroscopically distinct
states, of coherent oscillations, and of entangled quantum states of
several qubits. For Josephson devices first successful experiments 
have been performed. These systems can be fabricated by
established lithographic methods, and the control and measurement
techniques are quite advanced. They further exploit
the coherence of the superconducting state, which helps achieving
sufficiently long phase coherence times. 

Two alternative realizations of quantum bits
have been proposed, based on either charge or phase (flux) degrees of
freedom. In the former the charge in low-capacitance Josephson
junctions is used as quantum degree of freedom, with basis
states differing by the number of Cooper pair charges on an
island.  These devices combine the coherence of Cooper pair tunneling
with the control mechanisms developed for single-charge systems and
Coulomb-blockade phenomena.  The manipulations can be
accomplished by switching gate voltages~\cite{Our_PRL};
designs with controlled inter-qubit  couplings were
proposed~\cite{Averin,Our_Nature}.  Experimentally, the 
coherent tunneling of Cooper 
pairs and the related properties of quantum mechanical superpositions
of charge states has been demonstrated~\cite{Bouchiat_PhD,Nakamura}.
Most spectacular are 
recent experiments of~\citeasnoun{Nakamura_Nature}, where the quantum
coherent oscillations of a Josephson charge qubit prepared in a
superposition of eigenstates were observed in the time domain.  We
describe these systems, concepts and results in
Section~\ref{sec:JosChargeBit}. 

The alternative realization is based on the phase of a Josephson
junction or the flux in a ring geometry near a degeneracy point as
quantum degree of freedom \citeaffixed{Mooij,Blatter}{see e.g.}. In
addition to the earlier experiments, where  macroscopic quantum
tunneling had been observed \cite{Voss-Webb,Clarke_PRB,Lukens},
the groups in Delft and Stony Brook \cite{Delft_Cats,Friedman_Cats}
demonstrated recently by spectroscopic
measurements the flux qubit's eigenenergies, they  
observed eigenstates which are superpositions of
different flux states, and new efforts are made to observe the
coherent oscillation of the flux between degenerate
states~\cite{Mooij,Friedman_Cats,Rome}. We will discuss the quantum
properties of flux qubits in Section~\ref{sec:FluxQubits}. 

To make use of the quantum coherent time evolution it
is crucial to find systems with intrinsically long phase coherence
times and to minimize external sources of dephasing. The latter
can never be avoided completely since, in order to perform the necessary
manipulations, one has to couple to the qubits, e.g., by attaching
external leads. Along the same channels as the signal (e.g., gate 
voltages) also the noise  enters the system. However, by operating at
low temperatures and
choosing suitable coupling parameters, these dephasing effects can be
kept at an acceptable level. We provide estimates of
the phase coherence time in Section \ref{sec:Dephasing}. 

In addition to controlled manipulations of qubits, quantum
measurement processes are needed, e.g., to read out the final state of
the system. In our quantum mechanics courses 
we learned to express the measurement process
as a ``wave function collapse'', i.e.,  as  a
non-unitary projection, which reduces the quantum state of the qubit
to one of the possible eigenstates of the observed quantity
with state-dependent probabilities. However, in reality any measurement 
is performed by a device which itself is realized by a physical 
system, suitably coupled to the measured quantum system and with a
macroscopic read-out variable. Its presence, in general, disturbs the
quantum manipulations. Therefore the dissipative processes which
accompany the measurement should be switched on only when needed. 

An example is provided by a
normal-state single-electron transistor (SET) coupled 
capacitively to a single-Cooper pair box. This system is widely used as an
electro-meter in classical single-charge systems. We describe in
 Section~\ref{sec:Measurement} how a SET can also be used to read out the
quantum state of a charge qubit. For this purpose 
we study the time evolution of the density matrix
of the coupled system~\cite{Our_PRB}. During  quantum manipulations  
of the qubit the transport voltage of the SET is turned off, in which
case it acts only as an extra capacitor. 
To perform the measurement the transport voltage is turned on. In this
stage the dissipative current through the transistor dephases the the
state of the qubit
rapidly. This current also provides the macroscopic read-out signal for
the quantum state of the qubit. However, it requires a
longer `measurement time' until the noisy signal resolves different
qubit states. Finally, on the still longer `mixing time' scale, the
measurement process itself destroys the information about the initial
quantum state. 

Many results and observations made in the context of the normal state
single-electron transistor apply also 
to other physical systems, e.g.,  a superconducting SET (SSET) coupled
to a charge qubit~\cite{Averin_SQUID,EsteveSSET} or a dc-SQUID monitoring
as a quantum magneto-meter the state of a flux
qubit~\citeaffixed{Mooij,Friedman_Cats,Averin_SQUID}{see e.g.}. 
The results can also be compared to the nonequilibrium dephasing
processes discussed theoretically~\cite{Levinson,Aleiner,Gurvitz}
and demonstrated experimentally by \citeasnoun{Buks}. 

One of the motivations for quantum state engineering with
Josephson devices is their potential application as logic
devices and quantum computing. By exploiting the massive
parallelism of the coherent evolution of superpositions of
states quantum computers could perform certain tasks which no
classical  computer can do in acceptable
times~\cite{Bennett-PT95,Barenco_Review,DVD-Sci95,Aharonov}. 
In contrast to the development of physical realizations of qubits and
gates, i.e., the ``hardware'', the theoretical concepts of quantum
computing, the ``software'', are already rather advanced. As an
introduction,  and in order to clearly define the goals, we 
present in Appendix \ref{app:IdealModel} an ideal model
Hamiltonian with sufficient control to perform all the needed
manipulations. (We can mention that the Josephson junction devices
come rather close to this ideal model.) Then in Appendix
\ref{app:gates} we show by a few representative examples how these
manipulations can be combined for useful computations.

Various other physical systems have been suggested as possible realizations
of qubits and gates. We mention ions in electro-magnetic traps
manipulated by laser irradiation \cite{Zoller,Wineland}, NMR on
ensembles of molecules in liquids \cite{Chuang,Cory}, and cavity QED
systems~\cite{Kimble}. In comparison, the mentioned Josephson systems
are more easily embedded in electronic circuits and scaled up to large
registers.  Ultra-small quantum 
dots with discrete levels, and in particular, spin degrees of freedom
embedded in nano-structured materials are candidates as well.  
They can be manipulated by tuning potentials and barriers~\cite{Loss,Kane}.
In Appendix \ref{app:spins_dots} we describe these alternative solid state
realizations of qubits and look at their advantages and 
drawbacks. Because of the difficulties of controlled fabrication their
experimental realization is still at a very early stage.

\section{Josephson charge qubit}
\label{sec:JosChargeBit}

\subsection{Superconducting charge box as a quantum bit}
\label{subsec:ChargeBit}

In this section we describe the properties of low-capacitance
Josephson junctions, where the charging energy dominates over the
Josephson coupling energy, and discuss how they can be manipulated in a
quantum coherent fashion. Under suitable conditions they
provide physical realizations of qubits with two states
differing by one Cooper pair charge on a small island.  The necessary
one-bit and two-bit gates can be performed by controlling  applied
gate voltages and magnetic fields. Different designs will be presented which
differ in complexity, but also in the accuracy and flexibility 
of the manipulations.

\begin{figure}
\centerline{\hbox{\psfig{figure=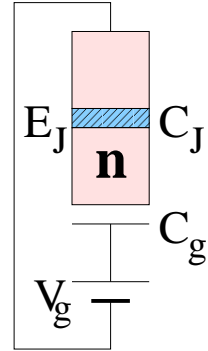,width=0.3\columnwidth}}}
\vskip 0.8cm
\caption[]{\label{Fig:BITS_IDEAL}
A Josephson charge qubit in its simplest design formed by a
superconducting single-charge box.} 
\end{figure}

The simplest Josephson junction qubit is shown in
Fig.~\ref{Fig:BITS_IDEAL}. It consists of a small superconducting
island (``box'') with $n$ excess Cooper pair charges
(relative to some neutral reference state),
connected by a tunnel junction with capacitance $\CJ$ and
Josephson coupling energy $\EJ$ to a superconducting electrode.  
A control gate voltage $\Vqb $ is coupled to the system via a gate
capacitor $\Cg$.
Suitable values of the junction capacitance, which can be fabricated
routinely by present-day technologies,  are in the range of
Femtofarad and below, $\CJ \le 10^{-15}$~F, while the gate capacitances
 can be chosen still smaller. The relevant energy scale,
the single-electron charging energy  $\EC   \equiv e^2/2(\Cg +
\CJ)$, which depends on the total capacitance of the island,
is then in the range of Kelvin~\footnote{Throughout this review we frequently 
use temperature units for energies.}
or above, $\EC \ge 1$~K. The
Josephson coupling energy $\EJ$ is proportional to the critical
current of the Josephson junction~\citeaffixed{Tinkham}{see e.g.}. 
Typical values considered here are in the range of 100~mK. 

We choose a material such that the superconducting energy gap $\Delta$
is the largest energy in the problem, larger even than the single-electron
charging energy.  In this case 
quasiparticle tunneling  is suppressed at low temperatures, and a
situation can be reached where no quasiparticle excitation is found on
the island~\footnote{In the ground state the superconducting
state is totally paired, which requires an even number of electrons
on the island. A state with an odd number of electrons necessarily
costs an extra quasi-particle energy
$\Delta$ and is exponentially suppressed at low $T$. This `parity effect' has 
been
established in experiments below a crossover temperature $T^{*} \approx
\Delta /(k_{\rm B}\ln N_{\rm eff})$, where $N_{\rm eff}$ is the number of
electrons in the system near the Fermi
energy~\cite{Parity,Lafarge,Schoen-Zaikin94,Tinkham}.  For a small
island, $T^{*}$ is typically one order of magnitude lower
than the superconducting transition temperature.}. Under these
conditions  only Cooper pairs tunnel -- coherently -- in the
superconducting junction, and the system is described by 
the Hamiltonian:
\begin{equation}
\label{Eq:1bit_Hamiltonian_Eqb}
        {\cal H} = 4\EC  (n - \nqb  )^2 -  \EJ \cos\Theta \ . 
\end{equation}
Here, $n$ is the number operator of (excess) Cooper pair charges  on
the island, and  $\Theta$,
the phase of the superconducting order parameter of the island, is its
quantum mechanically conjugate, $n = -i\hbar\,\partial/\partial(\hbar
\Theta)$.   The
dimensionless gate charge, $\nqb   \equiv \Cg \Vqb  /2e$, accounts for
the effect of the gate voltage and acts as a control parameter.
Here we consider systems where the charging energy is much larger than
the Josephson 
coupling energy, $\EC   \gg \EJ$.  In this regime a convenient basis
is formed by the charge states, parameterized by the number of Cooper
pairs $n$ on the island.  In this basis the Hamiltonian
(\ref{Eq:1bit_Hamiltonian_Eqb}) reads
\begin{eqnarray}
{\cal H} =&& \sum_n \Big\{ 4\EC(n - \nqb)^2 |n\rangle \langle n| 
\nonumber\\
&&  - \frac{1}{2}\EJ 
\Big(\ket{n}\bra{n+1} + \ket{n+1}\bra{n}\Big)\Big\} \; .
\end{eqnarray}

\begin{figure}
\centerline{\hbox{\psfig{figure=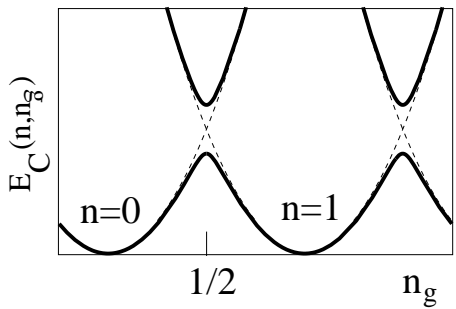,width=0.7\columnwidth}}}
\vskip 0.4cm
\caption[]{\label{Parabolas}
The charging energy of the superconducting electron box  is shown
as a function of  the gate charge $\nqb  $ for
different  numbers of extra Cooper pairs $n$ on the island
(dashed parabolas).  Near
degeneracy points the weaker Josephson coupling mixes the
charge states and modifies the energy of the  eigenstates (solid
lines). In the vicinity of these points 
the system  effectively reduces to a two-state quantum system.  }
\end{figure}

For most values of $\nqb$ the energy levels are dominated by the
charging part of the Hamiltonian.  However, when $\nqb$ is
approximately half-integer and the charging energies of two  adjacent
states are close to each other (e.g., at $\Vg=\Vdeg\equiv e/\Cg$), the
Josephson  tunneling mixes them strongly (see Fig.~\ref{Parabolas}).
We concentrate on such a voltage range near a degeneracy point where
only two charge states, say $n=0$ and $n=1$, play a role,  while all other
charge states, having a much higher energy, can be ignored.  In this
case the superconducting charge box (\ref{Eq:1bit_Hamiltonian_Eqb})
reduces to a two-state quantum system (qubit) with Hamiltonian which
can be written in spin-$1\over2$ notation as
\begin{equation}
\label{Eq:Magnetic_Hamiltonian}
        {\cal H}_{\rm ctrl}= -\frac{1}{2}  B_z \hat\sigma_z - {1\over
        2} B_x \hat\sigma_x \ . 
\end{equation}
The charge states $n=0$ and  $n=1$ correspond to the spin basis states
$\left|\uparrow  \right.\rangle \equiv \left(_0^1 \right)$ and
$\left|\downarrow\right.\rangle \equiv \left(_1^0 \right)$,  
respectively. The charging energy splitting, which is controlled by the gate
voltage, corresponds in spin notation to the $z$-component of the
magnetic field 
\begin{equation}
\label{Eq:ChargeBit-Bz}
B_z \equiv \Ech \equiv 4\EC (1-2\nqb) 
\, ,
\end{equation}
while the Josephson energy provides the $x$-component
of the effective magnetic field
\begin{equation}
\label{Eq:ChargeBit-Bx}
B_x \equiv \EJ
\, .
\end{equation}

For later convenience we rewrite the Hamiltonian as
\begin{equation}
\label{Magnetic_Hamiltonian'}
        {\cal H}_{\rm ctrl} = -\frac{1}{2}\Delta E(\eta)
        \left(\cos\eta\ \sigma_z+\sin\eta\ \sigma_x \right) \; ,
\end{equation}
where the mixing angle  $\eta\equiv \tan^{-1} (B_x/B_z)$ determines
the  direction of the effective magnetic field in the $x$-$z$-plane, and
the energy  splitting between the eigenstates is $\Delta E(\eta) =
\sqrt{B_x^2+B_z^2}=\EJ/\sin\eta$. 
At the degeneracy point, $\eta = \pi/2$, it reduces to $\EJ$.  The
eigenstates  are denoted in the following as $|0\rangle$  and
$|1\rangle$. They depend on the gate charge  $\nqb  $ as
\begin{eqnarray}
\label{Eigen_Basis}
        \ket{0}  &=& \;\;\;  \cos{\eta\over2} \ket{\uparrow}  +
\sin{\eta\over2} \ket{\downarrow} \nonumber \\ |1\rangle &=& -
\sin{\eta\over2} \ket{\uparrow} + \cos{\eta\over2} \ket{\downarrow} \ .
\end{eqnarray}

We can further express the Hamiltonian in the
basis of eigenstates.  To avoid confusion we introduce a second set of
Pauli matrices, $\bbox{\rho}$, which operate in the basis $|0\rangle$
and $\ket{1}$, while reserving the operators  $\bbox{\sigma}$ for the
basis of charge states $\ket{\uparrow}$ and $\ket{\downarrow}$.  By
definition  the Hamiltonian then becomes
\begin{equation}
\label{Eq:Ham_diagonalized}
H=- \frac{1}{2}\Delta
E(\eta)\rho_z \;.
\end{equation}

The Hamiltonian (\ref{Eq:Magnetic_Hamiltonian}) is similar to the
ideal single-qubit model (\ref{Eq:idealH-1qubit})
presented in Appendix \ref{app:IdealModel}. Ideally
the bias energy (effective magnetic field in
$z$-direction) and the tunneling amplitude 
(field in $x$-direction) are controllable. 
However, at this stage we can control -- by the gate voltage
-- only the bias energy, while  the
tunneling amplitude has a constant value set by the Josephson
energy.  Nevertheless, by switching the gate voltage we can perform
the required  one-bit operations \cite{Our_PRL}. If, for example, one 
chooses  the idle state far to the left from the degeneracy point, the
eigenstates  $\ket{0}$ and $\ket{1}$ are close to $\ket{\uparrow}$
and $\ket{\downarrow}$, respectively. Then,
switching the system suddenly to the degeneracy point for a time
$\Delta t$ and back produces a rotation in spin space,
\begin{equation}
\label{1bit_spin_flip}
U_{\rm 1-bit}(\alpha)=  \exp \left(i \frac{\alpha}{2} \sigma_x
\right) = \left( 
\begin{array}{cc}
\cos\frac{\alpha}{2}  & i\sin\frac{\alpha}{2} \\ 
i\sin \frac{\alpha}{2} & \cos\frac{\alpha}{2}
\end{array} 
\right) 
\; ,
\end{equation}
where $\alpha = \EJ \Delta t/\hbar$.  Depending on the value of
$\Delta t$, a spin flip can be produced,  or, starting from
$|0\rangle$,  a superposition of states with any  chosen weights can
be reached. (This is exactly the operation performed in the experiments of
\citeasnoun{Nakamura_Nature}; see
Subsection~\ref{subsec:Experiments_Charge}).  Similarly, a phase shift  
between the two
logical states  can be achieved  by changing the gate voltage $\nqb $
for some time by a small amount, which modifies the energy difference
between the ground and excited states.

Several remarks are in order:

\noindent
(1) Unitary rotations by $B_x$ and $B_z$
are sufficient for all manipulations of a single qubit. By using a
sequence of no more than three such elementary rotations we can
achieve any unitary transformation of a qubit's state.

\noindent
(2) The example presented above, with control of $B_z$ only, 
provides an {\sl approximate} spin flip
for the situation where the idle point is far from degeneracy and
$\EC\gg\EJ$.  But a spin flip in the logical basis can also be
performed {\sl exactly}.  It requires that we switch from the idle point
$\eta_{\rm idle}$ to the point where the effective magnetic field is
orthogonal to the idle one, $\eta=\eta_{\rm idle} + \pi/2$.  This
changes the Hamiltonian from $H=-\frac{1}{2}\Delta E(\eta_{\rm
idle})\rho_z$ to $H=-\frac{1}{2}\Delta E(\eta_{\rm idle}+\pi/2)\rho_x$.
To achieve this, the dimensionless gate charge $\nqb  $ should be
increased by $\EJ/(4\EC  \sin 2\eta_{\rm idle})$.  For the limit
discussed above, $\eta_{\rm idle} \ll 1$, this operating point is close
to the degeneracy point, $\eta=\pi/2$.
 
\noindent
(3) An alternative way to manipulate the qubit is to use resonant
pulses, i.e., ac-pulses with frequency close to 
the qubit's level-spacing. We do not describe this technique 
here as it is well known from NMR methods.

\noindent
(4) So far we were concerned with the time dependence during elementary
rotations.  However, frequently the 
quantum state should be kept unchanged for some time, for
instance, while other qubits are manipulated. Even in the idle state,
$\eta=\eta_{\rm idle}$, because the energies of the two 
eigenstates differ, their phases evolve relative to each
other. This leads to the `coherent oscillations', typical for a
quantum system in a superposition of eigenstates.  We have to keep
track of this time dependence with high precision and, hence, of the
time $t_0$ from the very beginning of the manipulations. The
time-dependent phase factors can be removed
from the eigenstates if all the calculations are performed in the
interaction representation, with zero-order Hamiltonian being the one
at the idle point.  
However, the
price for this simplification is an additional time dependence in
the Hamiltonian during operations, introduced by the transformation
to the interaction representation.  This point has been discussed
in more detail by \citeasnoun{Our_Book}.

\noindent
(5) The choice of the
logical basis of the qubit is by no means unique.  As follows from the 
preceding
discussion, we can perform $x$- and $z$-rotations in the charge 
basis, $\left|\uparrow \right.\rangle$, $\left|\downarrow \right.\rangle$, 
which
provides sufficient tools for any unitary operation.  On the other
hand, since we can perform {\it any} unitary transformation, we can
choose any other basis as logical basis as well. The Hamiltonian at
the idle point is diagonal in the eigenbasis (\ref{Eigen_Basis}),
while the controllable part of the Hamiltonian, the charging energy,
favors the charge basis.  The preparation procedure (thermal
relaxation at the idle point) is easier described in the eigenbasis,
while coupling to the meter (see Section~\ref{sec:Measurement}) is
diagonal in the charge basis.  So, the choice of the logical states 
remains a matter of convention.

\noindent
(6) A final comment concerns normal-metal single-electron systems.  
While they may serve as classical bits and logic
devices, they are ruled out as potential quantum logic devices. The
reason is that due to the large number of electron states involved, their
phase coherence is destroyed in the typical sequential tunneling processes.

\subsection{Charge qubit with tunable coupling}
\label{subsec:ChargeBit-w-SQUID}

\begin{figure}
\centerline{\hbox{\psfig{figure=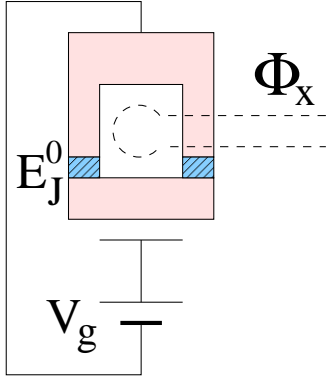,width=0.5\columnwidth}}}
\vskip 0.8cm
\caption[]{\label{Fig:BitwSQUID}
A charge qubit with tunable effective Josephson coupling. The single 
Josephson junction is replaced by a flux-threaded SQUID. The flux in
turn can be controlled by a current carrying loop placed on top
of the structure.}
\end{figure}

A further step towards the ideal model (\ref{Eq:idealH-1qubit}),
where the tunneling amplitude ($x$-component of the field) is
controlled as well, is the ability to tune the 
Josephson coupling. This is achieved by the design shown in
Fig.~\ref{Fig:BitwSQUID}, where the single
Josephson junction is 
replaced by two junctions in a loop configuration~\cite{Our_Nature}.
This dc-SQUID  is biased by an external flux $\Phi_{\rm x}$, which is
coupled into the system through an inductor loop.  If  the
self-inductance of the SQUID loop is low \cite{Tinkham}, the
SQUID-controlled qubit is described by a Hamiltonian of the form
(\ref{Eq:1bit_Hamiltonian_Eqb}) with modified potential energy:
\begin{eqnarray}
   - \EJ^0 \cos\left(\Theta + \pi\frac{\Phix}{\Phi_0}\right) &-& \EJ^0
   \cos\left(\Theta - \pi\frac{\Phix}{\Phi_0}\right) = \nonumber\\ &-&
   2\EJ^0 \cos\left(\pi\frac{\Phix}{\Phi_0}\right)\cos\Theta \; .
\label{EJPhi}
\end{eqnarray}
Here $\Phi_0 = hc/2e$ denotes the flux quantum.  We assumed that the two
junctions are identical~\footnote{While this cannot be guaranteed with high
precision in an experiment, we note that the effective Josephson coupling can
be tuned to zero exactly by a design with three junctions.}  with the same
$E^0_{\rm J}$.  The effective junction capacitance is the sum of individual
capacitances of two junctions, in symmetric cases $\CJ=2C^0_{\rm J}$.

When parameters are chosen such that only two charge states play a
role, we arrive again at the Hamiltonian (\ref{Eq:Magnetic_Hamiltonian}), 
but the effective Josephson coupling,
\begin{equation}
\label{Eq:EffJosCoupling}
B_x=\EJ(\Phi_{\rm x})=2\EJ^0 \cos\left(\pi\frac{\Phix}{\Phi_0}\right)
\, ,
\end{equation} 
is  tunable. Varying the external flux $\Phix$ by amounts of order
$\Phi_0$ changes the coupling between
$2\EJ^0$ and zero~\footnote{If the SQUID inductance is not small, the
fluctuations of the flux within the SQUID renormalize the energy
(\ref{EJPhi}). But  still, by symmetry arguments, at $\Phix=\Phi_0/2$
the effective Josephson coupling vanishes.}.

The SQUID-controlled qubit is, thus, described by the ideal single-bit
Hamiltonian 
(\ref{Eq:idealH-1qubit}), with field components $B_z(t)=\Ech(\Vqb(t))$
and $B_x(t)=\EJ(\Phix(t))$ controlled independently by the gate voltage and the
flux.  If we fix in the idle state $\Vg=\Vdeg$ and $\Phix=\Phi_0/2$, the
Hamiltonian is zero, $\Hqb^0=0$, and the state does not evolve in time.  Hence,
there is no need to control the total time from the beginning of the
manipulations, $t_0$.  If we change the voltage or the current, the modified
Hamiltonian generates rotations around the $z$- or $x$-axis, which are
elementary one-bit operations.  Typical time spans of single-qubit logic gates
are determined by the corresponding energy scales and are of order $\hbar/\EJ$,
$\hbar/\Ech$ for $x$- and $z$-rotations, respectively.  If at all times at most
one of the fields, $B_z(t)$ or $B_x(t)$, are turned on, only the time integrals
of their profiles determine the results of the individual operations.  Hence
these profiles can be chosen freely to optimize speed and simplicity of the
manipulations.

The introduction of the SQUID permits not only simpler and more accurate
single-bit manipulations, but it also allows us to control the two-bit
couplings, as we will discuss next.  Furthermore, it simplifies the
measurement procedure, which is more accurate at $\EJ=0$ (see
Section~\ref{sec:Measurement}).

\subsection{Controlled inter-qubit coupling}
\label{subsec:Controlled_2Bit}

\begin{figure}
\vskip 0.4cm
\centerline{\hbox{\psfig{figure=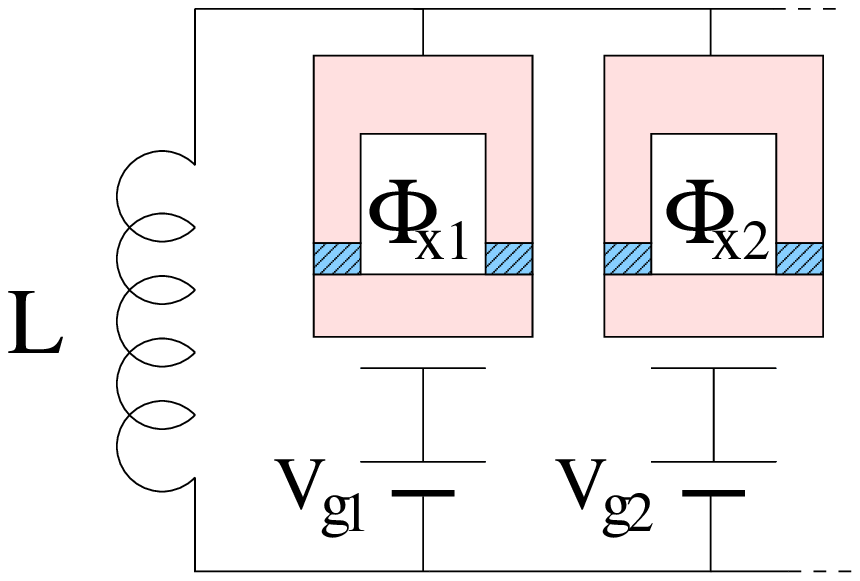,width=0.7\columnwidth}}}
\vskip 0.8cm
\caption[]{\label{Fig:ManyBits}
A register of many charge qubits coupled by the oscillator modes in
the $LC$-circuit formed by the inductor and the qubit capacitors.}
\end{figure}

In order to perform two-qubit logic gates we need to couple pairs of
qubits together and to control the interactions between them.  One
possibility is to connect the superconducting boxes ($i$ and 
$j$) directly, e.g., via a capacitor.  The resulting charge-charge interaction
is described by a Hamiltonian of the form (\ref{Eq:idealH}) with an
Ising-type coupling term $\propto \sigma_z^i\sigma_z^j$.  Such a
coupling allows an easy realization of a controlled-NOT operation.  On
the other hand, it has severe drawbacks. In order 
to control the two-bit interaction, while preserving the single-bit
properties discussed above, one needs a switch to turn the two-bit
interaction on and off. Any externally operated switch, however, connects 
the system to the dissipative external circuit, thus introducing
dephasing effects (see Section~\ref{sec:Dephasing}). They are
particularly strong if the switch is attached directly and unscreened
to the qubit, which would be required to control the direct capacitive
interaction. Therefore
alternatives were explored where the control fields are coupled only weakly to
the qubits.  A solution~\cite{Our_Nature} is shown in Fig.~\ref{Fig:ManyBits}.
All $N$ qubits are connected in parallel to a common $LC$-oscillator mode which
provides the necessary two-bit interactions.  It turns out that the possibility
to control the Josephson couplings by an applied flux, simultaneously allows us
to switch the two-bit interaction for each pair of qubits.  This
brings us close to the ideal model~(\ref{Eq:idealH}) with a coupling
term $\propto \sigma_y^i\sigma_y^j$.

In order to demonstrate the mentioned properties of the coupling we
consider the Hamiltonian of the chain (register of qubits) shown in 
Fig.~\ref{Fig:ManyBits}. It is 
\begin{eqnarray}
\label{2bit_Hamiltonian}
        {\cal H} = && \sum_{i=1}^N \left\{ {(2e n_i - \Cg  V_{{\rm g}i})^2\over
        2(\CJ+\Cg )} -  \EJ(\Phi_{{\rm x}i}) \cos\Theta_i 
\right\} \nonumber \\ + 
        && \frac{1}{2N\Cqb} \left(q-\frac{\Cqb}{\CJ}\sum_{i} 2e
        n_i\right)^2  + {\Phi^2\over 2L} \ .
\end{eqnarray}
Here $q$ denotes the total charge accumulated on the gate capacitors of the 
array of qubits.
Its conjugate variable is the phase drop $\phi$ across the inductor,
related to the flux by $\phi/2\pi=\Phi/\Phi_0$. Furthermore,
\begin{equation}
\label{Eq:Cqb}
\Cqb = \frac{\CJ\Cg}{\CJ + \Cg}
\end{equation}
is the capacitance of the qubit in the  external circuit.  

Depending on the relations among the parameters the Hamiltonian
(\ref{2bit_Hamiltonian}) can be reduced. We first consider
the situation where the frequency of the  
($q,\Phi$) oscillator, $\omega_{LC}^{(N)} = {1/\sqrt{N\Cqb  L}}$, is
higher  than typical frequencies of the qubit's dynamics:
\begin{equation}
\label{High_Frequency_Condition}
\hbar\omega_{LC}^{(N)}\gg \EJ,\Ech\; .
\end{equation}
In this case the oscillator modes are not excited, but still their virtual
excitation produces an effective coupling
between the qubits. To demonstrate this we eliminate the
variables $q$ and $\Phi$ and derive an effective description in terms
of the qubits' variables only.  As a first step we perform a
canonical transformation, $\tilde q = q - (\Cqb/ \CJ)\,\sum 2e n_i$ and
$\tilde \Theta_i = \Theta_i + 2\pi (\Cqb/ \CJ) \,(\Phi/\Phi_0)$, while
$\Phi$ and $n_i$ are unchanged. This step leads to the new
Hamiltonian (we omit the tildes): 
\begin{eqnarray}
\label{2bit_Hamiltonian_Cosine}
        {\cal H}& =  \displaystyle\frac{q^2}{2N\Cqb} +& {\Phi^2\over 2L}
        +\sum_{i} \Bigl\{{(2e n_i - {\Cg V_{{\rm g}i}})^2\over{2(\CJ+\Cg
        )}}
\nonumber \\
        &&- \EJ(\Phi_{{\rm x}i})   \cos\left(\Theta_i -  {2\pi\over
        \Phi_0}{\Cqb \over \CJ} \Phi \right)\Bigr\}  \ .
\end{eqnarray}
We assume that the fluctuations of $\Phi$ are weak
\begin{equation}
\label{Small_Fluctuation_Assumption}
        \frac{\Cqb}{\CJ} \sqrt{\langle\Phi^2\rangle}\, \ll \, \Phi_0 \
        , 
\end{equation}
since, otherwise, the Josephson tunneling terms in the Hamiltonian
(\ref{2bit_Hamiltonian_Cosine}) are washed out \cite{Our_PRL}. Assuming
(\ref{Small_Fluctuation_Assumption}) to be satisfied, we expand the
Josephson terms in Eq.~(\ref{2bit_Hamiltonian_Cosine})  up
to linear terms in $\Phi$. Then we can trace over the variables $q$
and $\Phi$. 
As a result we obtain an effective Hamiltonian, consisting of a sum
of $N$ one-bit Hamiltonians (\ref{Eq:1bit_Hamiltonian_Eqb}) and the coupling
terms
\begin{equation}
\label{COUPLING_TERM}
\Hcoup = -  {2\pi^2 L \over \Phi_0^2}\left({\Cqb \over
\CJ}\right)^2
\left[\sum_{i} \EJ(\Phi_{{\rm x}i})\sin\Theta_{i}\right]^2
\ .
\end{equation}
In spin-$1\over2$ notation this becomes~\footnote{While the expression
(\ref{HintGen}) is valid only in 
leading order in an  expansion in $E_{\rm J}^i/\hbar\omega^{N}_{LC}$,
higher terms also vanish  when the Josephson couplings are put to
zero. Hence, the decoupling in the idle periods persists.}
\begin{equation}
        \Hcoup =-\sum\limits_{i<j}\frac{\EJ(\Phi_{{\rm x}i})
        \EJ(\Phi_{{\rm x}j})}{E_L}\hat\sigma_y^i\hat\sigma_y^j \ +\
        {\rm const} 
\ ,
\label{HintGen}
\end{equation}
where we introduced the scale
\begin{equation}
E_L=  \left(\frac{\CJ}{\Cqb}\right)^2 {\Phi_0^2 \over \pi^2L}\;.
\end{equation} 

The coupling Hamiltonian (\ref{HintGen}) can be understood as the magnetic
free energy of the current-biased inductor $-L I^2/2$.  This current
is the sum of 
the contributions from the qubits with non-zero Josephson coupling, $I \propto
\sum_i \EJ^i(\Phi_{{\rm x}i}) \sin\Theta_i \propto \sum_i \EJ^i(\Phi_{{\rm
x}i}) \hat \sigma_y^i$.

Note that the strength of the interaction does not depend directly on
the number 
of qubits $N$ in the system.  However, the frequency of the
$(q,\Phi)$-oscillator $\omega^{(N)}_{LC}$ scales as $1/\sqrt{N}$.  The
requirement that this frequency should not drop below typical eigenenergies of
the qubit, ultimately, limits the number of qubits which can be coupled by a
single inductor.

The system with flux-controlled Josephson couplings $\EJ(\Phi_{{\rm
x}i})$ and the interaction 
of the form (\ref{HintGen}) allows us to perform all necessary gate
operations in a straightforward way. In the idle 
state all Josephson couplings are turned off and the interaction
(\ref{HintGen}) is zero. Depending on the choice of the idle state we
may also tune the qubits by their gate voltages to the
degeneracy points, which makes the Hamiltonian vanish, ${\cal H} =
0$. The interaction Hamiltonian remains zero during  one-bit 
operations, as long as we perform   one such operation at a time,
i.e., only for one qubit we have $\EJ^i = \EJ(\Phi_{{\rm x}i}) \ne 0$.
To perform a two-bit operation for any 
pair of qubits, say, $i$ and $j$, $\EJ^i$ and $EJ^j$ are
switched on simultaneously, yielding the Hamiltonian 
\begin{equation}
{\cal H} =- \frac{\EJ^i}{2}\hat\sigma_x^i - \frac{E_{\rm
J}^j}{2}\hat\sigma_x^j - 
\frac{\EJ^i\EJ^j}{E_L}\hat\sigma_y^i\hat\sigma_y^j \; .
\label{Our2bitHam}
\end{equation}
While (\ref{Our2bitHam}) is not identical to the form (\ref{Eq:idealH})
it equally well allows the relevant non-trivial two-bit operations, which,
combined  with the one-bit operations discussed above, provide a
universal set of gates.

A few comments should be added:

\noindent
(1) We note that typical time spans of two-bit operations are of the order
$\hbar E_L/E_{\rm  J}^2$. It follows from  the conditions
(\ref{High_Frequency_Condition})  and
(\ref{Small_Fluctuation_Assumption}) that the interaction energy is 
never much larger than  $\EJ$. Hence, at best 
the two-bit gate can be as fast as a single-bit operation.

\noindent
(2) It may be difficult to fabricate a nano-meter-scale inductor with
    the required inductance $L$, in particular, since it is not supposed to
    introduce stray capacitances. However, it is possible to realize
    such an element by a Josephson junction in the classical regime
    (with negligible charging energy) or an array of junctions. 

\noindent
(3) The design presented above does not permit performing singe- or two-bit
operations simultaneously on different qubits.  However, this becomes possible
in more complicated designs where parts of the many-qubit register are
separated, e.g., by switchable SQUIDs.

\noindent
(4) In the derivation of the qubit interaction presented here we
assumed a dissipation-less high-frequency oscillator mode. To minimize
dissipation effects, the circuit, including the inductor, should be made
of superconducting material. Still at finite frequencies some
dissipation will arise. To estimate its influence, the effect of an
Ohmic resistance $R$ in the circuit  has been analyzed by
\citeasnoun{Our_PRL}, with the result that the inter-qubit coupling
persists if the oscillator is underdamped, $R \ll \sqrt{L/N\Cqb}$. In
addition the dissipation causes dephasing. An estimate of the
resulting dephasing time can be obtained along the lines of the
discussion in Section~\ref{sec:Dephasing}. For a reasonably low-loss circuit
the dephasing due to the coupling circuit is weaker than the
influence of the external control circuit.

\begin{figure}
\vskip 0.4cm
\centerline{\hbox{\psfig{figure=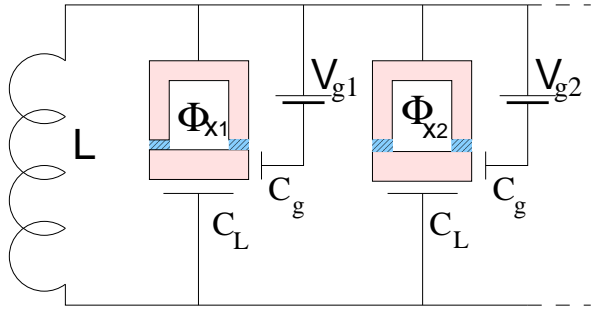,width=0.9\columnwidth}}}
\vskip 0.6cm
\caption[]{\label{Fig:ManyBits-NewDesign}
A register of charge qubits coupled to an inductor via separate capacitors
$C_L\sim\CJ$, independent from the gate capacitors $\Cg$.}
\end{figure}
\noindent
(5) The interaction energy~(\ref{HintGen}) involves via $E_L$ the
ratio of  $\CJ$ and $\Cqb$. The latter effectively screens 
the qubit from electromagnetic
fluctuations in the  voltage source's circuit, and hence should be
taken as low as possible (see
Section~\ref{sec:Dephasing}). Consequently, to achieve a reasonably high
interaction strength and hence speed of two-bit operations
a large inductance is needed. For 
typical  values of $\EJ\sim 100$~mK and $\Cg/\CJ\sim 0.1$ one needs an
inductance of  $L\ge 1\ \mu$H in order not to have the two-bit operation 
more than ten times slower than the  single-bit operation. However,
large values of the inductance are difficult to reach without
introducing large stray capacitances. To overcome this problem
\citeasnoun{Our_Fortschritte} suggested to use  separate gate capacitors
to couple the qubits to the inductor, as shown in
Fig.~\ref{Fig:ManyBits-NewDesign}. As long as the superconducting circuit of
the inductor is at most weakly dissipative, there is no need to screen
the qubit from the electromagnetic fluctuations in this
circuit, and one can choose $C_L$ as large as  $\CJ$ (still larger $C_L$ would
decrease the charging energy $E_C$ of the  superconducting box), which
makes the relevant capacitance ratio in  Eq.~(\ref{COUPLING_TERM}) of
order one. Hence, a fairly low inductance induces an interaction of
sufficient strength. For instance, for the  circuit parameters
mentioned above, $L\sim 10$~nH would suffice. At the same time,
potentially dephasing voltage fluctuations are screened by $\Cg\ll\CJ$.

\noindent
(6) So far we discussed manipulations on time scales much slower than the 
eigenfrequency of the $LC$-circuit, which leave the $LC$-oscillator
permanently  in the ground state. Another possibility is to use the
oscillator as a bus mode, similar to the techniques used for ion
traps. In this case an ac-voltage with properly chosen frequency
is applied to a qubit to entangle its quantum state  
with that of the $LC$-circuit (for instance, by exciting the 
oscillator conditionally on the qubit's state). Then by addressing
another qubit  
one can absorb the oscillator quantum, simultaneously exciting the
second qubit. As a result, a two-qubit unitary operation is performed.
This coupling via real excitations is a first-order process, as opposed 
to the second-order interaction (\ref{HintGen}). 
Hence, this method allows for faster two-qubit operations.
Apart from this technical advantage, the creation of entanglement between  
a qubit and an oscillator would by itself be a very interesting
experimental achievement \cite{Buisson_Hekking_Qubit_Oscillator}. 

\subsection{Experiments with Josephson charge qubits}
\label{subsec:Experiments_Charge}

Several of the concepts and properties described above have been
verified  in experiments. This includes the demonstration of
superpositions of charge states, the spectroscopic verification of the
spectrum, and even the  demonstration of coherent oscillations.

In a superconducting charge box the coherent tunneling of Cooper pairs
produces eigenstates which  are gate-voltage dependent superpositions
of charge states. This property  has been first observed, in a
somewhat  indirect way, in the dissipative current through
superconducting single-electron transistors. In this system single-electron
tunneling processes (typically treated  in perturbation theory) lead
to transitions between the eigenstates.  Since the eigenstates are
not pure charge states, the Cooper-pair  charge may also change in a
transition. In the resulting  combination of coherent Cooper-pair tunneling and
stochastic  single-electron tunneling the charge transferred is not
simply $e$ and the work done by the voltage source not simply
$eV$. (In an expansion in the Josephson coupling to
$n$-th order the charge $(2n+1)e$ is transferred.)  As a
result a dissipative current can be transferred at subgap voltages.  The
theoretical analysis predicted  a richly structured
$I$-$V$ characteristic at subgap voltages~\cite{Averin89,Geerligs,Siewert},
which has been qualitatively confirmed by  experiments
\cite{Geerligs,Parity,Hadley}. 

\begin{figure}
\centerline{\hbox{\psfig{figure=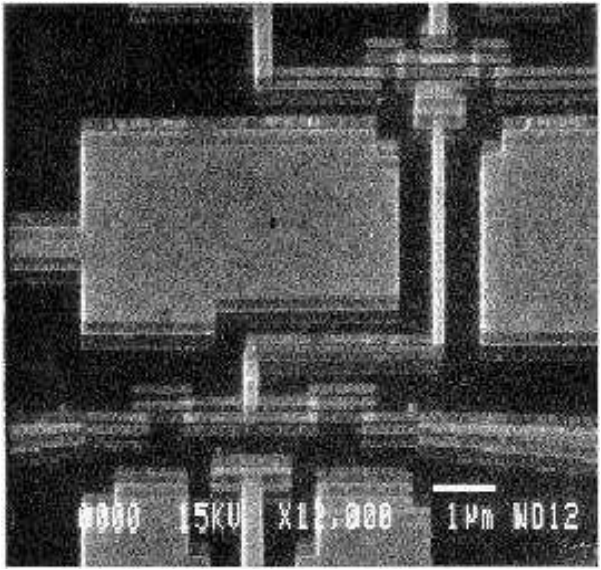,width=0.92\columnwidth}}}
\vskip 0.2cm
\caption[]{\label{Fig:BouchiatSetup}
Scanning electron micrograph of a Cooper-pair box coupled to a single-electron 
transistor used in the experiments of the Saclay group 
\cite{Bouchiat_PhD,BouchiatPhysScr}.}
\end{figure}

A more direct demonstration of eigenstates which arise as
superpositions of charge states was found in the Saclay
experiments~\cite{Bouchiat_PhD,BouchiatPhysScr}. In their setup
 (see Fig.~\ref{Fig:BouchiatSetup})
a single-electron transistor is coupled to a superconducting charge
box (in the same way as the measurement setup to be discussed in
Section~\ref{sec:Measurement}) and the expectation value of the charge
of the box was measured.  When the gate voltage is  
increased adiabatically this expectation value
increases in a series of rounded steps near half-integer
values of $\ng$.  At low temperatures the width of
this transition agrees quantitatively with the predicted ground
state properties of
Eqs.~(\ref{Eq:Magnetic_Hamiltonian},\ref{Eigen_Basis}).  At higher
temperatures, the  excited state  contributes, again as expected from
theory.

Next we mention the experiments of \citeasnoun{Nakamura} who
studied the superconducting charge box by spectroscopic means. When
exposing the system to radiation they found resonances  (in the
tunneling current in a suitable setup) at frequencies corresponding to the
difference in the energy between  excited and ground state, again in
quantitative agreement with the predictions of
Eq.~(\ref{Eq:Magnetic_Hamiltonian}). 

\begin{figure}
\centerline{\hbox{\psfig{figure=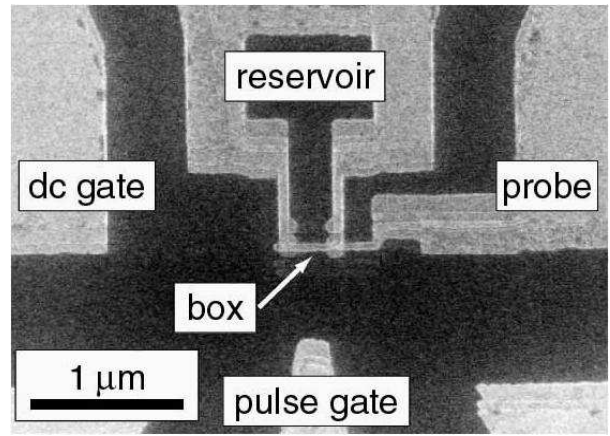,width=0.92\columnwidth}}}
\vskip3mm
\caption[]{\label{Fig:NakamuraSetup}
Micrograph of a Cooper-pair box with flux-controlled Josephson
junction and a probe junction  \cite{Nakamura_Nature}.}
\end{figure}

The most spectacular demonstration so far of the concepts of Josephson
qubits has been provided by \citeasnoun{Nakamura_Nature}. Their setup
is shown in Fig.~\ref{Fig:NakamuraSetup}. In the experiments the 
Josephson charge qubit is prepared far from the degeneracy point for
sufficiently long time to relax to the ground state. In this
regime  the ground state is close to a 
charge state, say, $\ket{\uparrow}$. Then the gate voltage is
suddenly switched to a different value. Let us, first, discuss the
case where it is switched precisely to the degeneracy point. Then the
initial state, a pure charge state, is an equal-amplitude
superposition of the ground state $\ket{0}$ and the excited state
$\ket{1}$. These two eigenstates have different energies, 
hence in time they acquire  different phase factors:
\begin{equation}
\label{coherentoscillation}
\ket{\psi(t)} = {\rm e}^{-iE_0\,t/\hbar}\ket{0} +  
{\rm e}^{-iE_1\,t/\hbar}\ket{1} \;.
\end{equation}
After a delay time $\Delta t$ the gate voltage is switched  back to
the original gate voltage.  Depending on the delay, the system then
ends up either in the ground state $\ket{\uparrow}$  [for
$(E_1-E_0)\Delta t/h = 2\nu\pi$ with $\nu$ integer], in the  excited state
$\ket{\downarrow}$ [for $(E_1 -E_0)\Delta t/h= (2\nu+1)\pi$], or in
general in 
a $\Delta t$-dependent superposition. The probability that, as a
result of this manipulation, the qubit is in the excited state
is measured by monitoring the current through a probe junction. 
In the experiments this current was averaged over many repeated
cycles, involving relaxation and switching processes,
 and the oscillatory
dependence on $\Delta t$ described above has been observed. 

In fact even more details of the theory have been quantitatively
confirmed.  For 
instance one expects and finds an oscillatory behavior also when the gate
voltage is switched to a point different from the degeneracy point,
with the frequency of oscillations being a function of this gate
voltage.  Secondly, the frequency of the coherent oscillations depends
on the Josephson coupling energy.  The latter 
can be varied, since the Josephson coupling is controlled by a flux-threaded
SQUID (see Fig.~\ref{Fig:BitwSQUID}).  Also this aspect has been verified
quantitatively.

The coherent oscillations with a period of roughly 100~psec could be
observed in the experiments of \citeasnoun{Nakamura_Nature} for at least
2~nsec~\footnote{In later experiments the same group reported phase
  coherence times as long as 5~nsec
  \cite{Nakamura_5nsec}.}.  This puts a lower limit on the 
phase coherence time  $\tau_\phi$ and, in fact, represents its first
direct measurement in the time domain.  Estimates show that a major
contribution to the dephasing is due to the measurement process by the
probe junction itself.  In the experiments so far the detector was
permanently coupled to the qubit and observed it continuously.  Still,
information about the quantum dynamics 
could be obtained since the coupling strength was optimized:  it was
weak enough not to destroy the  quantum time evolution too fast and
strong enough to produce 
a sufficient signal.  A detector which does not induce dephasing
during manipulations should significantly improve the operation of the
device.  In Section~\ref{sec:Measurement} we suggest to 
use a single-electron transistor, which performs a quantum measurement
only when switched to a dissipative state.

So far only experiments with single qubits have been demonstrated.
Obviously the next step is to couple two qubits and to create and
detect entangled states. Experiments in this direction have not been
successful yet, partially because of difficulties as, for instance, 
dephasing due to fluctuating background charges. However, the
experience with experiments with single qubits demonstrates that extensions
to coupled qubits should be possible as well.

\subsection{Adiabatic charge manipulations}
\label{subsec:Averin}

\begin{figure}
\vskip 0.3cm
\centerline{\hbox{\psfig{figure=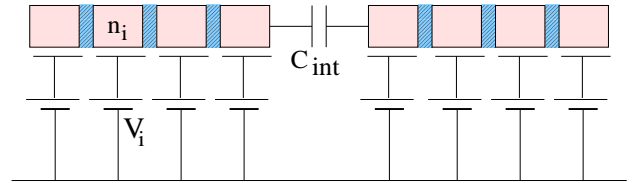,width=0.95\columnwidth}}}
\vskip 0.3cm
\caption[]{\label{Fig:Averin}
Two coupled qubits as proposed by \citeasnoun{Averin}.}
\end{figure}

Another qubit design, based on charge degrees of freedom in Josephson junction
systems was proposed by \citeasnoun{Averin}.  It also allows controlling the
two-bit coupling at the price of representing each qubit by a chain of
Josephson coupled islands.  The basic setup is shown in Fig.~\ref{Fig:Averin}.
Each superconducting island (with index $i$) is biased via its own
gate capacitor by a gate voltage $V_i$. The control of these voltages
allows moving the charges along the chain 
similar to the adiabatic pumping  of charges in junction arrays 
\citeaffixed{Pekola}{see e.g.}.  The capacitances of the Josephson junctions
as well as the gate capacitances are small enough so that the typical
charging energy prevails over the Josephson coupling. In this regime the
appropriate basis is that of charge states $\ket{n_1,n_2,\dots}$,
where $n_i$ is the number of extra Cooper pairs on island $i$.
There exist gate voltage configurations such that the two charge
states with the lowest energy are almost degenerate, while all other
charge states have much higher energy.  For instance, if all voltages
are equal except for the  voltages $V_m$ and $V_l$  at two sites, $m$
and $l$, one can achieve the situation where the states
$\ket{0,0,0,\dots}$ and $\ket{0,\dots,-1_m,0,\dots,1_l,\dots}$ are
degenerate.  The subspace of these  two charge states is used as the
logical Hilbert space of the qubit. They are coupled via the Josephson
tunneling across the $|m-l|-1$ intermediate junctions.

The parameters of the qubit's Hamiltonian can be tuned via the bias
voltages. Obviously, the bias energy $B_z(V_1,V_2,...)$ between these
two states can be changed via the local voltages,  $V_l$ and $V_m$.
Furthermore, the effective  tunneling amplitude $B_x(V_1,V_2,...)$ 
can be tuned by  adiabatic pumping of charges along the chain,
changing their distance $|m-l|$ and, hence, the effective Josephson
coupling, which depends  exponentially on this  distance.
(The Cooper pair must tunnel via $|m-l|-1$ virtual charge states with
much higher energy.) 

An inter-qubit interaction can be produced by placing a
capacitor between the edges (outer islands) of two qubits. If at
least one of the charges in each qubit is shifted close
towards this capacitor, the Coulomb interaction leads to an
interaction of the type $J_{zz}\sigma_z^1\sigma_z^2$. The resulting
two-bit Hamiltonian is of the form
\begin{equation}
\label{Eq:AVERIN_HAMILTONIAN}
{\cal H}=-\frac{1}{2}\sum_{j=1,2}
\left[
B_z^j(t)\sigma_z^j + B_x^j(t)\sigma_x^j
\right]
+ J_{zz}(t) \sigma_z^1\sigma_z^2
\ .
\end{equation}
For controlled manipulations of the qubit the coefficients of the
Hamiltonian are modified by adiabatic
motion of the charges along the junction array.
The adiabaticity is required to suppress
transitions between different eigenstates of the qubit system. 
While conceptually satisfying, this proposal appears
difficult to implement: It requires many gate voltages for each
qubit. Due to the complexity a high accuracy of the operations is
required. Its larger size as compared to simpler designs makes the
system more vulnerable to dephasing effects, e.g., due to
fluctuations of the offset charges.

Adiabatic manipulations of the Josephson charge qubit can lead to
Berry phases. \citeasnoun{Fazio_Berry} suggested that a Berry phase
can accumulate during suitable manipulations of a flux controlled
charge qubit with an asymmetric  dc-SQUID, and that it can be detected in  
an experiment similar to that of \citeasnoun{Nakamura_Nature}.
If the bare Josephson couplings of the SQUID loop are $\EJ^{1}$
and $\EJ^{2}$ the  effective Josephson  energy is given by
(cf. Eq.(\ref{EJPhi})) 
\begin{equation}
   - \EJ^1 \cos\left(\Theta + \pi\frac{\Phix}{\Phi_0}\right) - \EJ^2
   \cos\left(\Theta - \pi\frac{\Phix}{\Phi_0}\right) \; .
\end{equation}
Hence, the corresponding Hamiltonian of the qubit has all three components of
the effective magnetic field:  $B_x=(\EJ^1+\EJ^2)\cos(\pi\Phix/\Phi_0)$ and
$B_y=(\EJ^2-\EJ^1)\sin(\pi\Phix/\Phi_0)$, while $B_z$ is given by
Eq.(\ref{Eq:ChargeBit-Bz}).  With three non-zero field components, adiabatic
changes of the control parameters $\Vqb$ and $\Phix$ may result in $\bbox{B}$
enclosing a non-zero solid angle.  This results in a Berry phase shift,
$\gamma_{\rm B}$, between the ground and excited states.  In general, a dynamic
phase $\int\Delta E(t) dt$ is also accumulated in the process.  To single
out the Berry phase, \citeasnoun{Fazio_Berry} suggested to encircle the loop in
parameter space back and forth, with a NOT operation performed in between.
The latter exchanges the ground and excited state, and, as a result,
the dynamic phases accumulated during both paths cancel.  At the
same time the Berry phases add up to $2\gamma_{\rm  B}$.  This phase
shift can be 
measured by a procedure similar to that used by
\citeasnoun{Nakamura_Nature}:  the system is prepared in a charge state away
from degeneracy, abruptly switched to the degeneracy point where adiabatic
manipulations and the NOT gate are performed, and then switched
back.  Finally, the average charge is measured.  The probability to
find the qubit in the excited charge state, $\sin^2 2\gamma_{\rm B}$,
reflects the Berry phase.

The experimental demonstration of topological phases in Josephson
junction devices would constitute a  new class of ``macroscopic''
quantum effects in these systems. They can be performed with a single
Josephson qubit in a design as used by \citeasnoun{Nakamura_Nature}
and thus appear feasible in the near future.

\section{Qubits based on the flux degree of freedom}
\label{sec:FluxQubits}
In the previous Section we described the quantum dynamics of low
capacitance Josephson devices where the 
charging energy dominates over the Josephson energy, $E_{\rm C} \gg
E_{\rm J}$, and the relevant quantum degree of freedom
is the charge on superconducting islands.     
We will now review the quantum properties of superconducting devices
in the opposite regime, $E_{\rm J} \gg E_{\rm C}$, where
the flux is the appropriate quantum degree of freedom. These systems
 were proposed by \citeasnoun{Caldeira-Leggett} 
in the mid 80s as test objects to study various quantum mechanical
effects. This includes the `macroscopic quantum tunneling' of the
phase (or flux) as well as resonance tunneling. Both had been observed
in several experiments~\cite{Voss-Webb,Clarke_PRB,Clarke_Science,Lukens}. 
Another important quantum effect has been reported recently:
The groups in Stony Brook \cite{Friedman_Cats} and in Delft 
\cite{Delft_Cats} demonstrated in experiments the avoided level crossing  
due to coherent tunneling of the flux in a double well potential.
In principle, all other manipulations discussed in the previous section
should be possible with Josephson flux devices as well. They have the
added advantage not to be sensitive to fluctuations in the background
charges. However, attempts to observe `macroscopic quantum coherent
oscillations' in Josephson flux devices have not been successful
yet~\cite{LeggettMQC,Tesche}.  
   
\subsection{Josephson flux (persistent current) qubits}
\label{subsec:FluxQubits}

\begin{figure}
\vskip 0.3cm
\parbox{0.4\columnwidth}{{\LARGE\sf a}\vskip-3mm
\centerline{\hbox{\psfig{figure=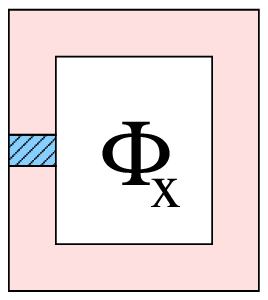,width=0.25\columnwidth}}}}%
\parbox{0.5\columnwidth}{{\LARGE\sf b}\vskip-3mm
\centerline{\hbox{\psfig{figure=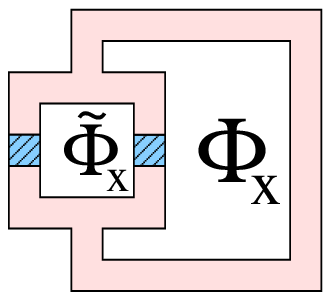,width=0.35\columnwidth}}}}
\vskip 0.4cm
\caption[]{\label{Fig:FluxQubit}
a) The rf-SQUID, a simple loop with a Josephson junction, forms the
simplest Josephson flux qubit.
b) An improved design of a flux qubit. The flux $\tilde\Phix$ in the smaller
loop controls the effective Josephson coupling of the rf-SQUID.} 
\end{figure}

We consider superconducting ring geometries interrupted by one or
several Josephson junctions. In these systems persistent currents
flow and magnetic fluxes are enclosed. The simplest design of
these devices is an rf-SQUID, which is formed by a loop with one
junction, as shown in Fig.~\ref{Fig:FluxQubit}~a.  The phase difference
across the junction is related to the flux $\Phi$ in the loop (in
units of the flux quantum $\Phi_0 = h/2e$) by 
$\varphi/2\pi = \Phi/\Phi_0 + \mbox{integer}$. 
An externally applied flux $\Phix$ biases the system. 
Its Hamiltonian, with Josephson coupling, charging energy, and
magnetic contributions taken into account, thus reads  
\begin{equation}
{\cal H} = -\EJ\cos\left(2\pi\frac{\Phi}{\Phi_0}\right)
+\frac{(\Phi-\Phix)^2}{2L}
+\frac{Q^2}{2\CJ}
\; .
\label{rf-HAM}
\end{equation}
Here $L$ is the self-inductance of the loop and $\CJ$ the capacitance
of the junction. The charge $Q=-i \hbar \partial/\partial\Phi$ on the
leads is canonically conjugate to the flux $\Phi$.

If the self-inductance is large, such that $\beta_L\equiv \EJ
/(\Phi_0^2/4\pi^2 L)$ is larger than 1  
and the externally applied flux $\Phix$
is close to $\Phi_0/2$, the first two terms in the Hamiltonian form a
double-well potential near $\Phi=\Phi_0/2$.  At low temperatures only
the lowest states in the two wells contribute. Hence the
reduced Hamiltonian of this effective two-state system
again has the form (\ref{Eq:Magnetic_Hamiltonian}), 
${\cal H}_{\rm ctrl}= -\frac{1}{2}  B_z \hat\sigma_z - {1\over
        2} B_x \hat\sigma_x$. 
The diagonal term $B_z$ is the bias, i.e., the asymmetry of
the double well potential, given for $\beta_L-1\ll1$ by 
\begin{equation}
B_z(\Phix)= 4\pi\sqrt{6(\beta_L-1)}\; \EJ\,
\left(\Phix/\Phi_0-1/2 \right) \;.
\end{equation} 
It can be tuned by the applied flux $\Phix$. The
off-diagonal term $B_x$ describes the tunneling amplitude between the
wells. It depends on the height of the barrier and thus on $\EJ$. This
Josephson energy, in 
turn, can be  controlled if the junction is replaced by a dc-SQUID, as
shown in Fig.~\ref{Fig:FluxQubit}~b, introducing the flux
$\tilde\Phix$ as another control variable~\footnote{See \cite{Mooij} for
suggestions how to control $\tilde\Phix$ independent of $\Phix$.}. 
With these two external control parameters the elementary single-bit
operations, i.e., $z$- and $x$-rotations can be performed, 
equivalent to the manipulations described for charge qubits in the
previous section. Also for flux qubits we can either perform the
operations by sudden switching of the external fluxes $\Phix$ and
$\tilde\Phix$ for a finite time, or we can use ac-fields and resonant
pulses. To enable coherent manipulations the parameter $\beta_L$ should be 
chosen larger than unity (so that two wells with well-defined levels
appear) but not much larger, since the resulting large separation of the wells 
would suppress the tunneling.

The rf-SQUID described above had been discussed in the mid 80s as a realization
of a 2-state quantum system.  Some features of macroscopic quantum behavior
were demonstrated such as `macroscopic quantum tunneling' (MQT) of the flux,
resonant tunneling and level quantization
\cite{Voss-Webb,Clarke_PRB,Clarke_Science,Lukens,Silvestrini97}.
However, only very recently coherent superpositions of macroscopically
different flux states have been demonstrated~\cite{Friedman_Cats,Delft_Cats}.

The group in Stony Brook \cite{Friedman_Cats}  probed spectroscopically
the superposition of {\it excited} states in different wells.  The
rf-SQUID used had self-inductance $L=240$~pH and
$\beta_L=2.33$.  A substantial separation of the minima of the
double-well potential (of order $\Phi_0$) and a high inter-well
barrier made the tunnel coupling between the lowest states in the
wells negligible.  However, both wells contain a set of higher
localized levels -- under suitable conditions one state in each well
--  with relative energies also controlled by $\Phix$ and
$\tilde\Phix$. Being closer to the top of the barrier these states 
mix more strongly and form  eigenstates,  which are
superpositions of localized flux states from different wells.
External microwave radiation was used to pump the system from a 
well-localized ground state in one well to one of these eigenstates.
The energy spectrum of these levels was studied  for
different biases $\Phix$, $\tilde\Phix$, and the properties of the
model~(\ref{rf-HAM}) were confirmed.  In particular, the level splitting
at the degeneracy point indicates a superposition of
distinct quantum states.  They differ in a macroscopic way:  the
authors estimated that two superimposed flux states differ in flux by
$\Phi_0/4$, in current by 2--3~$\mu$A, and in magnetic moment by
$10^{10}~\mu_{\rm B}$. 

The Delft group \cite{Delft_Cats} performed microwave spectroscopy experiments
on a similar but much smaller three-junction system described below.  The {\it
ground} states in two wells of the Josephson potential landscape were probed.
The obtained results verify the spectrum of the qubit and
the level repulsion at the degeneracy point expected
from the model Hamiltonian (\ref{Eq:Magnetic_Hamiltonian}) with the
parameters $B_x$, $B_z$ calculated from the potential
(\ref{Eq:Mooij-Potential}). Similar to the experiments of
\citeasnoun{Friedman_Cats}, this provides clear evidence for superpositions of
macroscopically distinct phase states.

In spite of this progress, attempts to observe `macroscopic quantum coherence',
i.e., the coherent oscillations of a quantum system prepared in a superposition
of eigenstates have not been successful so far \cite{LeggettMQC,Tesche}.  A
possible reason for this failure has been suggested recently by
\citeasnoun{Mooij}.  They argue that for the rf-SQUID designs considered so far
the existence of the double-well potential requires that $\beta_L>1$ which
translates into a sufficiently high product of the critical current of the
junction and its self-inductance.  In practice, only a narrow range of circuit
parameters is useful, since high critical currents require a relatively large
junction area resulting in a high capacitance which suppresses tunneling.  A
high self-inductance of the rf-SQUID can be achieved only in large loops.  This
makes the system very susceptible to external noise.

\begin{figure}
\vskip3mm
{\LARGE\sf a}\hskip0.4\columnwidth{\LARGE\sf b}
\vskip-3mm
\parbox{0.45\columnwidth}{%
\centerline{\hbox{\psfig{figure=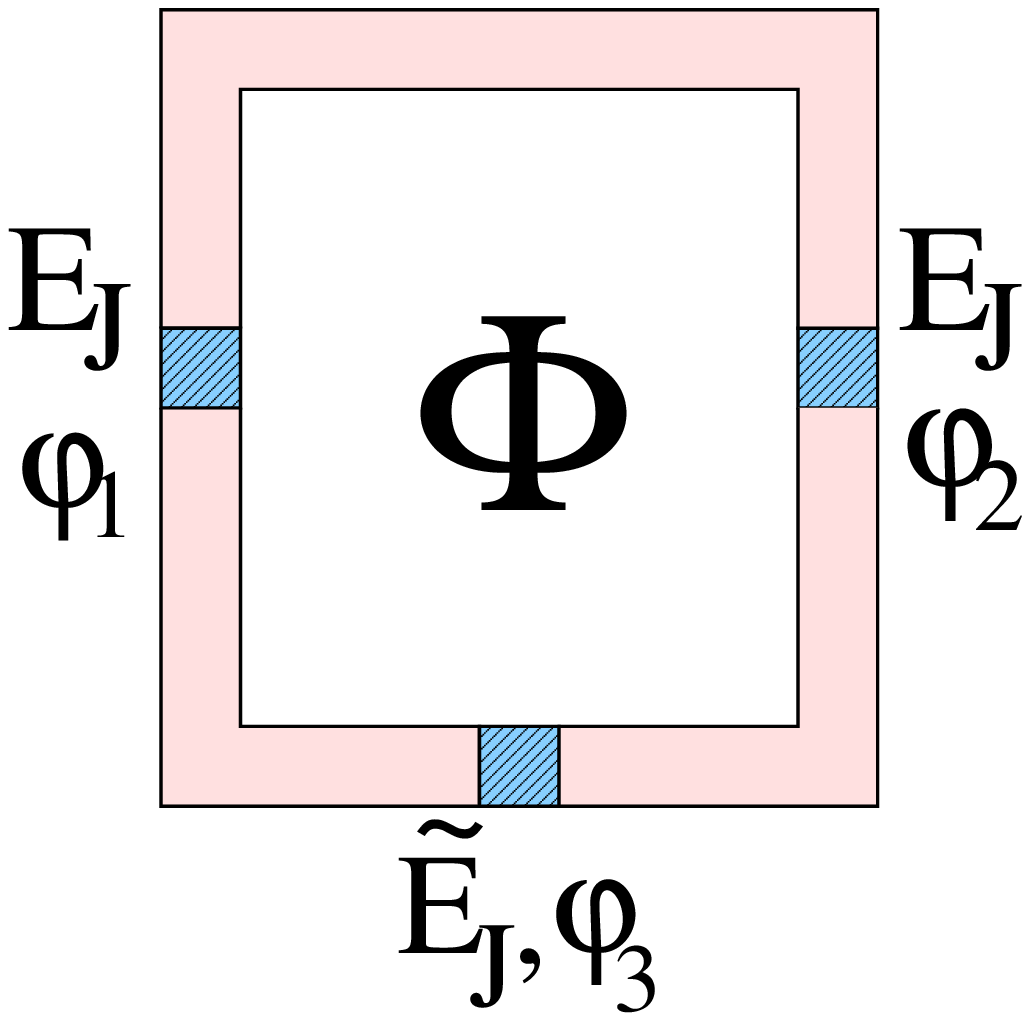,width=0.35\columnwidth}}}}
\parbox{0.5\columnwidth}{%
\centerline{\hbox{\psfig{figure=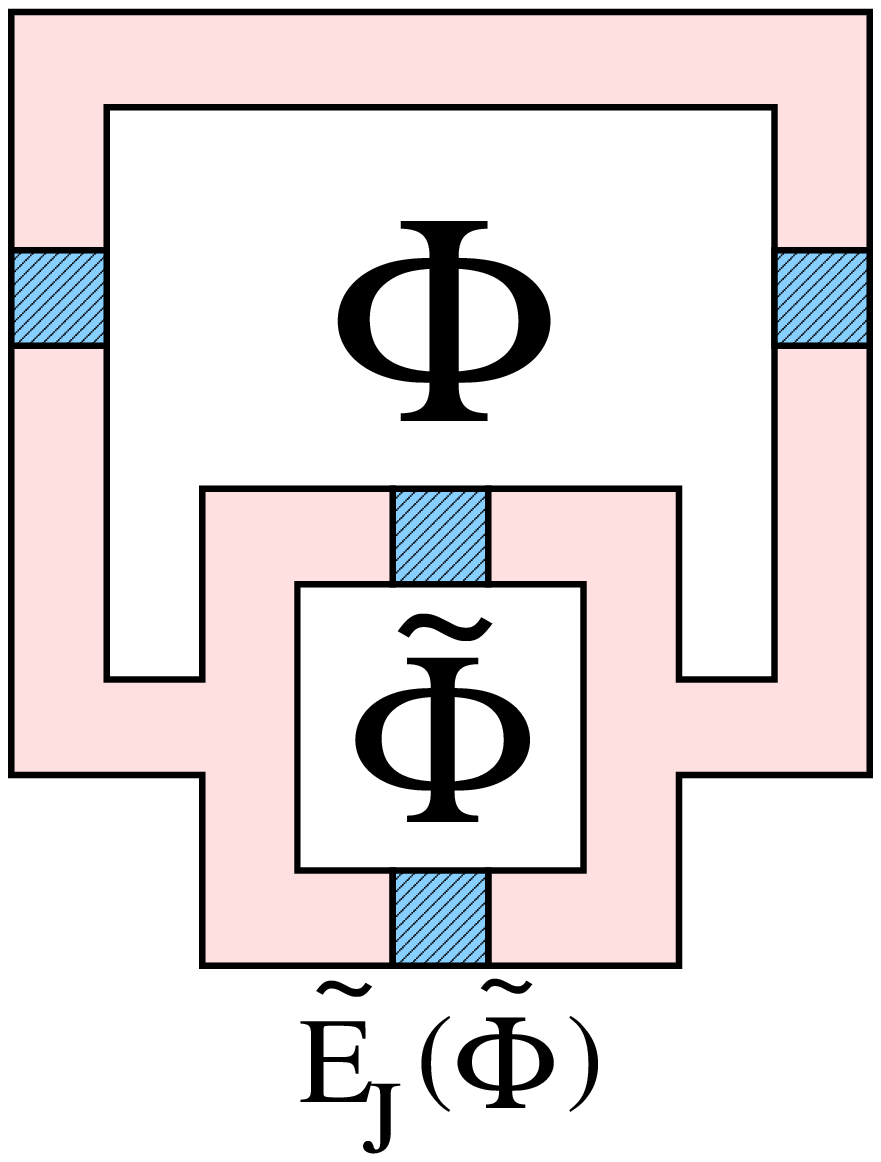,width=0.35\columnwidth}}}}

{\LARGE\sf c}
\centerline{\hbox{\psfig{figure=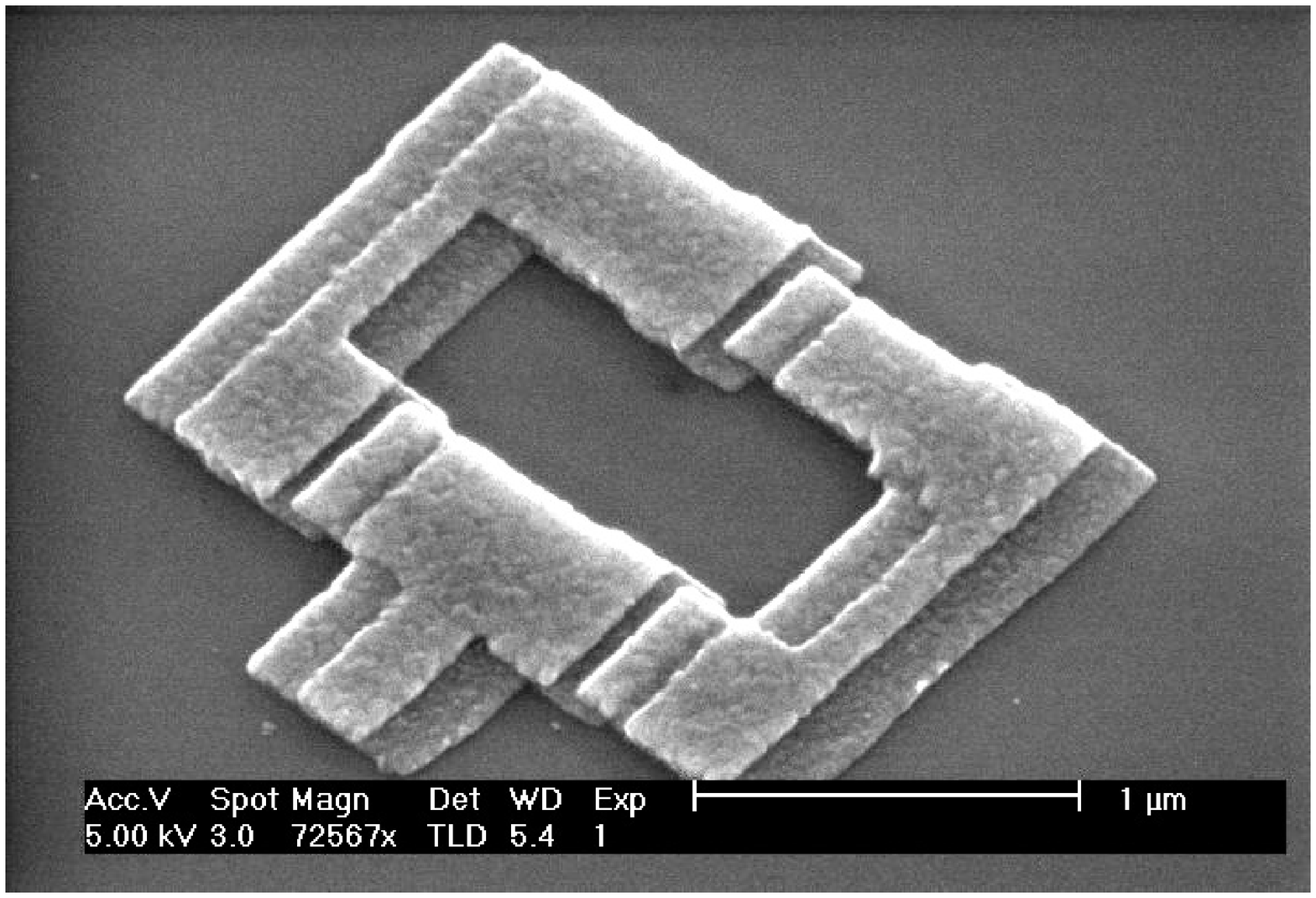,width=0.8\columnwidth}}}

\vskip 0.5cm
\caption[]{\label{Fig:Mooij3}
a) and c) A 3-junction loop as a flux qubit~\cite{Mooij}. The reduced size and 
lower inductance of this system as compared to earlier designs 
(e.g., Fig.\ \ref{Fig:FluxQubit}~a) reduce the coupling to
the external world and hence dephasing effects.
b) Multi-junction flux qubit with a controlled Josephson coupling~\cite{Mooij}.
Control over two magnetic fluxes, $\Phi$ and $\tilde\Phi$, allows one
to perform all single-qubit logic operations.
}
\end{figure}

To overcome this difficulty \citeasnoun{Mooij} and \citeasnoun{Blatter4} 
proposed to use a smaller superconducting loop with three or four junctions, 
respectively. Here we discuss the 3-junction circuit shown in 
Fig.~\ref{Fig:Mooij3}~a,c.
In this low-inductance circuit the flux through the loop remains close
to the externally applied value, $\Phi=\Phix$. Hence the phase
differences across the junctions are constrained by
$\varphi_1+\varphi_2+\varphi_3=2\pi\Phix/\Phi_0$,
leaving $\varphi_1$ and $\varphi_2$ as independent dynamical
variables. In the plane spanned by these two variables the Josephson
couplings produce a potential landscape given by
\begin{eqnarray}
U(\varphi_1,\varphi_2)=&&-\EJ\cos\varphi_1-\EJ\cos\varphi_2
\nonumber\\
&&
-\tilde\EJ\cos(2\pi\Phix/\Phi_0-\varphi_1-\varphi_2)
\, .
\label{Eq:Mooij-Potential}
\end{eqnarray}
If $\tilde\EJ/\EJ>0.5$, a double-well potential is formed
within each $2\pi\times 2\pi$ cell in the phase plane. For an optimal
value of $\tilde\EJ/\EJ\approx 0.7$--$0.8$ the cells are separated by high
barriers, while tunneling between two minima within one cell is still
possible. The lowest states in the wells
form a two-state quantum system, with two different current
configurations. \citeasnoun{Mooij} and \citeasnoun{MooijPRB} discuss
junctions with  $\EJ\sim 2$~K and $\EJ/\EC\sim 80$ and  loops of
micrometer size with 
very small self-inductance $L\sim 5$~pH (which can be neglected 
when calculating the energy levels). Typical qubit operation
parameters are the level splitting $B_z\sim 0.5$~K 
and the tunneling amplitude $B_x\sim50$~mK. For the optimal choice of
$\tilde\EJ/\EJ$ the  two minima differ in 
phases by an amount of order $\pi/2$.
Due to the very low inductance and the relatively low
critical current  $I_{\rm c}\sim 200$~nA this translates into
a flux difference of $\delta\Phi\sim L I_{\rm c}\sim 10^{-3}\Phi_0$. 
While this corresponds to a still `macroscopic' magnetic moment of 
$10^4$ to $10^5~\mu_{\rm B}$,  the two basis states 
are similar enough to make the coupling to external fluctuating  
fields  and hence the dephasing effects weak 
(for a further discussion, see Section~\ref{sec:Dephasing}). 
In this respect the new design is qualitatively superior to the simple
rf-SQUID. 


\subsection{Coupling of flux qubits}

\begin{figure}
\vskip3mm
\centerline{\hbox{\psfig{figure=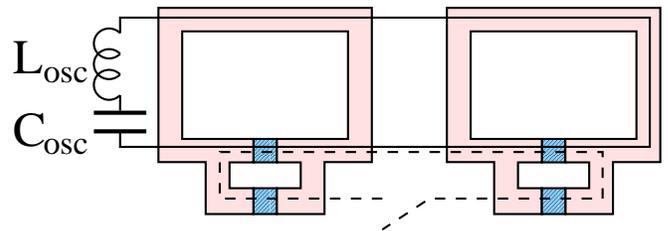,%
width=1\columnwidth}}}

\vskip 0.5cm
\caption[]{\label{Fig:CoupledFluxQubits}
Flux qubits coupled in two ways. The dashed line induces a direct inductive 
coupling. Alternatively, an inter-qubit coupling is provided by the
$LC$-circuit indicated by a solid line.
}
\end{figure}

In order to couple different flux qubits one can use a direct
inductive coupling \cite{Mooij,MooijPRB} as shown by the dashed line in
Fig.~\ref{Fig:CoupledFluxQubits}.  A mutual inductance between the
qubits can be established in different ways.  The dashed loop shown in
the figure couples the currents and fluxes in the lower parts of the
qubits.  Since fluxes through these loops control the barrier heights
of the double-well potentials, this gives rise to the interaction term
$\propto\hat\sigma_x^1 \hat\sigma_x^2$. Placing the loop differently
produces in addition contributions to the interaction Hamiltonian of the form
$\hat\sigma_z^1 \hat\sigma_z^2$.
The typical interaction energy is of order $MI_{\rm c}^2$ where $M$ is
the mutual inductance and $I_{\rm c}=(2\pi/\Phi_0)\EJ$ is the critical
current in the junctions.  For their design, \citeasnoun{Mooij}
estimate the typical interaction energy to be of order $0.01 \EJ\sim
50$~mK in frequency units, i.e.,  of the order of single-qubit energies.
For a typical rf-SQUID \cite{Friedman_Cats} this coupling can be even
stronger than the tunneling rate between the flux states of the SQUID.

In the simplest form this interaction is always turned on.  To turn it
off completely, one needs a switch which has to be controlled
by high-frequency pulses. The related coupling to the 
external circuit leads to
decoherence (see the discussion at the end of this Section).  An
alternative is to keep the interaction turned on constantly and use ac
driving pulses to induce coherent transitions between the levels of
the two-qubit system \citeaffixed{Our_PRL,Mooij}{cf.}.  
A disadvantage of this approach is that permanent couplings result in
an unwanted accumulation of relative phases between 
the 2-qubit states even in the idle periods. Keeping track 
of these phases, or their suppression by repeated refocusing pulses
(see Section~\ref{sec:Dephasing}), require a high precision and
complicate the operation.

A controllable inter-qubit coupling without additional switches is
achieved in the 
design shown by the solid line in Fig.~\ref{Fig:CoupledFluxQubits}
\cite{Our_JLTP}.  The coupling is mediated by an $LC$-circuit, with
self-inductance $L_{\rm osc}$ and capacitance $C_{\rm osc}$, which is coupled
inductively to each qubit. Similar to the design of the charge qubit
register in Section~\ref{subsec:Controlled_2Bit}, the coupling depends
on parameters of of individual qubits and can be controlled in this
way. The effective coupling can be found again by integrating out the fast 
oscillations in the $LC$-circuit. It can be understood in a simple way  
by noting that in the limit $C_{\rm osc}\to 0$ the qubits 
establish a voltage drop across the inductor, $V=\sum_i
M\dot\Phi_i/L$, and the Hamiltonian for the oscillator mode is ${\cal H}_{\rm
osc}=\Phi^2/2L_{\rm osc} + Q^2/2C_{\rm osc} - VQ$, with the charge
$Q$ being conjugate to the flux $\Phi$ through the $LC$-circuit.
Here $\Phi_i$ is the flux in the loop of qubit $i$, $L$ is the
self-inductance of the loop and $M$ is its mutual inductance with the
$LC$-circuit. Continuing as described in
Section~\ref{subsec:Controlled_2Bit} we obtain
the inter-qubit interaction term $-C_{\rm osc}V^2/2$.  In the
limit of weak coupling to the $LC$-circuit, we have
$\dot\Phi_i=\frac{i}{\hbar}[{\cal H}_i,\Phi_i]=\delta\Phi_i B_x^i
\hat\sigma_y^i/\hbar$, where $\delta\Phi_i$ is the separation between
two minima of the potential and $B_x^i$ is the tunneling amplitude.
Hence, the interaction is given by
\begin{equation}
{\cal H}_{\rm int}=-\pi^2 \left(\frac{M}{L}\right)^2
\sum\limits_{i<j} \frac{\delta\Phi_i\delta\Phi_j}{\Phi_0^2}
\frac{B_x^i B_x^j}{e^2/C_{\rm osc}}\; \hat\sigma_y^i\hat\sigma_y^j
\;.
\end{equation}

To turn off the interaction one should suppress the tunneling amplitudes  
$B_x^i$.  This can be done with exponential precision 
by increasing the height of the potential barrier via $\tilde \Phi_x$. Note, 
that in this case also unwanted fluctuations of $B_x^i$ and resulting
dephasing effects are exponentially suppressed. All needed single and
two-qubit manipulations can be performed by turning on the fields
$B_x^i$ and $B_z^i$, in complete analogy to what we discussed in
Section~\ref{subsec:Controlled_2Bit}. We also encounter the equivalent
drawbacks: the design shown in Fig.~\ref{Fig:CoupledFluxQubits}
does not allow simultaneous manipulations on different qubit pairs, 
and the conditions of high oscillator frequencies and weak
renormalization of qubit parameters by the coupling, similar  
to Eqs.(\ref{High_Frequency_Condition}), (\ref{Small_Fluctuation_Assumption}), 
limit the two-qubit coupling energy.
The optimization of this coupling requires $\sqrt{L_{\rm osc}/C_{\rm 
osc}}\approx R_{\rm K} (\delta \Phi/\Phi_0)^2 (M/L)^2$ and
$\omega_{LC}$ not far above the qubit frequencies. For rf-SQUIDs
\cite{Friedman_Cats} the resulting coupling can reach the same
order as the single-bit terms. On the other hand, for the design of
\citeasnoun{Mooij}, where two basis phase states differ only slightly
in their magnetic properties, the coupling term is much weaker than
the single-bit energies. 

\subsection{``Quiet'' superconducting phase qubits}

The circuits considered so far in this Section are vulnerable to
external noise. First, they need for their operation an external bias
in the vicinity of $\Phi_0/2$, which should be kept stable for the time of
manipulations.  In addition, the two basis flux states of the qubit
have different current configurations, which may lead to magnetic
interactions with the environment and possible crosstalk between
qubits.  To a large extent the latter effect is suppressed already in
the design of \citeasnoun{Mooij}. To further reduce these problems
several designs of so-called ``quiet'' qubits have been suggested 
\cite{Blatter,Blatter2,Zagoskin,Blais_Zagoskin_DW}.  They are based on
intrinsically doubly degenerate systems, e.g.,  Josephson junctions
with d-wave leads and energy-phase relation (e.g., $\cos2\phi$) with
two minima, or the use of $\pi$-junctions which removes the need for a
constant magnetic bias near $\Phi_0/2$.  As a result the relevant two
states differ only in their distribution of {\it internal\/} currents in
the Josephson junctions while {\it external\/} 
loops carry no current.  As a result the coupling of the qubit to the
electromagnetic environment is substantially reduced and coherence is
preserved longer. 

The mentioned designs are similar and we discuss them in parallel.
\citeasnoun{Blatter} suggested to use an SD tunnel junction with the
s-wave lead matched to the $(110)$ boundary of the d-wave
superconductor.  In this geometry the first harmonic,
$\propto\cos\varphi$, in the Josephson coupling vanishes due to
symmetry reasons, and one obtains a bistable system with the potential
energy $\EJ\cos2\varphi$ and minima at $\pm\varphi_0$
with $\varphi_0=\pi/2$.  (A similar current-phase relation was
observed recently by \citeasnoun{Ilichev} in a DD junction with a mismatch
angle of $45^\circ$.)
\citeasnoun{Zagoskin} proposed to use DND, or D--grain-boundary--D Josephson
junctions formed by two $d$-wave superconductors with different spatial
orientation of the order parameter, connected by a normal metal.
The energy-phase relation for such junctions also has two degenerate minima, 
at the phase differences $\pm\varphi_0$. The separation $2\varphi_0$
of these minima, and hence the tunneling amplitude, are controlled by
the mismatch angle of the d-wave leads. 

\begin{figure}
\centerline{\hbox{%
\psfig{figure=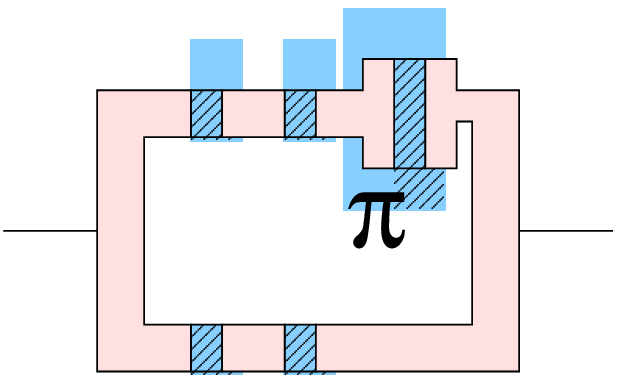,width=0.5\columnwidth}}}

\vskip 0.5cm
\caption[]{\label{Fig:Blatter5}
A five-junction loop, a basic bistable element of a 
``quiet'' superconducting qubits \cite{Blatter2}, is made of four
ordinary junctions and one stronger $\pi$-junction. In two stable
configurations the phase difference across this element is $\pm\pi/2$. 
}
\end{figure}

In a later development a `macroscopic analogue' of d-wave qubits was discussed 
\cite{Blatter2}. Instead of an SD-junction, it is based on a
five-junction loop, shown in Fig.~\ref{Fig:Blatter5}, which contains
one strong $\pi$-junction and four ordinary junctions. The presence of
the $\pi$-junction is equivalent to magnetically biasing the loop with
a half superconducting flux quantum. Four other junctions,
frustrated by the $\pi$-phase shift, have two lowest-energy states
with the phase difference of $\pm\pi/2$ between the external legs in the
figure. In this respect the 5-junction loop is similar to the
SD-junction discussed above and can be called a $\pi/2$-junction.  

In all these designs the bistability is a consequence of the 
time-reversal symmetry (which changes the signs of all the phases)
of the Hamiltonian.  Thus the degeneracy persists also in systems
containing different   Josephson junctions, although the phase
differences in the two lowest-energy states and their separation can
change. If charging effects with $\EC\ll\EJ$ are taken included,
one arrives at a double-well system with tunneling between the wells.
Such a qubit can be operated by connecting or disconnecting it from
external elements, as described below.

The first issue to be addressed is the question how to store the qubit's
state, i.e., how to freeze the evolution.  This can be achieved by
connecting the qubit in parallel to a large capacitor
\cite{Blatter}.  This makes the phase degree of freedom very massive,
thus suppressing the tunneling and restoring the needed degeneracy.
In order to perform a $\hat\sigma_x$-rotation the inter-well
tunneling is turned on by disconnecting the capacitor. This means a
switch is needed in the circuit.

The $\hat\sigma_z$-rotation or phase shift can be accomplished by
lifting the degeneracy between the wells.  This can be done by
connecting another, much stronger $\pi/2$-junction (DS-junction  
or 5-junction loop) and a weak ordinary, s-wave junction (with
Josephson energy $\propto\cos\varphi$) in series to the qubit, to form
a closed loop. This again requires a switch. The  
auxiliary $\pi/2$-junction shifts the phase differences of the
potential minima of the  qubit to $0$ and $\pi$. Hence the s-junction
is in the ground state or frustrated depending on the qubit's phase
drop. The corresponding energy difference produces the needed phase
shift between two qubit's states. 

To perform two-qubit manipulations and control the entanglement
\citeasnoun{Blatter} proposed to form a loop, connecting in series two qubits
and one s-junction with weak Josephson coupling $\EJ^s\ll\EJ$.  The phase
state of each qubit is characterized by the phase difference of
$\pm\varphi_0$, i.e., the total phase drop on the qubits is equal to
$\pm2\varphi_0$ or $0$ depending on whether the qubits are in the same
state or in a different ones.  When the connection between the qubits and
the s-junction is turned on, this phase drops across the s-junction,
and its energy differs by
$\EJ^s(1-\cos2\varphi_0)$ for the states $\ket{00}$, $\ket{11}$ as compared
to the states $\ket{01}$, $\ket{10}$. The net effect is an Ising-type
interaction between the pair of qubits, which allows
performing unitary two-qubit transformations.

Another way of operation was discussed by
\citeasnoun{Blais_Zagoskin_DW}.  They suggested to use a magnetic force
microscope tip for single-bit manipulations (local magnetic field lifts the
degeneracy of two phase states) and for the read-out of the phase state.  The
tip should be moved towards or away from the qubit during manipulations.  The
short time scales of qubit operation make this proposal difficult to realize.

Even in ``quiet'' designs, in both SD and DD systems, there are microscopic
persistent currents flowing inside the junctions which differ for the two
logical states \cite{Blatter2,Zagoskin}.  These weak currents still
couple to the outside world and to other qubits, thus spoiling the
ideal behavior. Furthermore, all the designs mentioned require
externally operated switches to connect and disconnect qubits. We
discuss the associated problems in the following  subsection.

To summarize, the ``quiet'' designs require rather complicated
manipulations as well as circuits with many junctions, including
$\pi$-junctions or d-wave  
junctions, which are difficult to fabricate in a controlled and
reliable way. In  addition, many constraints imposed on the circuit
parameters (in particular, on   
the hierarchy of Josephson couplings) appear difficult to satisfy.
In our opinion the ``quiet'' phase qubit designs belong to a higher complexity
class than the previously discussed charge and flux qubits, and their
experimental realization may remain a challenge for some time.

\subsection{Switches}

Switches may be used in variety of contexts in quantum nano-circuits.
They are needed, e.g., for a direct capacitive coupling between charge
qubits or magnetic  coupling of flux qubits.  They are also a major
tool for controlling the dynamics of ``quiet'' qubits.  Ideal switches
should decouple qubits from the environment and at the same time let
through control signals.  They should operate on the very fast time
scale of the qubit dynamics and have a high ``switching ratio'', that
is the ratio of the interaction with the switch in the on- or
off-state. Such switches are hard to realize.  In this subsection we
compare the characteristics of several Josephson-junction-based
switches and the associated problems.

Possible switches are dc-SQUIDs as well as SSETs  (single-Cooper-pair
transistors) in a mode where they act as
Josephson junctions with an externally 
controlled coupling.  Then the switching ratio is the ratio of the minimal and
maximal values of the coupling.  In a dc-SQUID with Josephson energies of its
junctions equal to $\EJ^1$ and $\EJ^2$, this ratio is
$(\EJ^1-\EJ^2)/(\EJ^1+\EJ^2)$. It reached a value below 1\% in the
experiment of \citeasnoun{Lukens}. However, fast switching of the bias
flux may be difficult to perform. In a SSET the effective 
coupling is controlled by a gate voltage, which can be switched
fast. However, the switching ratio of order  
$\EJ/\EC$ ($\EJ$ and $\EC$ are characteristics of the SSET) is
hardly below several percent. These non-idealities lead
to unwanted interactions when the switch is supposed to be
disconnected. 

Since a dc-SQUID requires an external bias to be operated as a switch,
\citeasnoun{Blatter2} suggested a similar construction with the bias
provided by $\pi/2$-junctions instead of an external magnetic field.
Namely, one can insert two $\pi/2$-junctions into one arm of the
SQUID-loop.  Depending on whether the phase drops across these
junctions are equal or opposite, they simulate an external bias of a
half flux quantum or no bias.  Accordingly, the Josephson couplings of
two s-junctions in the SQUID add up or cancel each other.  The
switching is realized via a voltage pulse which drives one
$\pi/2$-junction between $+\pi/2$- and
$-\pi/2$-states. \citeasnoun{Blatter2} also 
suggest to use an array of $n$ such switches, reducing the overall
Josephson coupling in the off-state by a factor
$[(\EJ^1-\EJ^2)/\EC]^n$.  Unfortunately, in the on-state the overall coupling
through the array  is also reduced with growing $n$, although this
reduction may be weaker than in the off-state, i.e., the switching ratio
increases with $n$. Still the quality of the switch in the on-state is
reduced.  Moreover, to operate the switch one would need to send 
simultaneously voltage pulses to $n$ intermediate elements which
complicates the operation. 
Note that this design is reminiscent of the qubit design proposed by
\citeasnoun{Averin} which is presented in
Subsection~\ref{subsec:Averin}.
However, while \citeasnoun{Blatter2} suggest to control the coupling, $\propto
(\EJ/\EC)^n$, by controlling $\EJ$, \citename{Averin} proposes to change the 
distance $n$ of the tunneling process.
 
While switches of the type as described above may be useful in first
experiments with simple quantum
nano-circuits, further work is needed before they can be used in more 
advanced designs which require high precision of manipulations and phase 
coherence for a long time.

\section{Environment and dissipation}
\label{sec:Dephasing}

\subsection{Identifying the problem}

For an ideal quantum system the time evolution is described by deterministic,
reversible unitary operations.  The concepts of quantum state engineering and
computation heavily rely on this quantum coherence, with many potential
applications requiring a large number of coherent
manipulations of a large number of qubits. On the other hand,
for any real physical quantum system the time evolution may be
disturbed in various ways, and the number of coherent manipulations is
limited. Possible sources of errors are inaccuracies in the preparation
of the initial state, inaccuracies 
in the manipulations (logic gates), uncontrolled couplings between qubits,
undesired excitations out of the two-state Hilbert space \cite{Saro_Leakage},
and -- unavoidable in devices which are to be controlled externally --
interactions with the environment.  Due to the coupling to the
environment the quantum state of the qubits gets entangled with the
environmental degrees of freedom. As a consequence the phase coherence 
is destroyed after a time scale called the 
dephasing time. In this Section we will describe the influence of the
environment on the qubit.  We determine how the dephasing time depends
on system parameters and how it can be optimized.

Some of the errors can be corrected by
`software' tools. One known from NMR and, in particular, NMR-based quantum
logic operations \citeaffixed{Chang_Book}{see e.g.} are the {\it
refocusing\ } 
techniques.  They serve to suppress the effects of undesired terms in
the Hamiltonian, e.g., deviations of the single-bit field terms from
their nominal values or uncontrolled interactions like stray direct
capacitive couplings of charge qubits or inductive couplings
of flux qubits.  As an example we consider the error due to  
a single-bit term $\delta B_x \sigma_x$, which after some time has
produced an unwanted rotation by $\alpha$.  Refocusing is based
on the fact that a $\pi$-pulse about the $z$-axis reverses the influence of
this term, i.e., $U_z(\pi) U_x(\alpha) U_z(\pi) = U_x(-\alpha)$.
Hence, fast repeated inversions of the bias $B_z(t)$ (with 
$|B_z|\gg \delta B_x$) eliminate the effects associated with  $\delta B_x$. 
The technique can also be applied to enhance the
precision of non-ideal control switches:  one first turns off the
coupling term to a low value and then further suppresses it by
refocusing. The examples demonstrate that refocusing requires 
very fast repeated switchings with a period much shorter than the
elementary operation time. This may make it hard to implement.

It was therefore a major breakthrough when the concepts of
quantum error-correction were discovered
\citeaffixed{SteaneBook,PreskillBook}{see e.g.}.  When applied they
should make it possible, even in the presence of dephasing processes
-- provided that the dephasing time is not too short -- to perform coherent
sequences of quantum manipulations of arbitrary length. 
The price to be paid is an increase in system size
(by roughly an order of magnitude), and a large number of steps are
needed for error correction before another computational step can be
performed (increasing the number of steps by roughly 3 orders of
magnitude). This imposes constraints on the dephasing time. The
detailed analysis shows that error correction can be successful if the
dephasing time is of the order of $10^4$ times longer than the time
needed for an elementary logic gate. 

In the Josephson junction systems, discussed here, the environment 
is usually composed of resistive elements in the circuits needed for 
the manipulations and the measurements. They produce voltage and current
noise. In many cases such fluctuations are Gaussian distributed 
with a Johnson-Nyquist power spectrum, coupling linearly
to the quantum system of interest. They can thus be described by a
harmonic oscillator bath with suitable frequency spectrum and coupling
strength~\cite{LeggettRMP,WeissBook}. 
For charge qubits, for instance, fluctuations in the gate voltage
circuit, coupling to $\sigma_z$, as well as the fluctuations in the
current, which control the Josephson energy and couple to
$\sigma_x$, can be described in this way \cite{Our_PRL}. 
In this section we will first describe these noise sources and the
dephasing introduced in this way. We later comment on other noise
sources such as telegraph noise, typically with a $1/f\,$ power spectrum
due to switching two-level systems (e.g.,
`background charge' fluctuations), or the shot noise resulting from the
tunneling in a single-electron transistor coupled to a qubit for the
purpose of a measurement. 

Depending on the relation between typical frequencies of the 
coherent (Hamiltonian) dynamics and the dephasing rates
we  distinguish two regimes.  In the first,
``Hamiltonian-dominated'' regime, where the controlled part of 
the qubit Hamiltonian $\Hcontr = -(1/2)\bbox{B}\bbox{\sigma}$,
governing the deterministic time evolution and logic gates, is large, 
it is convenient to describe
the dynamics in the eigenbasis of $\Hcontr$.  The coupling 
to the environment is weak, hence the environment-induced transitions
are slow.  One can then distinguish two stages:  a) {\it
dephasing\ } processes, in which the relative phase between the
eigenstates becomes random; b){\it energy relaxation\ } processes, in
which the occupation probabilities of the eigenstates change.

In the other, ``environment-dominated'' regime $\Hcontr$ is too weak to support
its eigenstates as the preferred basis.  The qubit's dynamics in this situation
is governed by dissipative terms and depends on details of the structure of the
coupling to the environment.  In general the evolution is complicated, and the
distinction between relaxation and dephasing may be impossible.

Both regimes may be encountered during
manipulations.  Obviously, the Hamiltonian should dominate when a coherent
manipulation is performed.  On the other hand, if in the idle state the
Hamiltonian vanishes (a very useful property as outlined in
Section~\ref{subsec:ChargeBit}, \ref{subsec:ChargeBit-w-SQUID}), the
environment-dominated regime is realized.  One has to ensure that the phase
coherence rate in this regime is still low enough.

\subsection{Spin-boson model}

Before we proceed discussing specific physical systems, we 
recall what is known about the spin-boson model, which has been 
studied extensively \citeaffixed{LeggettRMP,WeissBook}{see reviews
by}. It models the environment as an oscillator bath coupled to one
component of the spin. The Hamiltonian reads 
\begin{equation}
{\cal H}=\Hcontr + \sigma_z\sum_a \lambda_a x_a+
{\cal H}_{\rm B}
\; ,
\label{Eq:SpinBoson}
\end{equation}
where
\begin{eqnarray}
\Hcontr&=&-\frac{1}{2}B_z\;\sigma_z-
\frac{1}{2}B_x\;\sigma_x\\
&=&
-\frac{\Delta E}{2} (\cos\eta\;\sigma_z+\sin\eta\;\sigma_x)
\end{eqnarray}
is the controlled part of the Hamiltonian (cf. 
Eqs.~\ref{Eq:Magnetic_Hamiltonian} and \ref{Magnetic_Hamiltonian'}), while 
\begin{equation}
{\cal H}_{\rm B}=
\sum_a \left(
\frac{p_a^2}{2m_a} + \frac{m_a\omega_a^2x_a^2}{2}\right)
\label{Eq:FreeBosons}
\end{equation}
is the Hamiltonian of the bath. The bath operator
$X=\sum_a\lambda_a x_a$ couples to $\sigma_z$. 
In thermal equilibrium one finds for the Fourier transform of the
symmetrized  correlation function of this operator
\begin{equation}
\label{Eq:X-J}
\langle X^2_\omega \rangle
\equiv
\frac{1}{2} \left\langle \{ X(t), X(t') \} \right\rangle_\omega
= \hbar J(\omega) \coth \frac{\omega}{2 k_{\rm B}T}
\;,
\end{equation}
where the bath spectral density is defined by
\begin{equation}
J(\omega) \equiv {\pi\over 2}\sum_a \frac{\lambda_a^2}{m_a\omega_a}
\delta(\omega-\omega_a)
\,.
\label{Eq:CL-spectrum}
\end{equation}
This spectral density has typically a power-law behavior at low
frequencies \cite{LeggettRMP}. Of particular interest is Ohmic dissipation, 
corresponding to a spectrum
\begin{equation} 
\label{Eq:Linear_Spectrum}
J(\omega)=\frac{\pi}{2}\;\alpha\hbar\omega \; ,
\end{equation} 
which is linear at low frequencies up to some high-frequency 
cutoff $\omega_c$.
The dimensionless parameter $\alpha$ reflects the 
strength of dissipation. 
Here we concentrate on weak damping, $\alpha \ll 1$, 
since only this regime is relevant for quantum state engineering. 
But still the Hamiltonian-dominated  and the 
environment-dominated regimes are  both possible depending on the
ratio between the energy scale $\Delta E= \sqrt{B_z^2+B_x^2}$,
characterizing the coherent evolution, and the dephasing rate (to be
determined below).  

The Hamiltonian-dominated regime is realized when 
$\Delta E \gg \alpha k_{\rm B} T$.
In this regime it is natural to describe the evolution of the system 
 in the eigenbasis (\ref{Eigen_Basis}) which diagonalize $\Hcontr$:
\begin{equation}
\label{Eq:Spin_Boson_Eigen_Basis}
{\cal H} = -{1\over 2}\Delta E \rho_z + 
(\sin\eta\;\rho_x + \cos\eta\;\rho_z) \; X
+
{\cal H_{\rm B}}
\ . 
\end{equation} 
Two different time scales characterize the 
evolution~\cite{Weiss1,Weiss2,WeissBook}.  
On a first, dephasing time scale $\tau_\varphi$ the off-diagonal 
(in the preferred eigenbasis) 
elements of the qubit's reduced density matrix decay to zero.
They are represented by the expectation values of the  operators
$\rho_{\pm} \equiv (1/2) (\rho_x \pm i \rho_y)$. Dephasing leads 
to the following time-dependence (at long times):
\begin{equation}
\label{Eq:Rho_pm_dephasing} 
\langle \rho_{\pm}(t) \rangle = 
\langle \rho_{\pm}(0) \rangle\; e^{\mp i\Delta E t}\; e^{-t/\tau_\varphi}
\ .
\end{equation} 
On the second, relaxation time scale $\tau_{\rm relax}$ the diagonal
entries tend to their thermal equilibrium values:
\begin{equation}
\label{Eq:Rho_z_relaxation}
\langle \rho_z(t) \rangle = \rho_z(\infty) +
[\rho_z(0) - \rho_z(\infty)]\;e^{-t/\tau_{\rm relax}} \ , 
\end{equation}
where $\rho_z(\infty)= \tanh(\Delta E/2k_{\rm B}T)$.

The dephasing and relaxation times were originally evaluated for the
spin boson model in a path integral
technique~\cite{LeggettRMP,WeissBook}. The rates 
are~\footnote{Note that in the literature usually the evolution of
$\langle \sigma_z(t) \rangle$ has been studied.
To establish the connection to the results
(\ref{Eq:relaxation},\ref{Eq:dephasing}) one has to substitute   
Eqs.~(\ref{Eq:Rho_pm_dephasing},\ref{Eq:Rho_z_relaxation}) 
into the identity $\sigma_z = \cos\eta\;\rho_z + \sin\eta\;\rho_x$
Furthermore, we neglect renormalization 
effects, since they are weak for $\alpha \ll 1$.}  
\begin{eqnarray}
\tau_{\rm relax}^{-1}&=&\pi\alpha\;\sin^2\eta\;
\frac{\Delta E}{\hbar}
\coth\displaystyle\frac{\Delta E}{2k_{\rm B}T}
\label{Eq:relaxation}
\; ,
\\
\tau_\varphi^{-1}&=&
\frac{1}{2}\;\tau_{\rm relax}^{-1} +
\pi\alpha\;\cos^2\eta\;
\frac{2k_{\rm B}T}{\hbar}
\label{Eq:dephasing}
\; .
\end{eqnarray}

In some cases these results can be derived in a simple way, which we
present here to illustrate the origin of different terms.
As is apparent from the Hamiltonian (\ref{Eq:Spin_Boson_Eigen_Basis})
the problem can be mapped on the dynamics of a spin-$1/2$ 
particle in the external magnetic 
field $\Delta E$ pointing in $z$-direction and a fluctuating field in
the $x$-$z$-plane. The $x$-component of this fluctuating field, with
magnitude proportional to $\sin\eta$, induces transitions between the
eigenstates (\ref{Eigen_Basis}) of the unperturbed system. Applying
the Golden Rule for this term one obtains readily the relaxation rate 
(\ref{Eq:relaxation}). 

The longitudinal component of the fluctuating field, proportional to
$\cos\eta$ does not induce relaxation processes.  It does, however,
contribute to dephasing 
since it leads to random fluctuations of the eigenenergies and, thus, to a
random relative phase between the two eigenstates.  As an example we
analyze its effect on the dephasing rate in an exactly solvable limit.

The unitary operator
\begin{eqnarray}
\label{Eq:U_canonical}
U \equiv \exp\left(-i \sigma_z {\Phi\over 2} \right)
\; \; \mbox{with}\; \;
\Phi \equiv  \sum_a \frac{2\lambda_a p_a}{\hbar m_a \omega_a^2}
\end{eqnarray}     
transforms the Hamiltonian (\ref{Eq:SpinBoson}--\ref{Eq:FreeBosons}) to
a rotating spin-frame~\cite{LeggettRMP}:
\begin{eqnarray} 
\label{Eq:New_Hamiltonian}
\tilde {\cal H} &=& U{\cal H}U^{-1}= 
-(1/2)\Delta E \cos\eta\; \sigma_z
\nonumber \\ 
&&-(1/2)\Delta E \sin\eta 
\left(\sigma_{+} e^{-i\Phi} + {\rm h.c.}\right)
+ {\cal H}_{\rm B}
\ .
\end{eqnarray}
Here we recognize that in the limit $\eta = 0$ the spin and
the bath are decoupled, which allows an exact treatment.  
The  trivial time evolution in this frame,
$\sigma_\pm(t) = \exp(\mp i\Delta E t) \sigma_\pm(0)$, translates
in the laboratory frame to
\begin{equation}
\label{Eq:New_to_Old}
\sigma_\pm(t) = e^{\mp i\Phi(t)} e^{\pm i\Phi(0)} e^{\mp i\Delta E t} 
\sigma_\pm(0)
\ .
\end{equation}
To average over the bath we need the correlator
\begin{equation}
\label{Eq:P(t)}
P(t) \equiv \langle e^{i\Phi(t)}\,e^{-i\Phi(0)} \rangle = 
\langle e^{-i\Phi(t)}\,e^{i\Phi(0)} \rangle
\end{equation}
which was studied extensively by many authors 
\cite{LeggettRMP,P(E)_Panyukov_Zaikin,P(E)_Odintsov,P(E)_Nazarov,P(E)_Devoret}. 
It can be expressed as $P(t)=\exp[K(t)]$, where
\begin{eqnarray}
\label{K(t)}
        &&K(t) =  
        {4\over \pi\hbar} \int_0^{\infty} d\omega \, {J(\omega)\over\omega^2} 
        \nonumber \\
        &&\times
        \left[\coth\left({\hbar\omega\over2 k_B T}\right)(\cos\omega t-1)
        -i\sin\omega t\right] \ . 
\end{eqnarray}
For the Ohmic bath (\ref{Eq:Linear_Spectrum}) for $t>\hbar/2k_{\rm
  B}T$ one has ${\rm Re} K(t) 
\approx -(2 k_{\rm B}T/\hbar)\,\pi\,\alpha\,t$. Thus we reproduce
Eq.~(\ref{Eq:Rho_pm_dephasing}) 
with $\tau_\varphi$ given by (\ref{Eq:dephasing}) in the limit $\eta = 0$.
While it is not so simple to derive the general result for arbitrary $\eta$,
it is clear from Eqs.~(\ref{Eq:relaxation}) and (\ref{Eq:dephasing}) that the 
effects of the perpendicular ($\propto\sin\eta$) and longitudinal 
($\propto\cos\eta$) terms 
in (\ref{Eq:Spin_Boson_Eigen_Basis}) add up independently.    

In the environment-dominated regime, $\Delta E  \ll \alpha k_{\rm B}T$, the 
qubit's Hamiltonian is too weak to fix the basis, while the coupling to 
the bath becomes the dominant part of the total Hamiltonian. 
Therefore one should discuss the problem in the eigenbasis of the 
observable $\sigma_z$ to which the bath is coupled. The spin can tunnel 
incoherently between the two eigenstates of $\sigma_z$. To find the
tunneling rate one can again use the canonical transformation
(\ref{Eq:U_canonical}) leading to the   
Hamiltonian (\ref{Eq:New_Hamiltonian}). In the Golden Rule approximation one 
obtains~\cite{LeggettRMP} the following relaxation rate (for $\Delta E  \ll 
\alpha k_{\rm B}T$):
\begin{eqnarray}
\label{Eq:Tau_Rel_Zeno}
\tau_{\rm relax}^{-1}&=&
{2\pi\over\hbar} {\Delta E^2\sin^2\eta\over 4}
\left[\tilde P(\Delta E\cos\eta)+\tilde P(-\Delta E\cos\eta)\right]
\nonumber \\
&\approx&{\Delta E^2 \sin^2\eta \over 2\pi\hbar\alpha k_{\rm B}T}
\ ,
\end{eqnarray}  
where $\tilde P(..)$ is the Fourier transform of $P(t)$.
To find the dephasing rate we  
use again Eq.~(\ref{Eq:New_to_Old}) and obtain 
\begin{equation}
\label{Eq:Tau_Deph_Zeno}
\tau_\varphi^{-1} \approx 2\pi\alpha k_{\rm B}T/\hbar
\ .
\end{equation} 
In this regime the dephasing is much faster than the relaxation.
Moreover, we observe that
$\tau_{\rm relax}^{-1} \propto (\alpha k_{\rm B}T)^{-1}$ while 
$\tau_{\rm \varphi}^{-1} \propto\alpha k_{\rm B}T$.
This implies that the faster is the dephasing, the slower is the relaxation.
Such a behavior is an indication of the Zeno (watchdog) 
effect~\cite{Harris_Stodolsky}: the environment frequently 
``observes'' the state of the spin, thus preventing it from tunneling.

\subsection{Several fluctuating fields and many qubits}

Next we consider the general case of a qubit coupled to several
fluctuation sources (baths) via different spin components and below we
discuss a  many-qubit system. It is
described by the following generalization of the spin-boson model:
\begin{equation}
\label{Eq:Many_Spin_Boson}
{\cal H}=\Hcontr + \sum_i \bbox{\sigma n}_i
\left(\sum_a \lambda_a^i x_a^i \right)+
\sum_i {\cal H}_{\rm B}^i
\; ,
\end{equation}
where the index $i$ labels the different (Ohmic) baths and the 
unit vectors $\bbox{n}_i$ determine the spin components
to which the baths are coupled.  
The Hamiltonian-dominated regime is realized  
when $\Delta E \gg \sum_i \alpha_{i} k_{\rm B} T$, where 
$\alpha_{i}$ correspond to the bath $i$. In this case one should divide the 
fluctuations of all baths into transverse and longitudinal ones, as in 
Eq.~(\ref{Eq:Spin_Boson_Eigen_Basis}), with each bath being characterized 
by the angle $\eta_i$ between $\bbox{n}_i$ and the field direction
($\cos\eta_i  = \bbox{B n}_i/|\bbox{B}|$). 
The transverse fluctuations will add up to the relaxation rate as
\begin{equation}
\tau_{\rm relax}^{-1}=\sum_i \pi\alpha_i\;\sin^2\eta_i\;
\frac{\Delta E}{\hbar}
\coth\displaystyle\frac{\Delta E}{2k_{\rm B}T}
\label{Eq:relaxation_sum}
\; ,
\end{equation}
while the longitudinal fluctuations lead to the dephasing rate
\begin{equation}
\tau_\varphi^{-1}=
\frac{1}{2}\;\tau_{\rm relax}^{-1} + \sum_i
\pi\alpha_i\;\cos^2\eta_i\;
\frac{2k_{\rm B}T}{\hbar}
\label{Eq:dephasing_sum}
\; .
\end{equation} 

In the simplest environment-dominated situation 
one of the baths ($i=i_0$) is much stronger 
than all others, 
$\alpha_{i_0} \gg \sum_{i \ne i_0} \alpha_i$,
and satisfies $\alpha_{i_0}k_{\rm B}T\gg\Delta E$. 
Then, the dephasing is described in the eigenbasis of $\bbox{\sigma
  n}_{i_0}$ corresponding to this bath. The rate is
\begin{equation}
\label{Eq:max_dephasing}
\tau_\varphi^{-1} \approx 2\pi\alpha_{i_0} k_{\rm B}T/\hbar
\ .
\end{equation} 
The relaxation rate in this basis may be estimated using the Golden 
Rule:
\begin{equation}
\tau_{\rm relax}^{-1} \approx
{\Delta E^2 \sin^2\eta_{i_0} \over 2\pi\hbar\alpha_{i_0}\; 
k_{\rm B}T} +
\sum_{i\ne i_0} 2\pi \alpha_i \sin^2 \chi_i\; {k_{\rm B}T\over\hbar}
\ ,
\end{equation}
where $\cos\chi_i = (\bbox{n}_i \bbox{n}_{i_0})$.
This rate is smaller than the dephasing rate 
(\ref{Eq:max_dephasing}). Note that the relaxation rates due to 
the other than the strongest baths do not show the Zeno effect.

In the most complicated case of the environment-dominated regime 
with several baths coupled to different spin components and characterized
by constants $\alpha_i$ of the same order it is difficult to 
make quantitative predictions. However, we expect that the time
scale for a quantum state destruction (either dephasing or relaxation) will be 
longer than $\left(\sum_i 2\pi\alpha_i k_{\rm B}T/\hbar\right)^{-1}$.


So far we were concerned with dissipative effects 
in a single qubit. However, any but the simplest applications of
quantum state engineering make use of many coupled qubits and 
entangled states, i.e., states
whose properties can not be reduced to the single-bit ones. Therefore
the question arises how dissipation affects multi-qubit systems
and entangled states. 

As a first step we analyze the effect of dissipation on an $N$-qubit system
during an idle period when the single-bit terms and two-bit interactions are
switched off, $\Hcontr=0$.  We assume that each qubit is coupled to an
independent oscillator bath, with no correlations between the baths,
and we  choose a basis where this coupling is diagonal, i.e., the bath
is coupled to the $\sigma_z$-component.  
In this case the time evolution operator, governing the density matrix, 
$\hat{\rho}_{\nu_1\dots \nu_N;\mu_1\dots \mu_N}(t)$, factorizes. 
We perform for each qubit $i$ the unitary rotation
(\ref{Eq:U_canonical}) with the result:
\begin{equation}
\rho_{\nu_1\dots \nu_N;\mu_1\dots \mu_N}(t)\propto
\prod_{i=1}^N \left\langle
e^{i\Phi_i(t)(\nu_i-\mu_i)}e^{-i\Phi_i(0)(\nu_i-\mu_i)}
\right\rangle
\, .
\label{Eq:rho_ab}
\end{equation} 
Averaging over the baths yields  in the long-time limit the following
time dependence of the density matrix: 
\begin{equation}
\rho_{\nu_1\dots \nu_N;\mu_1\dots \mu_N}(t)\
\propto
\prod_{\{i:\nu_i\ne\mu_i\}}
\exp\left(-t/\tau_\varphi^i\right)
\, .
\label{manyqubits}
\end{equation}
Here the product is over all qubits $i$ which have off-diagonal entries
in the density matrix. This form shows that 
the dissipation has the strongest effect on those entries of
$\hat{\rho}$ which are off-diagonal 
with respect to each qubit. For instance, the dephasing rate of
$\rho_{0\dots0;1\dots1}(t)$ is $1/\tau_\varphi=\sum_i
1/\tau_\varphi^i$. It scales linearly with the number of qubits.

The result (\ref{manyqubits}) applies independent of the initial state,
i.e., equally  for product states or entangled states of
the multi-qubit system. However, it is valid only if the controlled
parts of the Hamiltonian are switched off. 
The question how dissipation influences the
dynamics of entangled states in general situations, e.g., during logic
operations when the many-qubit Hamiltonian is non-zero, remains
open. Further work is needed to analyze this interesting and important problem.

\subsection{Dephasing in charge qubits}

\begin{figure}
\centerline{\hbox{\psfig{figure=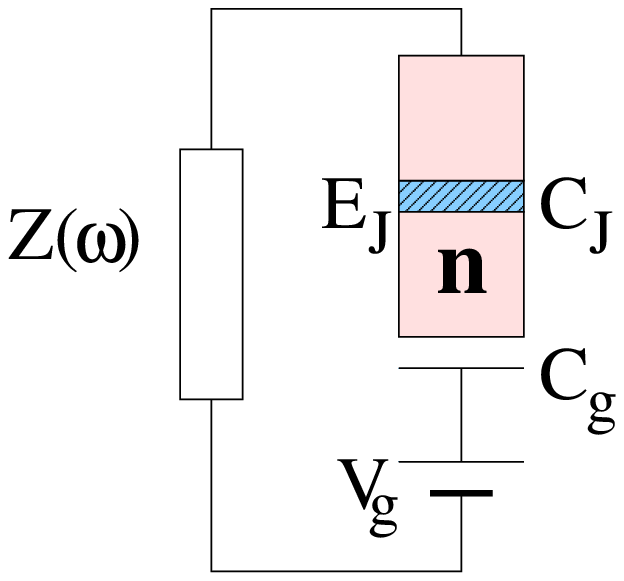,width=0.5\columnwidth}}}
\vskip 0.4cm
\caption[]{\label{Fig:BITS_CIRCUIT}
A qubit in an electro-magnetic environment characterized by the impedance
$Z(\omega)$.}
\end{figure}

We now turn to the specific case of a Josephson charge qubit coupled to the
environment and determine how the
dephasing and relaxation rates depend on system parameters.  
The system is sensitive to various electro-magnetic
fluctuations in the external circuit and the substrate, as well as to background
charge fluctuations.  We first estimate the effect of fluctuations
originating from the circuit of the voltage sources.
In Fig.~\ref{Fig:BITS_CIRCUIT} the equivalent circuit of a qubit 
coupled to an impedance $Z(\omega)$ is shown.
The latter has intrinsic voltage fluctuations 
with Johnson-Nyquist power spectrum.  
When embedded in the circuit
shown in Fig.~\ref{Fig:BITS_CIRCUIT} but with $E_{\rm J}=0$,
the voltage fluctuations between the terminals of $Z(\omega)$ are 
characterized by the spectrum
\begin{equation}
\langle \delta V  \delta V \rangle_\omega =
{\rm Re}\{Z_{\rm t}(\omega)\} \hbar \omega 
\coth\left( \frac{\hbar \omega}{2k_{\rm B}T}\right) \ .
\label{dVdV_Connected}
\end{equation}
Here $Z_{\rm t}(\omega) \equiv \left[i\omega \Cqb + 
Z^{-1}(\omega)\right]^{-1}$ is the total impedance 
between the terminals of $Z(\omega)$ and
$\Cqb$ is the capacitance (\ref{Eq:Cqb}) of the qubit in the circuit.
Following \citeasnoun{Caldeira-Leggett} we model 
the dissipative element $Z(\omega)$ by a bath of harmonic oscillators
described by the Hamiltonian ${\cal H}_{\rm B}$ as in 
Eq.~(\ref{Eq:FreeBosons}) \cite{Our_PRL}. 
The voltage fluctuations between the 
terminals of $Z(\omega)$ are represented by $\delta V = 
\sum_\alpha \lambda_\alpha x_\alpha$, while the 
spectral function (\ref{Eq:CL-spectrum}) has to be chosen as
$J(\omega) = \omega {\rm Re}\{Z_{\rm t}(\omega)\}$ in order to reproduce the 
fluctuation spectrum (\ref{dVdV_Connected}).

To derive the Hamiltonian we introduce an auxiliary variable, the charge $q$ on 
the gate capacitor, and add a term which couples $q$ to the bath (a natural 
choice since the current through the impedance is $\dot q$). For simplicity of 
derivation it is convenient also to add a small inductance in series with 
the impedance $Z$ to provide a `mass' for the $q$-mode. 
Allowing for a time-dependent external voltage $\Vqb(t)$ and integrating 
out $q$, we find the Hamiltonian:
\begin{eqnarray}
{\cal H} &=& {[2en-\Cg\Vqb(t)]^2\over 2 (\CJ+\Cg)}  
- \EJ \cos\Theta +  \sum_a \frac{p_a^2}{2m_a} 
\nonumber \\
&+&\sum_a 
\frac{m_a\omega_a^2}{2} 
\left[x_a-{\lambda_a\over m_a\omega_a^2}
\left\{2en{\Cqb \over \CJ} + \Cqb\Vqb(t)\right\} 
\right]^2
\ .
\nonumber \\
\label{dissHamUsualForm}
\end{eqnarray}
The voltage $\Vqb(t)$ in the last term (coupling to the bath) 
is relevant only if it depends on time and is usually dropped in the 
literature. Its role is to provide the $R C$ time delay. This means that 
when $\Vqb$ is changed the qubit 
feels the change only after the $R C$ time. In the systems described
here this delay is much shorter than the other relevant time 
scales and, thus, need not to be discussed further. 

To be specific  we will 
concentrate in the following on the fluctuations due to an Ohmic 
resistor $Z(\omega)=R_V$ in the 
bias voltage circuit. In the two-state approximation, 
using the relations $n = (1/2)(1+\sigma_z)$ and 
$\cos\Theta = (1/2)\sigma_x$, we thus arrive at the Hamiltonian of the 
spin-boson model (\ref{Eq:SpinBoson})
with $B_z$ and $B_x$ given by Eq.~(\ref{Eq:ChargeBit-Bz}) and 
(\ref{Eq:ChargeBit-Bx}), respectively. 
The dimensionless parameter $\alpha_z^{\rm ch}$, characterizing the
effect of fluctuations coupling to $\sigma_z$ of the charge qubit, is
given by 
\begin{equation}
\alpha_z^{\rm ch}=\frac{4R_V}{R_{\rm K}}
\left(\frac{\Cqb}{\CJ}\right)^2
\; .
\label{gammaV}
\end{equation}
The circuit resistance is compared to the quantum 
resistance $R_{\rm K}=h/e^2 \approx 25.8~{\rm k}\Omega$. Since the
parameter $\alpha_z^{\rm ch}$ directly relates the dephasing rate to typical 
energy scales, its inverse determines the number of coherent single-qubit 
manipulations which can be performed within the dephasing time.  
From Eq.~(\ref{gammaV}) we see that in order to keep the dissipative
effects of external voltage fluctuations  weak 
one has to use a voltage source with  low resistance and 
choose the gate capacitance $\Cg\approx\Cqb\ll\CJ$ as low as possible. 
The latter screens out the voltage fluctuations, 
at the expense that  one has to apply larger gate voltages for 
the manipulations.
At high frequencies discussed here the typical impedance of the
voltage circuit  is $R_{\rm V} \approx 50~\Omega$, and one obtains
$\alpha_z^{\rm ch}\approx 10^{-2} (\Cg/\CJ)^2$. Taking the ratio
\footnote{\citeasnoun{Nakamura_Nature} reached an even smaller ratio
  for the qubit, but the probe circuit introduced a high stray capacitance.} 
$\Cg/C_{\rm J}=10^{-2}$ one  
can reach the dissipation as weak as $\alpha_z^{\rm ch} \approx
10^{-6}$, allowing in principle for $10^6$ coherent single-bit manipulations.

The fluctuations of the externally controlled flux through the SQUID loop of
a charge qubit with tunable Josephson coupling (see
Fig.~\ref{Fig:BitwSQUID}) also destroy the phase coherence. As can be
seen from Eq.~(\ref{EJPhi}) these fluctuations couple to  
$\cos\Theta \propto \sigma_x$. Analogously to the voltage fluctuations 
considered above, their strength can be expressed by the effective
impedance of the current circuit which supplies the flux. For an
estimate we take this impedance $R_I$ to be purely real. At typical
high frequencies of the qubit's operation  
it is of order of the vacuum impedance, $R_I\sim 100$~$\Omega$. 
In terms of this resistance the effect of fluctuations in the flux is
characterized by the parameter 
\begin{equation}
\label{Eq:alphaI_ch}
\alpha_x^{\rm ch}=
\frac{R_{\rm K}}{4R_I}
\left(\frac{M}{\Phi_0}
\frac{\partial\EJ(\Phix)}{\partial\Phix}
\right)^2
\, ,
\end{equation}
where the flux-controlled Josephson coupling $\EJ(\Phix)$ is given by
Eq.(\ref{Eq:EffJosCoupling}).  The effect is weak for low values of the
mutual inductance. For $M \approx 0.01 - 0.1$~nH and 
$E_{\rm J}^{0} \approx 0.1$~K we obtain $\alpha_x^{\rm ch}
\approx 10^{-6}  - 10^{-8}$. The dephasing and relaxation times, if
only flux fluctuations need to be considered, are thus given
by Eqs.~(\ref{Eq:relaxation}),  
(\ref{Eq:dephasing}), (\ref{Eq:alphaI_ch}), but with a substitution
$\tan\eta =  B_z/B_x$ since the noise terms couple to $\sigma_x$. When
both the gate voltage and the flux fluctuate a multi-bath situation
described by Eq.~(\ref{Eq:Many_Spin_Boson}) is realized. 

For typical parameters, e.g., for those of the experiments of 
\citeasnoun{Nakamura_Nature}, one can estimate the dephasing time 
associated with the noise in the external circuit to be of the 
order of 100~ns. These experiments  allow a
direct probing of the phase coherence. Coherent 
oscillations have been observed for
5~ns.  Hence the theoretical estimate appears to be of the
right order~\footnote{In the experiments of 
\citeasnoun{Nakamura_Nature} actually much
of the dephasing can be attributed to the measurement device,
a dissipative tunnel junction which was coupled permanently to the qubit.
Its tunneling resistance was optimized to be large enough
not to destroy the qubits quantum coherence completely, but low enough
to allow for a measurable current. Single-electron tunneling
processes, occurring on a time scale of the order of 10~ns, destroy 
the state of the qubit (escape out of the two-state Hilbert space) 
 thus putting an upper limit on the
time when coherent time evolution can be observed. For a more
detailed discussion of the experiment and the measurement process we
refer to the article by \citeasnoun{Bruder-qubit}.}.

Another important source of decoherence in charge qubits are 
fluctuations of the background charge. It was found experimentally that 
they lead to $1/f$ noise at low frequencies. Typically, their contribution to
fluctuations of the effective gate  voltage, $\Qg=\Cg\Vqb$, is of
order $S_Q(\omega)=10^{-8}e^2/\omega$. These fluctuations limit the
time of coherent evolution and can hardly be improved by present-day
experimental techniques 
\footnote{See, however, recent work of \citeasnoun{Zorin_Isolated_Islands} where
the $1/f$-noise was suppressed by fabricating a metallic island on top of an 
electrode instead of placing it on the substrate.}. 
Fortunately, for the ideas of 
quantum state engineering, the configurations of background
charges change rarely, on a typical slow relaxation time scale which
can reach minutes or  even hours. Thus during each  
cycle of manipulations these charges  provide random static gate
voltages, which do not destroy, e.g., coherent
oscillations.  Their effect can be suppressed using refocusing
techniques. On the other hand, when averaging over many experimental
runs (as done by \citeasnoun{Nakamura_Nature}), then an ensemble-average is
formed and the dephasing  rate can be extracted.

When the qubit is coupled to a measurement device, which necessarily implies a
coupling to a macroscopic variable with dissipative dynamics, the feedback
introduces fluctuations and causes dephasing.  We will discuss this explicitly
in the next section for the case where a Josephson charge qubit is coupled to a
dissipative single-electron transistor.  We find that the shot noise of the
tunneling current in the SET introduces dephasing and relaxation processes.

\subsection{Dephasing in flux qubits}

Flux qubits have the advantage that they are practically insensitive to the
background charge fluctuations and are advertised for this reason~\cite{Mooij}.
Still their phase coherence can be destroyed by a number of effects as well.
Some sources of dissipation for flux qubits have been discussed by
\citeasnoun{MooijDecoh} and estimates have been provided for the
parameters of the circuits of \citeasnoun{Mooij} and \citeasnoun{MooijPRB}.
This includes the effect of  background charge fluctuations 
($\tau_\varphi\approx0.1$~s) as well as quasiparticle
tunneling in the superconductor with a non-vanishing subgap conductance
($\tau_\varphi\approx 1$~ms). 
The effect of nuclear spins in the substrate producing
fluctuating magnetic fields is similar to the effect of background charges on 
charge qubits. While the static random magnetic field may induce
substantial changes of  the qubit frequencies of order $\delta\Omega_{\rm 
nucl}\approx30$~MHz 
(which can be suppressed, in principle, by using refocusing pulses), 
they cause no dephasing for the qubits until  
a typical nuclear spin relaxation time $T_1$, which can reach minutes.  
Other sources of dephasing studied by 
\citeasnoun{MooijDecoh} include the electro-magnetic radiation
($\tau_\varphi\approx10^3$~s), much weaker than in typical rf-SQUID designs, 
and unwanted dipole-dipole magnetic couplings between qubits, which for the 
inter-qubit distance of 10~$\mu$m produces substantial effects after a 
relatively short time $\tau\approx 0.2$~ms. Variations of the design were 
suggested to reduce the latter effect.

An important source of dissipation of flux qubits are the
fluctuations in the external circuit which supplies fluxes through the
loops. They can be analyzed along the same lines as presented
above. Since they couple to $\sigma_z$, the relevant parameter is
$\alpha_z^{\rm fl}$. It is fixed by the impedance $Z_I$ of the 
current source in the input loop providing the flux bias,
and the mutual inductance of the input and the qubit's loop, $M$: 
\begin{equation}
\alpha_z^{\rm fl}=
\frac{a}{4}
R_{\rm K}
\mathop{\rm Re}Z^{-1}_I
\left(\frac{4\pi^2 \EJ M}{\Phi_0^2}\right)^2
\;.
\label{Eq:alphazfl}
\end{equation}
The numerical prefactor in Eq.~(\ref{Eq:alphazfl}) is
$a=6(\beta_L-1)/\pi^2$ for the rf-SQUID and
$a=b^2\sqrt{4b^2-1}/\pi^2$ for the design of \citeasnoun{Mooij} 
with $b\equiv\tilde\EJ/\EJ$ being the ratio of critical currents for
the junctions in  the loop (see Section~\ref{subsec:FluxQubits}).
The dephasing is slow for small loops and junctions with low critical
currents. Indeed, the argument in the bracket in (\ref{Eq:alphazfl}) can be
represented as the  
product of the `screening ratio' $M/L$ and the quantity
$\beta_L = 4\pi^2L\EJ/\Phi_0^2$.  While the latter should be 
larger than one for the rf-SQUID, it can be much smaller for the 
design of \citeasnoun{Mooij}, resulting in a slower dephasing. 

The impedance $Z_I(\omega)$ and the bath spectrum 
$J(\omega)=\hbar\omega\mathop{\rm Re}Z_I^{-1}(\omega)$ are
frequency-dependent.  They can lead to resonances,
e.g., if the current source is attached to the qubit 
via low-loss lines. Care has to be taken in experiments to avoid these
resonances. In addition, the dephasing and relaxation times have to be
estimated for this situation. The analysis
of the Hamiltonian (\ref{Eq:Spin_Boson_Eigen_Basis}) with general bath
spectrum shows that they can be defined as the 
times when the quantities $\cos^2\eta\, \langle(\int_0^t  
X(\tau)d\tau)^2\rangle$ and $\sin^2\eta\, \langle|\int_0^t
X(\tau)e^{\pm i\Delta E\tau} d\tau|^2\rangle$ reach values of order
one, respectively.
The first quantity can be expressed as $t \int d\omega \langle X^2_\omega 
\rangle \tilde\delta(\omega)$ involving the function 
$\tilde\delta(\omega)\equiv 2\sin^2(\omega t/2)/(\pi t \omega^2)$ 
which is peaked at $\omega=0$ and has the width 
$t^{-1}$. Recalling that $\langle X^2_\omega\rangle$ is related to the bath 
spectrum (\ref{Eq:X-J}), one finds that only the low-frequency part of
the impedance 
$Z_I(\omega)$, with $\omega<\tau_\varphi^{-1}$, determines the dephasing 
rate. On the other hand, the relaxation rate depends on the values of
$Z_I(\omega)$ at frequencies in the vicinity of $\omega = \Delta
E/\hbar$ in a range of width $\tau_{\rm relax}^{-1}$.

The choice of a high dc resistance of the remote current source
strongly suppresses the fluctuations at low frequencies.  Using 
Eqs.~(\ref{Eq:alphazfl}, \ref{Eq:dephasing}) we can estimate the
dephasing rate for the parameters of \citeasnoun{Mooij}. Assuming
$Z_I(\omega \approx 0)\sim 1$~M$\Omega$ and $M\approx
L$ we find $\alpha_z^{\rm fl}\sim 10^{-9}$, which implies a
negligible dephasing.  At higher frequencies of order $\Delta E$, even
if resonances are suppressed, $Z_I(\omega \approx \Delta E)$ is of
order 100~$\Omega$, leading to $\alpha_z^{\rm fl}\sim 10^{-5}$. This
determines  the relaxation rate via Eq.~(\ref{Eq:relaxation}).  For
\possessivecite{Mooij} parameters we estimate a relaxation time of $\sim
5~\mu$s at the degeneracy point.

The effect of fluctuations of the $\sigma_x$-term in the Hamiltonian (in the
flux circuit of the dc-SQUID-loop which controls the Josephson
coupling) can be described in a similar way~\cite{Our_Nature}. 
The effect of these fluctuations is relatively weak for the operation regimes 
discussed by  \citeasnoun{Mooij}.

While a further analysis of the dephasing effects in flux qubits may
be needed, the 
above-mentioned estimates of $\tau_\varphi$ suggest that the observation of
coherent flux oscillations is feasible in the near future.  However,
an open problem remains the fact that the {\it
observation\ } requires an efficient quantum detector, e.g.,  a quantum
magneto-meter.  We discuss this issue in the next section.

\section{The quantum measurement process}
\label{sec:Measurement}

\subsection{General concept of quantum measurements}
 
Quantum state engineering requires controlled quantum manipulations
but also quantum measurement processes. They are needed, e.g., at the
end of a computation to read out the final results, or even in the
course of the computation for the purpose of error correction. The
problem of quantum 
measurement has always attracted considerable attention, 
and it still stirs controversy. In most of the literature
on quantum information theory the measurement process is expressed
simply as a ``wave function collapse'', i.e.,  as  a
non-unitary projection, which reduces the quantum state of a qubit
to one of the possible eigenstates of the observed quantity
with state-dependent probabilities. 

On the other hand, in reality any measurement 
is performed by a device which itself is a physical 
system, suitably coupled to the measured quantum system and with a
macroscopic read-out variable. This is accounted for in the approaches
described below, which are based on the concepts of the dissipative
quantum mechanics \cite{LeggettRMP,WeissBook,Zurek_Physics_Today}.
The qubit and the
measuring device are described as coupled quantum systems. Initially
there exist no  correlations between the qubit and the meter but, due to
the coupling, such correlations (entanglement) emerge in time. This
is precisely the concept of measurement of \citeasnoun{von_Neumann}.
Furthermore,  the meter is actually a dissipative  system, coupled to
the environment. The dissipation  eventually reduces the entanglement
between the qubit and the meter to  classical correlations between
them. This is exactly what is needed for the measurement.

Several groups have studied problems related to quantum
measurement processes in mesoscopic systems.  
\citeasnoun{Aleiner}, \citeasnoun{Levinson}, and \citeasnoun{Gurvitz} 
investigated the effect of a dissipative conductor, whose conductance
depends on  the state of  quantum system, onto this system
itself. Such a configuration is realized in recent experiment
\cite{Buks,Sprinzak}, where a quantum dot is
embedded in one of the paths of a `which path' interferometer. The
flow of a dissipative current through a nearby quantum point contact, with
conductance which depends on the charge on the dot, amounts to a
`measurement' of the path chosen by the electrons in the
interferometer. And indeed the flow of a current through the point 
contact leads to 
an observable reduction of the flux-dependence of the current through the
interferometer.  On the other hand, no real measurement was performed.
The experiments merely demonstrated that the amount of dephasing can be
controlled by a dissipative current through the QPC, but they did not
provide information about the path chosen by each
individual electron.

In what follows we will discuss systems in which one is not only able
to study the dephasing but also, e.g., by measuring a dissipative current,
to extract information about the quantum state of the qubit.
As an explicit example we investigate a single-electron transistor
(SET) in the sequential 
tunneling regime coupled capacitively to a charge qubit
\cite{Our_PRB,Schoelkopf}. Our purpose is not philosophical but rather
practical.  We do not search for 
an ideal measurement device, instead we describe the properties of a
realistic system, known to work in the classical regime, and investigate
whether it can serve as quantum measurement device. At the same time
we  keep in mind that 
we should come as close as possible to  the projective measurement
picture assumed in quantum algorithms. 
With the same goal, namely, to establish their potential use
as quantum detectors, various other mesoscopic
devices have been studied recently. These include a quantum   
point contact (QPC) coupled to a quantum dot \cite{Gurvitz,Korotkov},
a single-electron transistor in the co-tunneling regime 
\cite{Averin_SET_Cotunneling,Maassen_SET_Cotunneling}, or a
superconducting SET  (SSET) \cite{Averin_SQUID,EsteveSSET}. For a flux
qubit a dc-SQUID, or suitable modifications of it, can serve as a
quantum detector. 

Technically the measurement process is described by the time evolution
of the reduced density matrix of the coupled system of qubit and
meter. To analyze it we first derive a Bloch-master equation  for the
time-evolution of the system of a Josephson charge qubit and
a dissipative single-electron transistor (SET) coupled to it. This
derivation demonstrates how the unitary time evolution 
of system plus detector can lead to a quantum measurement.
It also allows us to follow the dynamics of the system and detector 
and to analyze the mutual influence of their variables. During the
measurement the qubit looses its phase coherence on a short dephasing
time scale  $\tau_\varphi$
\footnote{In this section  $\tau_\varphi$ denotes
the dephasing time during a measurement. It is usually 
much shorter than the dephasing time during the controlled
manipulations discussed in the previous sections. from the context it
should be clear which situation we refer to.}. 
This means that the off-diagonal elements
of the qubit's density matrix (in a preferred basis, which depends on
relative strengths of the coupling constants) vanish, while
the diagonal ones still remain unchanged. At the same time the
information about the initial state  of the qubit is  transferred to
the macroscopic state of the detector (the current in the SET).  Under
suitable conditions, after another time scale $\tau_{\rm meas}$, this 
information can be read out. Finally, on a longer, mixing (or
relaxation) time scale $\tau_{\rm mix}$ the detector acts back onto the
qubit and destroys the information about the initial state. The
diagonal entries of the density matrix tend to their stationary
values. (These are either determined by the detector or 
they are thermally distributed, depending on the relative strength of
the measurement device and residual interactions with the bath.) One has 
to chose parameters such that this back-action does not change the occupation
probabilities of the qubit's state before the information is actually
read out, $\tau_{\rm mix}\gg\tau_{\rm meas}\ge\tau_\varphi$.

The different times scales characterizing the
measurement process by a SET, show up  also in
equilibrium properties, e.g., in the  noise spectrum of the
fluctuations in the SET. This can be extracted from the time evolution
of the coupled density matrix as well.

Ideally, the meter is coupled to the qubit only during the
measurement. In practice, however, this option is hard to realize for
mesoscopic devices.  Rather the meter and the qubit are coupled
permanently but the former is kept in a non-dissipative state. To
perform a measurement, the meter is switched to the dissipative regime
which, as we mentioned above, is an important requirement for 
the effective quantum measurement.  For instance, in a SET, due to the Coulomb
blockade phenomena, no dissipative currents are flowing in the meter as long as
the transport voltage is switched off.  Applying a voltage bias above
the Coulomb threshold induces a dissipative current through the SET,
which leads to dephasing and, at the same time, provides the
macroscopic read-out variable.

Various ways of operation can be used, depending on details of a particular
setting.  For instance, one can switch the meter abruptly into the dissipative
regime and monitor the response of dissipative currents to the qubit.  Another
possibility is to change the bias gradually until the system switches into the
dissipative regime.  The value of the bias at which the switching occurs
provides information about the qubit's state.  Conceptually, these techniques
are similar.  For definiteness, below we discuss the former operation strategy.

\subsection{Single-electron transistor as a quantum electrometer}

\begin{figure}  
\centerline{\hbox{\psfig{figure=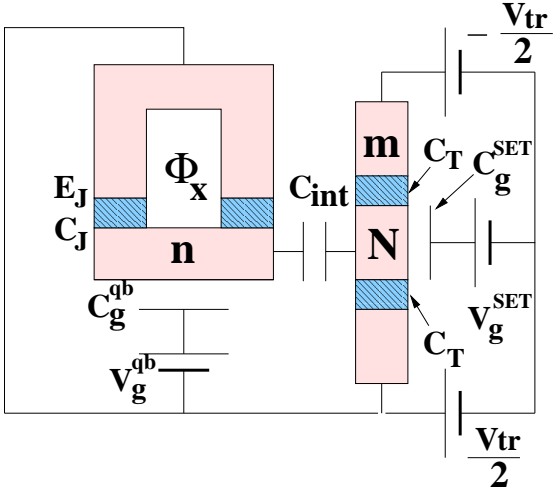,width=0.85\columnwidth}}}
\vspace{5mm}
\caption[]{\label{CIRCUIT}
The circuit consisting of a qubit plus a single-electron transistor
used as a measuring device.}
\end{figure}

Since the relevant quantum degree  of freedom of a Josephson charge
qubit is the charge of its island, the natural choice of measurement
device is a single-electron transistor (SET).  This system is shown in
Fig.~\ref{CIRCUIT}.  The left hand part is the qubit, with state
characterized by the number of extra Cooper pairs $n$ on the island,
and controlled by its gate voltage $\Vgqb$.  The right hand  part
shows a normal island between two normal leads,  which form the
SET. Its charging state is characterized by the  number of extra
single-electron charges  $N$ on the middle island. It is controlled by
gate and transport voltages, $\VgSET$ and $V_{\rm tr}$, and further,
due to the capacitive coupling to the qubit, by the state of the
latter.  A similar setup has been studied in the experiments of
\citeasnoun{Bouchiat_PhD} and \citeasnoun{BouchiatPhysScr}, 
where it was used to demonstrate that the ground state of a single Cooper 
pair box is a coherent superposition of different charge states.

During the quantum manipulations of the qubit the transport voltage
$V_{\rm  tr}$ across the SET transistor is kept zero and the gate
voltage of the SET, $\VgSET$, is  chosen to tune the island away from
degeneracy 
points.  Therefore at low temperatures  Coulomb blockade effects
suppress exponentially a dissipative current flow in the system, and the
transistor merely modifies the  capacitances of the
system~\footnote{More precisely, the leading contribution are 
co-tunneling processes, which are weak in high-resistance junctions.}.
To perform a measurement one tunes the SET by $\VgSET$ to  the vicinity
of its degeneracy point and applies a small transport voltage $V_{\rm
tr}$.  The resulting normal current through the transistor depends on
the  charge configuration of the qubit, since different charge states
induce different voltages on the island of the SET. While these
properties are well established as operation principle of a SET as
electro-meter in the classical limit, it remains to be demonstrated
that they also allow resolving different quantum states of the
qubit. For this purpose we have to discuss various noise  factors,
including the shot 
noise associated with the tunneling current and the measurement
induced transitions between the states of the qubit. They can be
accounted for by analyzing the time evolution of the density matrix of
the combined system. 
 
The Hamiltonian of the combined system consists of the  parts
describing the  qubit, the SET and the interaction between them: 
\begin{equation}
{\cal H} = {\cal H}_{\rm ctrl} + {\cal H}_{\rm SET} +  {\cal H}_{\rm
int} \ .
\end{equation}
Except for a redefinition and renormalization of
parameters, the qubit's part is identical to that of the bare qubit
(\ref{Eq:Magnetic_Hamiltonian}),
$$
{\cal H}_{\rm ctrl}= -\frac{1}{2}
B_z \hat\sigma_z - {1\over 2} B_x \hat\sigma_x \, ,
$$
with $B_z = 4\ECqb(1-2\ngqb)$ and $B_x = \EJ(\Phi_x)$.
For a detailed derivation and precise definition of the parameters 
see Appendix~\ref{app:SET}.
Here it is sufficient to know that the qubit's Hamiltonian can be
controlled by gate voltages and the flux through the SQUID.
After diagonalization of (\ref{Eq:Ham_diagonalized}), 
$\Hcontr = -(1/2)\Delta E(\eta)\;\rho_z$, the
characteristic energy scale of the qubit,
$\Delta E(\eta)=\sqrt{B_z^2+B_x^2}$, becomes apparent.

The Hamiltonian of the SET reads
\begin{eqnarray}
\label{SET_HAMILTONIAN}
        {\cal H}_{\rm SET} &=& \ECSET (N - \NgSET)^2 \nonumber \\ 
        &+&
        {\cal H}_{\rm L} + {\cal H}_{\rm R}  + {\cal H}_{\rm I} +
        {\cal H}_{\rm T}  \ ,
\end{eqnarray}
where the transistor's charging energy is given by
(\ref{CHARGING_ENERGY2}) and the gate charge $\NgSET \equiv
-eV_{N} / 2\ECSET$, defined in
Eq.~(\ref{Effective_Voltages}), can be controlled by $\VgSET$.  
The three  terms ${\cal H}_{\rm L}$,
${\cal H}_{\rm R}$, and ${\cal H}_{\rm I}$ describe microscopic
degrees of freedom of noninteracting electrons in  the two leads and
the middle island  of the SET transistor:
\begin{equation}
\label{FREE_LEADS_H}
        {\cal H}_r =  \sum_{k\sigma} \epsilon_{k\sigma}^{r}
        c_{k\sigma}^{r\dag}  c_{k\sigma}^{r \phantom{\dag}} \ \ (r=
        {\rm L,R,I})\ .
\end{equation}
The index $r={\rm L,R,I}$ labels the electrodes ``left'', ``right'' 
(viewed from a suitable angle) and island, 
$\sigma$ labels transverse channels including the spin,
while $k$ refers to the wave vector within one channel. Note that
similar terms should have been written for the electrode and  island
of the qubit; however, for this superconducting non-dissipative
element  the microscopic degrees of freedom can be integrated
out~\cite{AES82,Schon_Zaikin_Review}, resulting in the ``macroscopic''
quantum description presented in  Sections~\ref{sec:JosChargeBit} and 
\ref{sec:FluxQubits}.   In
this limit the tunneling terms reduce to the  Josephson
coupling  ${\cal H}_{\rm J} = - E_{\rm J}\cos\Theta$, expressed in a
collective variable describing the coherent transfer  of Cooper pairs,
$e^{i\Theta} |n\rangle = |n+1\rangle$.

The normal-electron tunneling in the SET transistor is described by
the standard tunneling Hamiltonian, which  couples the microscopic
degrees of freedom:
\begin{eqnarray}
\label{TUNNEL_HAMILTONIAN}
        {\cal H}_{\rm T} = && \sum\limits_{kk'\sigma} T^{L}_{kk'\sigma}
        c^{\rm L\dag}_{k\sigma} c_{k'\sigma}^{\rm I \phantom{\dag}}
        e^{-i\phi}
\nonumber \\ 
        &+&   \sum\limits_{k'k''\sigma}  T^{R}_{k'k''\sigma} c^{\rm
        R\dag}_{k''\sigma}  c_{k'\sigma}^{\rm I \phantom{\dag}}
        e^{-i\phi} e^{i\psi} + {\rm h.c.} \ .
\end{eqnarray} 
To make the charge transfer explicit, (\ref{TUNNEL_HAMILTONIAN})
displays two ``macroscopic''  operators, $e^{\pm i\phi}$ and $e^{\pm
i\psi}$.  The first one describes changes of the charge on the
transistor island  due to the tunneling: $e^{i\phi} |N\rangle =
|N+1\rangle$. If the total number of electrons on the island is large
it may be treated  as an independent degree of freedom. We further 
include the operator $e^{\pm i\psi}$, which acts on $m$, the  number
of electrons 
which have tunneled through the SET transistor,  $e^{i\psi} |m\rangle
= |m+1\rangle$.  Since the chemical potential of the right lead is
controlled, $m$ does not appear in any charging  part of the
Hamiltonian.  However, we have  to keep track of it, since it is the
measured quantity, related to the current through the SET transistor.

Finally, ${\cal H}_{\rm int}= \ECint N (2 n - 1)$ describes the capacitive
interaction between the charge on the qubit's island and the SET
island.  In detail it originates from the mixed term in
Eq.~(\ref{CHARGING_ENERGY}), where  $\ECint$ is given by
Eq.~(\ref{CHARGING_ENERGY2}).  In the 2-state approximation for the
qubits, $n = (1+\sigma_z)/2$, we obtain
\begin{equation}
\label{INTERACTION_HAMILTONIAN}
        {\cal H}_{\rm int} = N \dHint \equiv 
        N \ECint\,\sigma_z  
        \ .
\end{equation} 
The operator $\dHint$, introduced here for later convenience,
is the part of the interaction Hamiltonian which acts in the qubit's 
Hilbert space.

\subsection{Density matrix and description of measurement} 

The total system composed of qubit and SET is described by a total
density matrix $\hat\varrho(t)$. We can reduce it, by tracing out
the microscopic electron states of the left and right leads and of the
island, to
\begin{equation}  
\hat{\rho}(t) = {\rm Tr}_{\rm L,R,I}\{\hat{\varrho}(t)\} \,.  
\end{equation}
This reduced density matrix
$\hat{\rho}(i,i';N,N';m,m')$  is still a matrix in the indices $i$,
which label the quantum states of the qubit  $|0\rangle$ or
$|1\rangle$, in $N$, and in $m$. In the following we will assume that
initially, as a result of previous quantum manipulations, the
qubit is prepared in some quantum state and it is disentangled from the SET,
i.e., the initial density matrix of the whole system may
be written as a product 
$\hat{\rho}_0 = \hat\rho_0^{\rm qb} \otimes \hat{\rho}_0^{\rm SET}$.
At time $t=0$ we switch on a transport voltage of the
SET and follow the resulting time evolution of the
density matrix of the whole system. 
For specific questions we may further reduce the total density matrix
in two ways, either one providing complementary information 
about the measurement process.

The first, common procedure \cite{Gurvitz} is to trace over $N$
and $m$. This yields a reduced density matrix of the qubit
$\hat\rho_{ij} \equiv \sum_{N,m} \hat{\rho}(i,j;N,N;m,m)$.
At $t=0$ one has $\hat\rho_{ij} = (\hat\rho_0^{\rm qb})_{ij}$. 
Depending on the relation among the energy scales and  coupling strengths 
(see Section \ref{sec:Dephasing})
a preferred basis may exist in which
the dynamics of the diagonal and the off-diagonal elements of
$\hat\rho_{ij}$ decouple. We study how fast the off-diagonal  elements
(in that special basis) vanish after the SET is  switched to the
dissipative state, i.e., we determine the rate of dephasing induced by
the SET.  And we determine how fast the diagonal elements 
change their values. This process we call `mixing'. 

\begin{figure}
\centerline{\hbox{\psfig{figure=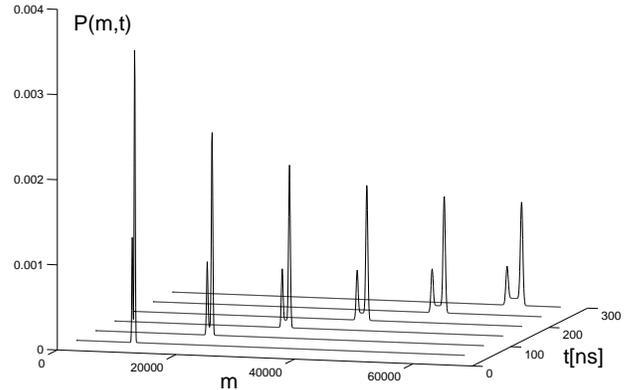,width=0.95\columnwidth}}}
\caption[]{\label{PLOT0.1}
$P(m,t)$, the probability that $m$ electrons have tunneled during time
$t$ (measured in nanoseconds). 
The initial amplitudes of the
qubit's states are $|a|^2 = 0.75$,  $|b|^2 = 0.25$. 
$E_{\rm J} = 0.1$~K, the remaining parameters are given later in 
the text.}
\end{figure}

The rates of `dephasing' and `mixing' refer to the quantum properties of
the measured system in the presence of the
measurement device. These quantities have also been analyzed in
Refs.\ \cite{Levinson,Aleiner,Gurvitz,Buks}. They do not tell us, however,  
anything about the quantity measured in an experiment, namely, the
current flowing through the SET.  Therefore the second way to reduce
the  density matrix is important as well. By
tracing the density matrix over the qubit's variables and the state
$N$ of the island, 
\begin{equation}
\label{P(m)}
        P(m,t) \equiv \sum_{i,N} \hat{\rho}(i,i;N,N;m,m)(t) \, ,
\end{equation}
we obtain the probability distribution for the  experimentally
accessible number of electrons $m$ which have tunneled through the SET
during time $t$. Its detailed analysis will be presented below. In
order to provide a feeling we first present and describe some
representative results: At $t=0$ no
electrons have tunneled, so $P(m,0) = \delta_{m,0}$.  Then, as
illustrated in Fig.~\ref{PLOT0.1}, the peak of the distribution moves
to nonzero values of $m$ and, simultaneously, widens due to shot
noise. If the two states of the preferred basis correspond to different
tunneling currents, and hence different $m$-shifts, and if the
mixing in this basis is sufficiently slow, then after some time the peak
splits into two with weights corresponding to the initial 
values of the diagonal elements of $\hat\rho_0^{\rm qb}$ in the
preferred basis (we will denote these values $|a|^2$ and $|b|^2=1-|a|^2$).  
Provided that after sufficient 
separation of the two peaks their weights are still close to the
original values, a good quantum measurement can be performed by
measuring $m$. After a longer time, due to transitions between the
states of the preferred basis (mixing) the two peaks transform into a
broad  plateau.  Therefore there is an optimum
time  for the measurement, such that, on one hand, the two peaks are
separate  and, on the other hand, the mixing has not yet influenced
the process.

Having introduced the relevant time scales, we can mention that the which-path 
interferometry \cite{Buks} can be thought of as a very short measurement 
process. Each particular electron -- the observed  quantum
system -- spends only a short time within the dot. This time of
interaction with  the meter may be shorter than the dephasing
time. Therefore the coupling leads only to a slight 
suppression of the interference pattern.

\subsection{Master equation}
\label{subsec:ME}

\begin{figure}  
\centerline{\hbox{\psfig{figure=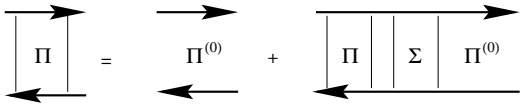,width=0.8\columnwidth}}}
\vskip 0.8cm
\caption[]{\label{Fig:DYSON}
The Dyson-type equation governing the time evolution of the density
matrix.  It is equivalent to the generalized master equation
(\ref{MASTER_EQUATION}).  The `self-energy' diagrams $\Sigma$
describe the transitions due to tunneling in the SET transistor.}
\end{figure}

The time evolution of the density matrix leads to  Bloch-type master
equations with coherent terms. Examples of this type  have recently
been analyzed in various contexts
\cite{Nazarov,Schoeller_PRB,Stoof,GurvitzPrager,Gurvitz}. 
~\citeasnoun{Schoeller_PRB} developed a diagrammatic technique 
which provides a formally exact master equation for a SET as
an expansion in the tunneling  term ${\cal H}_{\rm T}$, while all
other terms, including the charging energy, constitute the zeroth
order Hamiltonian ${\cal H}_0$.  
The time evolution of the reduced density matrix, $\hat\rho(t) =
\hat\rho(0)\Pi(0,t)$ is expressed by a propagator $\Pi(t',t)$, which
is expanded and displayed diagrammatically and finally
summed in a way reminiscent  of a Dyson equation. Examples are
shown in Fig.~\ref{Fig:DYSON} and Appendix~\ref{app:Diagrams}.  
In contrast to equilibrium many-body
expansions, since the time dependence of the density matrix is
described by a forward and a backward time-evolution operator, there
are two propagators, which are represented by  two horizontal lines
(Keldysh contour). The two bare lines describe the coherent time
evolution of the system. They are coupled due to the tunneling in the
SET. The sum of  all distinct
transitions defines a `self-energy' diagram $\Sigma$.  Below we will
present the rules for calculating  $\Sigma$ and present a suitable
approximate form. The Dyson equation is equivalent to a 
Bloch-master equation for the density matrix, which reads
\begin{equation}
\label{MASTER_EQUATION}
        {d\hat{\rho}(t)\over dt} -  {i\over\hbar}[\hat{\rho}(t),
        {\cal H}_0] = \int_0^t dt' \hat{\rho}(t') \Sigma(t-t') \ .
\end{equation}

In principle, the density matrix
$\hat{\rho}(i,i';N,N';m,m') \equiv \hat{\rho}^{i',N',m'}_{i, \; N,\; m}$
is a matrix in all three indices $i$, $N$, and $m$, and 
the  (generalized) transition rates due to single-electron tunneling
processes (in general of arbitrary order), $\Sigma^{i',N',m'
\rightarrow \bar{i'},
\bar{N'},\bar{m'}} _{i\phantom{'},N\phantom{'},m\phantom{'}
\rightarrow \,  \bar{i}, \; 
\bar{N},\; 
\bar{m}}(t - t')$,
connect these diagonal and off-diagonal states.  But a closed set of
equations for the time evolution of the system can be
derived \cite{Schoeller_PRB} which involves only the elements of $\hat\rho$
diagonal in $N$.  The same is true 
for the matrix structure in $m$.  I.e., we need to consider only the
following elements of the density matrix $\DM{N}{ij}(m) \equiv
\hat{\rho}^{i,N,m}_{j,N,m}$. Accordingly, of all the transition rates
we need to calculate only the corresponding matrix elements  
$\Sigma^{i',N',m' \rightarrow i,
N,m}_{j',N',m' \rightarrow j,N,m}(\Delta t)$.

If the temperature is low and the applied  transport voltage not too
high, the leading tunneling processes in the SET are sequential
transitions between two adjacent charge states,  say, $N=0$ and
$N=1$.  We concentrate here on this case (to avoid confusion with  the
states of the qubit we continue using the notation $N$ and $N+1$).
The transition rates can be calculated diagrammatically in the
framework of the real-time Keldysh contour technique. The derivation
is presented  in Appendix~\ref{app:Diagrams}. A simplification arises,
since the coherent part of the time evolution of the density matrix, 
which evolves on the time scale set by the qubits energies, is slow
compared to the `tunneling time', given by the inverse of the energy
transfer in  
the tunneling process (see arguments in square brackets in 
Eqs.~\ref{GammaL_check},~\ref{GammaR_check} below). As a result
the self-energy effectively reduces to a delta-function, 
$\Sigma(\Delta t)\propto\delta(\Delta t)$, and the Bloch-master
equation (\ref{MASTER_EQUATION}) reduces to a {\it Markovian} dynamics.

The resulting master  equation  is
translationally invariant in $m$-space. Hence a Fourier transformation
is appropriate,
$\DM{N}{ij}(k)\equiv \sum_m e^{-\i km} \DM{N}{ij}(m)$. As a result
Eq.~(\ref{MASTER_EQUATION_FINAL}) factorizes in $k$-space and we  get
a finite rank ($8 \times 8$) system of equations for each  value of
$k$. This system may be presented in a compact form if we combine the eight 
components of the density matrix into a pair $(\HDM{N}, \HDM{N+1})$ of the 
$2\times2$-matrices
$\hat\rho^{ij}_{N}(k)$, corresponding to $N$ and $N+1$:
\begin{eqnarray}
\hbar\frac{d}{dt}\left(\begin{array}{c}\HDM{N\phantom{+1}}\\
\HDM{N+1}\end{array}\right)
&+&\left(\begin{array}{c} i[\Hcontr,\HDM{N}]\\[3pt]
i[\Hcontr + \dHint,\HDM{N+1}]
\end{array}\right)
\nonumber
\\
&=&
\left(\begin{array}{cc}
-\check\Gamma_{\rm L} \ & e^{-\i k}\check\Gamma_{\rm R}\\
\phantom{-}\check\Gamma_{\rm L} \ & -\check\Gamma_{\rm R}
\end{array}\right)
\left(\begin{array}{c}\HDM{N\phantom{+1}}\\
\HDM{N+1}\end{array}\right)
\; .
\label{Eq:MasterEq}
\end{eqnarray}
The operator $\dHint \equiv  \ECint \sigma_z = \ECpar \rho_z +
\ECper\rho_x$ was introduced in Eq.~(\ref{INTERACTION_HAMILTONIAN}).
The tunneling
rates in the left and right junctions are represented by
operators $\check\Gamma_{\rm L}$ and $\check\Gamma_{\rm R}$, acting on the
qubit's density matrix:
\begin{eqnarray}
\check\Gamma_{\rm L}\HDM{N}&\equiv&
\Gamma_{\rm L}\HDM{N}
+\pi\alpha_{\rm L}\;[\dHint,\HDM{N}]_+
\; ,
\nonumber \\
\check\Gamma_{\rm R}\HDM{N+1}&\equiv&
\Gamma_{\rm R}\HDM{N+1}
-\pi\alpha_{\rm R}\;[\dHint,\HDM{N+1}]_+
\; .
\label{GLR}
\end{eqnarray}
Here $\alpha_{\rm L/R}\equiv R_{\rm K}/(4\pi^2R^{\rm T}_{\rm L/R})$ 
is the tunneling conductance of the left/right junction, measured
in units of the resistance quantum $R_{\rm K}=h/e^2$.  The tunneling
rates in the junctions are determined by the potentials $\mu_{L/R}$ of
the leads and the induced charge $\NgSET$ on the SET's island:
\begin{eqnarray}
\Gamma_{\rm L} &=& 2\pi\alpha_{\rm L}[\mu_{\rm L}-(1-2\NgSET)\ECSET]
\label{GammaL_check}\\
\Gamma_{\rm R} &=& 2\pi\alpha_{\rm R}[(1-2\NgSET)\ECSET-\mu_{\rm R}]
\label{GammaR_check}\\
\Gamma & \equiv&  \Gamma_{\rm L}\Gamma_{\rm R}/(\Gamma_{\rm L}+\Gamma_{\rm R}) 
\,.
\end{eqnarray}
They combine into the
parameter $\Gamma$ which gives the conductance of the SET in the
classical (e.g., high voltage) regime. As a
result of the last terms  in the Eqs.~(\ref{GLR}) the effective rates
and thus the current in the SET are sensitive to the state of the
qubit,  which makes the measurement possible.

\begin{figure}
\centerline{\hbox{\psfig{figure=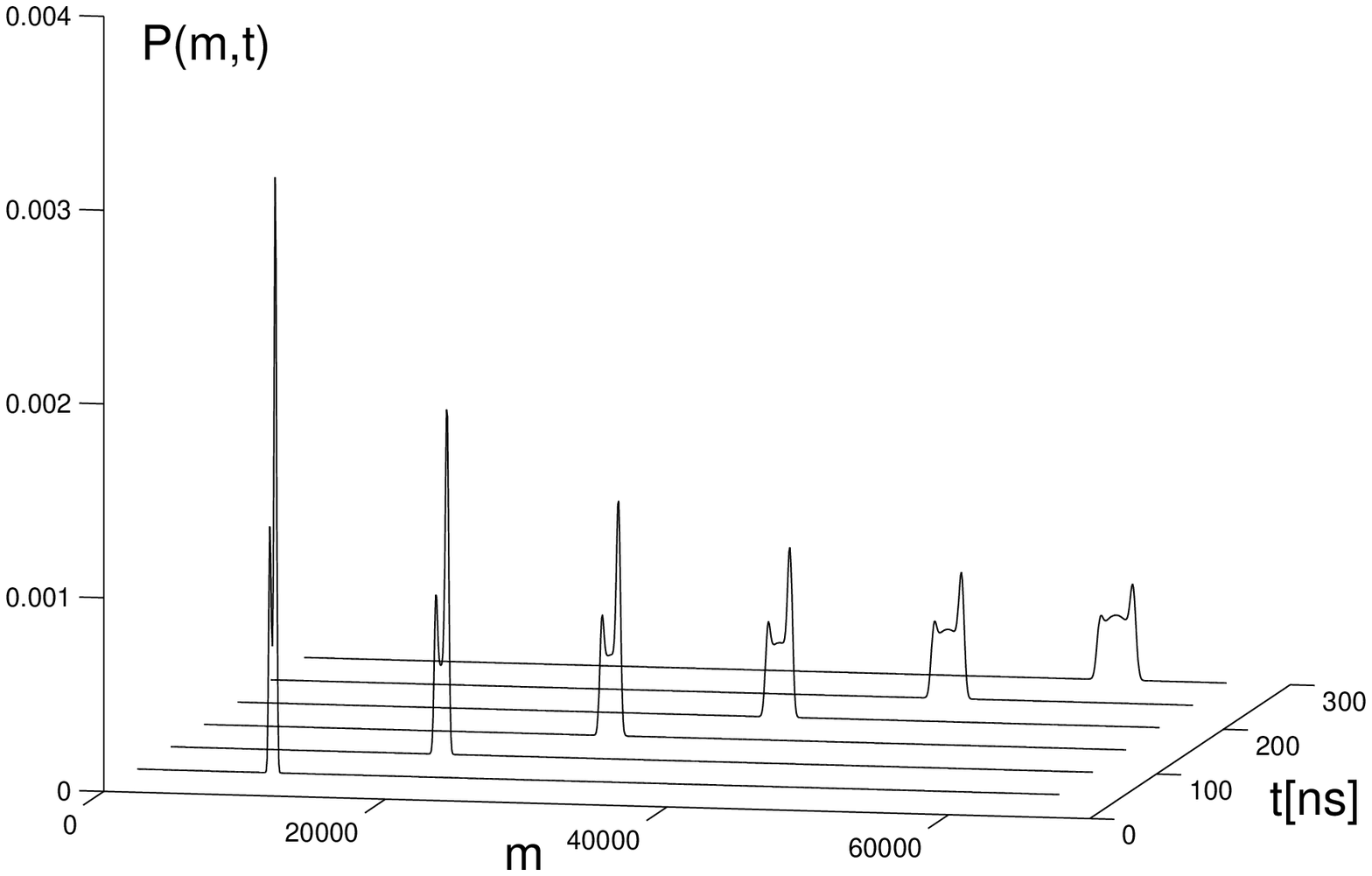,width=0.95\columnwidth}}}
\vspace{3mm}
\caption[]{\label{PLOT0.3}
$P(m,t)$, the probability that $m$ electrons have tunneled during time
$t$. The parameters are the same as in Fig.~\ref{PLOT0.1} except that
$E_{\rm J} = 0.25$~K.}
\end{figure}

To provide a feeling for the types of solutions
we present here numerical solutions of Eq.~(\ref{Eq:MasterEq}) for the
following  system parameters: $B_z = 2$~K, $\ECint = 0.25$~K,
$\alpha_{\rm L} = \alpha_{\rm R} = 0.03$,
$\Gamma_{\rm L} = 1.8$~K, and
$\Gamma_{\rm R}=7.8$~K. We plot the results  for $\EJ = 0.1$~K
(Fig.~\ref{PLOT0.1}) and for  $\EJ = 0.25$~K (Fig.~\ref{PLOT0.3}). We
see that for  the smaller value of $\EJ$ (the eigenbasis of $\Hcontr$
is closer to the charge basis) the probability  distribution $P(m,t)$
develops a two-peak structure.  The weights of the peaks are
equal to the initial  values of the diagonal elements of the qubit's
density  matrix. For the larger value of $\EJ$ the two-peak structure
can also be seen but the valley between the  peaks is filled. This
indicates that the mixing  transitions take place on the same time scale
as that of peak separation, and no good measurement  can be
performed. In Fig.~\ref{P(m/t,t)} the probability  distribution
$P(m,t)$ is plotted for longer times.  (In order to cover many decades
of the time it is necessary to rescale the $m$ axis as well.) 
The parameters are the same as in
Fig.~\ref{PLOT0.1} ($\EJ = 0.1$~K). 
The figure displays clearly the measurement stage and the 
third stage on the mixing time scale. The valley 
between the two peaks fills up and a single, broad 
(note the rescaled $m$ axis) peak develops.

As one can see from (\ref{Eq:MasterEq}) the Hamiltonian of the qubit
switches  between $\Hcontr$ and $\Hcontr+\dHint$ after each tunneling
event in  the SET. This leads, in general, to a complicated
dynamics. Above we showed representative numerical results. 
Next we will analyze Eq.~(\ref{Eq:MasterEq})  perturbatively which
provides insight into the physics of the measurement process.

\begin{figure}
\vspace{5mm}
\centerline{\hbox{\psfig{figure=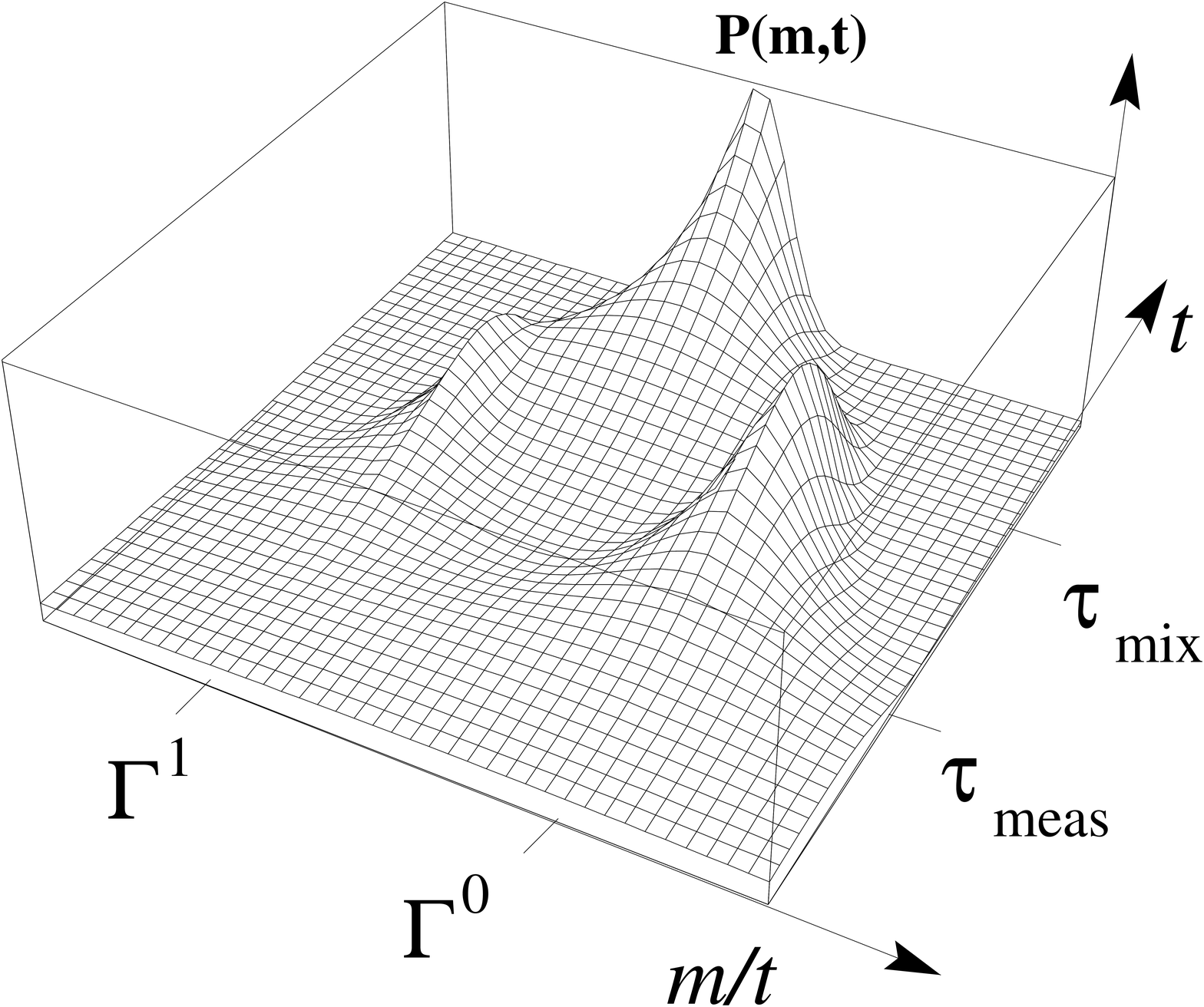,width=0.85\columnwidth}}}
\vspace{5mm}
\caption[]{\label{P(m/t,t)}
$P(m,t)$ obtained in the two-mode approximation (\ref{P(m,t)_FINAL}), plotted 
versus $t$ (on a logarithmic scale) and $m/t$.  The
initial probabilities of the qubit's states are $|a|^2 = 0.75$, $|b|^2=0.25$.}
\end{figure}

In general, the two Hamiltonians $\Hcontr$ and $\Hcontr + \dHint$ are not 
close and their respective eigenbases may be quite different. 
In this situation a better choice for the qubit's Hamiltonian would be 
the ``average'' Hamiltonian 
\begin{equation}
\label{eq:Average_Hamiltonian}
\Hav \equiv \Hcontr + \Nav \dHint
\ ,
\end{equation}  
where $\Nav \equiv \Gamma_{\rm L}/(\Gamma_{\rm L}+\Gamma_{\rm R})$ is the steady 
state 
average of $N$. (The left-right asymmetry follows from the specific choice 
of the sequential tunneling regime: $N=0 \rightarrow N=1$ in the 
{\it left} junction and $N=1 \rightarrow N=0$ in the 
{\it right} junction.)
Relevant energy scales are, then, the level
splitting of the ``average'' Hamiltonian,
\begin{equation}
\Hav = -\frac{1}{2}\DEav \rho_z\;,
\end{equation}
the capacitive coupling 
energy $\ECint$, and the bare tunneling rates $\Gamma_{L/R}$. For 
definiteness we consider the case  
\begin{equation}
\label{Eq:WhiteNoiseCondition}
\ECint \ll \Gamma_{\rm L} + \Gamma_{\rm R}
\ .
\end{equation}
I.e., we exclude  the regimes of
too strong coupling or too weak tunneling  in the SET, which could
be treated similarly to what is presented below. The main simplification 
of the regime (\ref{Eq:WhiteNoiseCondition}) is that the spectrum of 
fluctuations of the electron number
$N$ on the SET island is white in a wide enough range of frequencies,
characterized by the zero frequency noise power 
$S_N(\omega \rightarrow 0) \equiv 2 
\langle \delta N_{\omega\rightarrow 0}^2 \rangle$. When the
SET is detached from the qubit and switched to the  dissipative
regime, this spectrum is given by
$S_N(\omega\rightarrow 0) = 4\Gamma/(\Gamma_{\rm L} + \Gamma_{\rm R})^2$.
The related back-action noise randomizes the relative phase 
between the charge states of the qubit. In the absence of other
sources for dynamics  the rate of this process is 
\begin{equation}
\label{Eq:Gamma_phi_detached}
\tau_{\varphi0}^{-1} = \ECint^{2} S_N(\omega\!\to\!0)
=  { 4 \ECint^{2} \Gamma \over (\Gamma_{\rm L} + \Gamma_{\rm R})^2 }
\ . 
\end{equation}  
It is an important scale for the distinction between different
sub-regimes of (\ref{Eq:WhiteNoiseCondition}), which will be described
in the following.

\subsection{Hamiltonian-dominated regime}
\label{subsec:Ham_Dom_Regime}

Let us first consider the case where the qubit's ``average'' Hamiltonian 
dominates over the dephasing due to the back-action of the SET,  
$\DEav \gg \tau_{\varphi0}^{-1}$.
Then, a perturbative analysis in the eigenbasis
of $\Hav$ is appropriate, where the operator $\dHint$ 
can be rewritten as 
$\dHint = \ECpar \rho_z + \ECper \rho_x$, where
$\ECpar \equiv \Eint\cos\eta_{\rm av}$, 
$\ECper \equiv \Eint\sin\eta_{\rm av}$, and 
$\sin\eta_{\rm av} \equiv \EJ/\DEav$.
In the following we treat the off-diagonal part of $\dHint$, i.e.,
$\ECper \rho_x$, (which leads to mixing, see below)
perturbatively.   
In zeroth order, i.e., for $\ECper =0$, the time evolution of 
$(\DM{N}{ij}(k),\DM{N+1}{ij}(k))$ with different $(i,j)$  are
decoupled from each other. For diagonal elements, $i=j=0$ or $1$, we obtain 
\begin{eqnarray}
\frac{d}{dt}\left(\begin{array}{cc}\DM{N\phantom{+1}}{00/11} \\[2pt]
\DM{N+1}{00/11}\end{array} \right)
= \left( \begin{array}{cc} -\Gamma^{0/1}_{\rm L} & e^{-\i k}
  \Gamma^{0/1}_{\rm R}\\[2pt] \phantom{-}\Gamma^{0/1}_{\rm L} & - 
\Gamma^{0/1}_{\rm R}
  \end{array} \right)
\left(\begin{array}{cc}\DM{N\phantom{+1}}{00/11} \\[2pt]
\DM{N+1}{00/11}\end{array} \right)
\; ,\label{Eq:Diagonal_Zeroth_Order_ME}
\end{eqnarray} 
where $\Gamma_{L}^{0/1}=\Gamma_{L} \pm 2\pi\alpha_{L}\Epar$ and 
$\Gamma_{R}^{0/1}=\Gamma_{R} \mp 2\pi\alpha_{R}\Epar$.
We obtain four eigenmodes ($\propto e^{\lambda t}$), two for each
element ($00$ or $11$).
Most interesting are small values $k\ll1$. In this case, 
two modes with eigenvalue $\lambda^{ii}_-(k)\approx
-\i\G{i} k-\frac{1}{2}f^{i}\G{i} k^2$ and eigenvectors $V_-^{ii}(k)$
given by  $\DM{N}{ii}(k)/\DM{N+1}{ii}(k)\approx \G{i}_{\rm
  R}/\G{i}_{\rm L}$, describe waves in $m$-space propagating  with 
group velocity 
\begin{equation}
\label{Eq:Gamma01}
\G{i} \equiv
\frac{\G{i}_{\rm L}\G{i}_{\rm R}}{\G{i}_{\rm L}+\G{i}_{\rm R}}
\;.
\end{equation}
The wave-packets widen with time as
$\sqrt{2f^{i}\G{i} t}$ due to shot-noise effects. The so called Fano 
factors, 
\begin{equation}
\label{Eq:Fano_Factors}
f^{i} \equiv ({\G{i}_{\rm L}}^2+{\G{i}_{\rm R}}^2)/(\G{i}_{\rm
  L}+\G{i}_{\rm R})^2  \, ,
\end{equation}
determine the suppression of the shot noise in the sequential 
tunneling regime.   
The second pair of eigenmodes
decay fast, $\lambda^{ii}_+(k)\approx -(\G{i}_{\rm L} + \G{i}_{\rm R})$. For
their eigenvectors we obtain $\DM{N}{ii}(k)/\DM{N+1}{ii}(k) \approx
-1$. This fast decay  means that after a few tunneling events,
on the  time scale $(\G{i}_{\rm L}+\G{i}_{\rm R})^{-1}$, detailed balance
$\DM{N}{ii}(k)/\DM{N+1}{ii}(k)\approx \G{i}_{\rm R}/\G{i}_{\rm L}$ is
established. 
For larger values of $k$ ($k \sim 1$) one can check that both 
$\DM{N}{ii}(k)$ and $\DM{N+1}{ii}(k)$ decay fast.

For the off-diagonal elements we obtain 
\begin{eqnarray}
\frac{d}{dt}\left(\begin{array}{cc}\DM{N\phantom{+1}}{01/10} \\[2pt]
\DM{N+1}{01/10}\end{array} \right)
&&= \left[ \left( \begin{array}{cc} -\Gamma_{\rm L}  & e^{-\i k} \Gamma_{\rm 
R}\\[2pt]
 \phantom{-}\Gamma_{\rm L} & - \Gamma_{\rm R} \end{array} \right) \right.
\label{Eq:Off_Diagonal_Zeroth_Order_ME} \\
 &&\left.  \pm \left( \begin{array}{cc} i\Delta E_N  & 0 \\[2pt] 0
 & i\Delta E_{N+1} \end{array} \right)
\right]
\left(\begin{array}{cc}\DM{N\phantom{+1}}{01/10} \\[2pt]
\DM{N+1}{01/10}\end{array} \right)
\; ,
\nonumber
\end{eqnarray}
where
$\Delta E_N = \DEav + 2 \Nav \ECpar$, and
$\Delta E_{N+1} = \DEav - 2(1-\Nav)\ECpar$. 
Again, there are two pairs of eigenmodes with 
eigenvalues such that 
$\lambda^{10}_\pm(k)=[\lambda^{01}_\pm(-k)]^*$.  
One pair of eigenmodes with ${\rm Re} \lambda_+^{01/10}(k\ll 1) \approx 
- (\Gamma_{\rm L} + \Gamma_{\rm R})$ decay fast. For the second pair 
$\lambda_{-}^{10/01}(k \ll 1) \approx \pm i\DEav -\tau_\varphi^{-1}$ where
\begin{equation}
\tau_\varphi^{-1} \approx
\frac{4\Gamma\Eint^{\parallel\;2}}{(\Gamma_{\rm L}+\Gamma_{\rm R})^2} =
{\ECpar}^2 S_N(\omega\rightarrow 0)=
\tau_{\varphi0}^{-1}\cos^2\eta_{\rm av}
\ .
\label{Eq:Gamma_phi}
\end{equation} 
Thus after detailed balance is
established, the $\lambda_{-}^{10}$ and  $\lambda_{-}^{01}$
modes describe coherent
oscillations of the off-diagonal matrix elements  with frequency
of order $\DEav$ and decay rate  $\tau_\varphi^{-1}$. For larger values of
$k$ the decay times are of the same order  or shorter than those for 
$k \approx 0$.

The fast decaying diagonal modes $\lambda_+^{00/11}$ do not 
contribute to $P(m,t)$, since for these modes 
$\DM{N}{ii}(k) + \DM{N+1}{ii}(k) \approx 0$. 
Thus there are only two modes contributing, with eigenvalues
$\lambda_-^{00}(k)$ and $\lambda_-^{11}(k)$. Starting from the initial 
occupation probabilities   
$|a|^2$ and $|b|^2$ we obtain   
\begin{equation}
\label{Eq:P(k,t)NOMIXING}
P(k,t)
\approx
|a|^2 e^{\lambda_{-}^{00}(k)\;t} + |b|^2 e^{\lambda_{-}^{11}(k)\;t}
\ ,
\end{equation}
where $P(m,t) \equiv \int dk/(2\pi) \, P(k,t)\,e^{\i km}$. 
This form describes the evolution of the
distribution $P(m,t)$ from the initial $\delta(m)$ at $t=0$  into
two peaks. The peaks shift to positive $m$-values linear in $t$ 
with velocities $\Gnull$ and
$\Gone$, and they grow in widths as $\sqrt{2f^{i}\G{i}t}$. 
This time dependence implies that only after a
certain time, which we denote as ``measurement time'' 
 $\tau_{\rm meas}$, 
the two peaks emerge from the broadened distribution. 
The associated rate is 
\begin{equation}
\tau_{\rm meas}^{-1} =
\left(
        \frac{\Gnull-\Gone}{\sqrt{2f^{0}\Gnull}+\sqrt{2f^{1}\Gone}}
\right)^2
\;.
\label{Eq:Gamma_Measurement}
\end{equation}
 In the linear response regime,
when $\Gnull$ and $\Gone$ are close, we obtain
\begin{equation}
\label{eq:G_mes_linear}
\tau_{\rm meas}^{-1} = {(\Delta I)^2 \over 4 S_I} 
\ ,
\end{equation}
where $S_I$ is the zero frequency power of the shot 
noise in the SET and $\Delta I = e(\Gnull - \Gone)$.
The weights of the peaks are given by the initial  weights of the
eigenstates of $\Hav$, $|a|^{2}$  and $|b|^{2}$. Measuring the
charge $m$  thus constitutes a perfect quantum
measurement~\cite{Our_PRB}.

Next we analyze the effect of the perturbation $\ECper \rho_z$ in
the master equation (\ref{Eq:MasterEq}). It appears in both 
the coherent (LHS) and incoherent (RHS) parts  of
Eq.~(\ref{Eq:MasterEq}) and leads to the mixing. As usual the  perturbation
has the strongest effect when it lifts a (near) degeneracy, 
since in this case the eigenvectors within the degenerate
subspace may change substantially. Therefore we first treat the  two
near degenerate modes, i.e.,  we restrict ourselves
to the two-dimensional subspace spanned by
$V_-^{00}(k=0)$ and $V_-^{11}(k=0)$, and check how
the  degeneracy between these two modes is lifted in
second order of perturbative expansion.  To account for this we
approximate the diagonal part of the density matrix as
\begin{equation}
\label{Eq:Diagonal_Approximation} 
\hat\rho_{\rm diag}(k,t) \approx 
A^{0}(k,t)V_-^{00}(k=0) + A^{1}(k,t)V_-^{11}(k=0)
\ ,
\end{equation}
(the LHS should be understood as an eight-column vector
consisting of all matrix elements of $\hat\rho_{\rm diag}$) 
for which we obtain an effective (reduced) master equation:
\begin{equation}
\label{Eq:REDUCED_MASTER_EQUATION}
\frac{d}{dt}\left(\begin{array}{cc} A^{0}(k) \\
A^{1}(k) \end{array} \right) = M_{\rm red}
\left(\begin{array}{cc} A^{0}(k) \\
A^{1}(k) \end{array} \right)
\;,
\end{equation} 
where
\begin{equation}
\label{Eq:M_RED}
M_{\rm red} =
\left(\begin{array}{cc}
\lambda^{00}_{-}(k) & 0
\\
0 & \lambda^{11}_{-}(k)
\end{array}
\right) 
+ {1\over {2\tau_{\rm mix}}}
\left(\begin{array}{cc}
-1 & \phantom{+}1
\\
\phantom{+}1 & -1
\end{array}
\right)
\ . 
\end{equation}
The second term in (\ref{Eq:M_RED}) results from the perturbative
expansion and, indeed, lifts the degeneracy between the two modes.
The mixing rate is obtained in second order perturbation expansion as
\begin{equation}
\label{Eq:GMIX}
\tau_{\rm mix}^{-1} =
\frac{4\Gamma {\ECper}^2}{\DEav^{2}
+(\Gamma_{\rm R} + \Gamma_{\rm L})^2} = {\ECper}^2 S_N\left(\omega = 
\frac{\DEav}{\hbar}\right)
\; .
\end{equation}
In the approximation (\ref{Eq:Diagonal_Approximation}) we have
$A^i(k,t)=\sum_N\DM{N}{ii}(k)$. At $t=0$ these are occupation
probabilities of the eigenstates of $\Hav$, $A^0(k,t=0)=|a|^2$,
$A^1(k,t=0)=|b|^2$, for all $k$. 
It is straightforward to diagonalize Eq.~(\ref{Eq:M_RED}) to find the
eigenvalues. While the sum of the  occupation probabilities is
conserved, $A^0(k=0,t)+A^1(k=0,t)=1$, the difference decays,
$A^0(k=0,t)-A^1(k=0,t)\propto \exp(-t/\tau_{\rm mix})$. Thus both 
occupations tend to 1/2 in the long-time limit.  This implies that the
evolution of $P(m,t)$ is given by Eq.~(\ref{Eq:P(k,t)NOMIXING})
only for times $t \ll \tau_{\rm mix}$, and in order to perform a
measurement one must have $\tau_{\rm mix}\gg \tau_{\rm meas}$.  As
$\tau_{\rm meas}^{-1} \propto {\ECpar}^2 \propto \cos^2\eta_{\rm av}$ and  
$\tau_{\rm mix}^{-1} \propto {\ECper}^2 \propto \sin^2\eta_{\rm av}$, 
a good  measurement may always be achieved by choosing
$\tan^2\eta_{\rm av}$ small enough.

Another effect of the mixing perturbation is that the dephasing 
rate, i.e., the rate of vanishing of the off-diagonal elements 
of the density matrix is changed as 
\begin{equation}
\tau_\varphi^{-1} = \tau_{\varphi0}^{-1}\cos^2\eta_{\rm av} + 
{1/2\tau_{\rm mix}}
\ .
\end{equation} 
This is analogous to the relation between the dephasing (\ref{Eq:dephasing})
and relaxation (\ref{Eq:relaxation}) rates in the spin-boson model.
It reflects the fact that the relaxation of the diagonal elements 
of the density matrix leads also to additional suppression of the 
off-diagonal ones. Thus even when $\eta_{\rm av}=\pi/2$,
the off-diagonal elements vanish with the rate $1/2\tau_{\rm mix}$.  

The long-time behavior of the qubit and detector, excluding the
period of the initial dephasing, is dominated by the two slowly
decaying modes. In this regime we obtain from
Eqs.~(\ref{Eq:REDUCED_MASTER_EQUATION},\ref{Eq:M_RED})
the reduced time evolution operator
\begin{equation} 
\label{U_RED}
U_{\rm red}(k,t) \equiv \exp[M_{\rm red}(k)\,t]
\;.
\end{equation} 
Its Fourier transform $U_{\rm red}(m,t)$ yields the expression for
the distribution 
\begin{equation}
\label{P(m,t)_U}
P(m,t) = (1,\,1) \cdot U_{\rm red}(m,t) \cdot
\left(\begin{array}{cc}
|a|^2 \\ |b|^2
\end{array}
\right)
\;.
\end{equation}
This Fourier transform can be performed analytically
\cite{Our_Current_PRL}, and we  arrive at
\begin{equation}
\label{P(m,t)_FINAL}
P(m,t)=
\sum_{m'=-\infty}^{\infty} {\tilde P}(m-m',t)
\;
\frac{%
e^{-{m'}^2/2f\bar\Gamma t}
}
{\sqrt{2\pi f\bar\Gamma t}}
\;,
\end{equation}
where
\begin{eqnarray}
&&\tilde P(m,t)=
P_{\rm pl}\left(2{m-\bar \Gamma t \over \delta\Gamma \, t}\;,\;
\frac{t}{2\tau_{\rm mix}} \right)
\nonumber \\
&&
+ e^{-t/2\tau_{\rm mix}} \left[|a|^2\delta\left(m-\Gamma^{0} t\right)
+|b|^2\delta\left(m-\Gamma^{1} t\right)\right]
\label{Eq:Ptilde}
\end{eqnarray}
and
\begin{equation}
\label{Eq:Gamma0/1}
\Gamma^{0/1}\equiv\bar\Gamma\pm\delta\Gamma/2
\;.
\end{equation}
We observe that the solution is constructed from two 
delta-peaks, smeared by the convolution with the shot-noise Gaussian,
and a plateau between them, which for $|x|<1$ is given by
\begin{eqnarray}
\label{Ppl}
&&P_{\rm pl}(x,\tau)=e^{-\tau}
\frac{1}{2\,\delta\Gamma\,\tau_{\rm mix}}
\left\{
I_0\left(\tau
\sqrt{1-x^2}\right)
\right.
\nonumber \\
&&+\left.
\left[1+x (|a|^2-|b|^2)\right]
I_1\left(\tau\sqrt{1-x^2}\right)/\sqrt{1-x^2}
\right\} \,,
\end{eqnarray}
while $P_{\rm pl}=0$ for $|x|>1$. Here $I_0$, $I_1$ are the 
modified Bessel functions. 
We work in the limit of weak qubit-detector coupling, where the Fano factors 
(\ref{Eq:Fano_Factors}) are close, and we denote them simply by $f$.
For short times, $t \ll \tau_{\rm mix}$, the peaks dominate, while the
plateau is 
low. At longer times the initial peaks disappear while the plateau 
is transformed into a single central peak (see Fig.~\ref{P(m/t,t)}).

We complement the results for the charge distribution function
$P(m,t)$ by a derivation of the distribution function for the values
of the current.  The measured quantity is actually the current
averaged over a certain time interval $\Delta t$, i.e., 
$I_{\Delta t} \equiv [m(t+\Delta t)-m(t)]/\Delta t \equiv 
\Delta m/\Delta t$. Accordingly the quantity of interest $P(I_{\Delta
t},t)$ can be expressed by the joint probability that $m$ electrons
have tunneled at time $t$ and $m+\Delta m$ electrons at a later time
$t+\Delta t$: 
\begin{eqnarray}
\label{P(m1,m2)_MARKOVIAN}
P(I_{\Delta t},t) & = & \sum _m P(m,t;m+\Delta m,t+\Delta t)
\nonumber \\
& =  & \sum _m {\rm Tr'}\,\left[
U(\Delta m,\Delta t)\,U(m,t)\hat\rho_{0}
\right] \, .
\end{eqnarray}
Here we made use of the Markovian approximation (see Section~\ref{subsec:ME}).
The trace is taken over all degrees of freedom except $m$, and also
$\hat\rho_0$ refers to their initial state.
In the two-mode approximation 
(\ref{Eq:REDUCED_MASTER_EQUATION},\ref{Eq:M_RED}),
sufficient at long times,
we replace $U$ again by $U_{\rm red}$ and obtain
\begin{eqnarray}
&&P(I_{\Delta t},t)=(1\;1)\cdot U_{\rm red}(\Delta m, \Delta t)\cdot
\nonumber \\
&&\cdot \left\{ {1\over 2}
\left(\begin{array}{cc}
1 \\ 1
\end{array}
\right)
+ {|a|^2-|b|^2\over 2}e^{-t/\tau_{\rm mix}}
\left(\begin{array}{cc}
\phantom{+}1 \\ -1
\end{array}
\right) 
\right\}
\label{Eq:PIt_Final}
\end{eqnarray}

\begin{figure}  
\hskip0.05\columnwidth
\centerline{\hbox{\psfig{figure=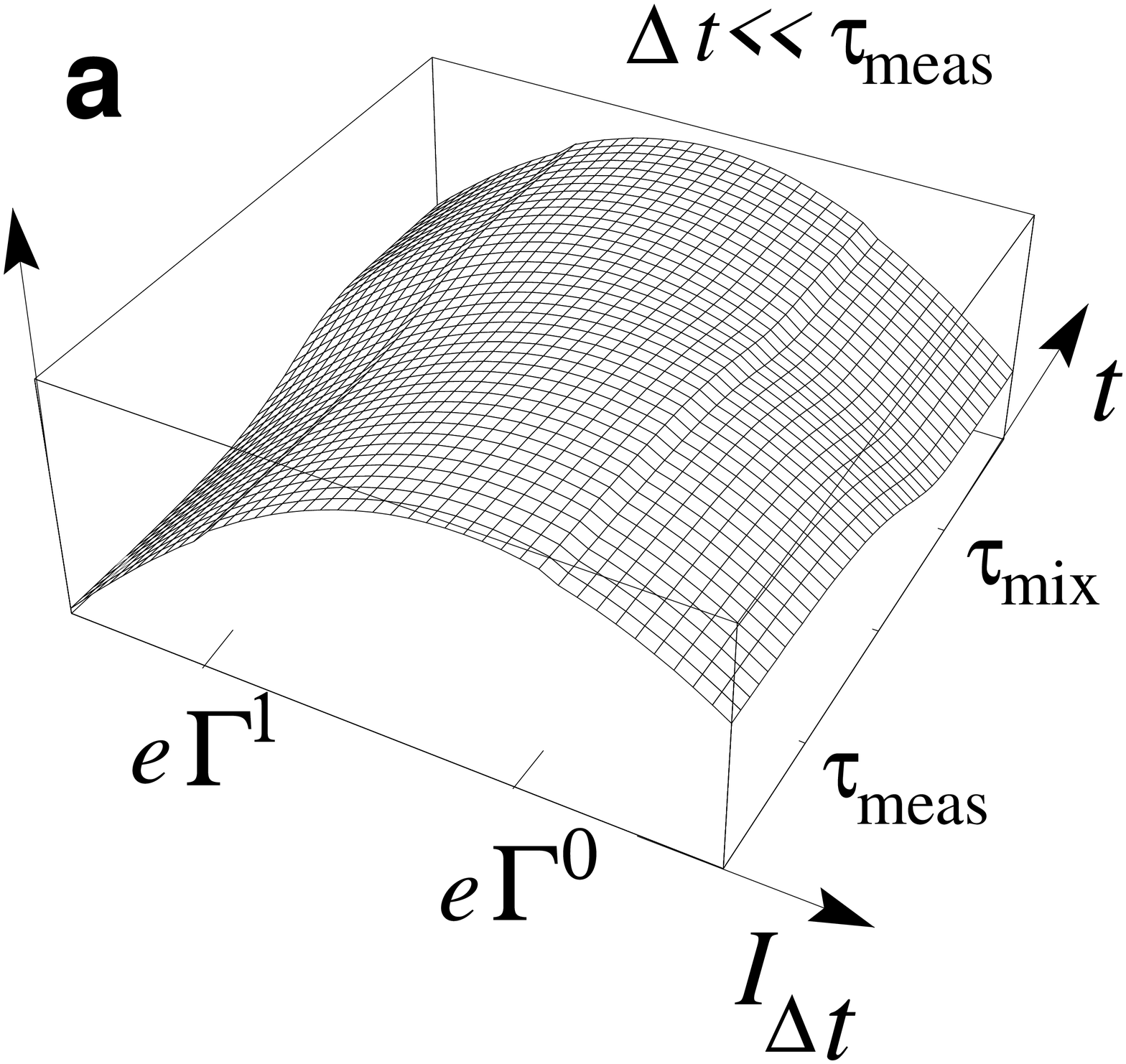,width=0.45\columnwidth}%
\hskip0.05\columnwidth
\psfig{figure=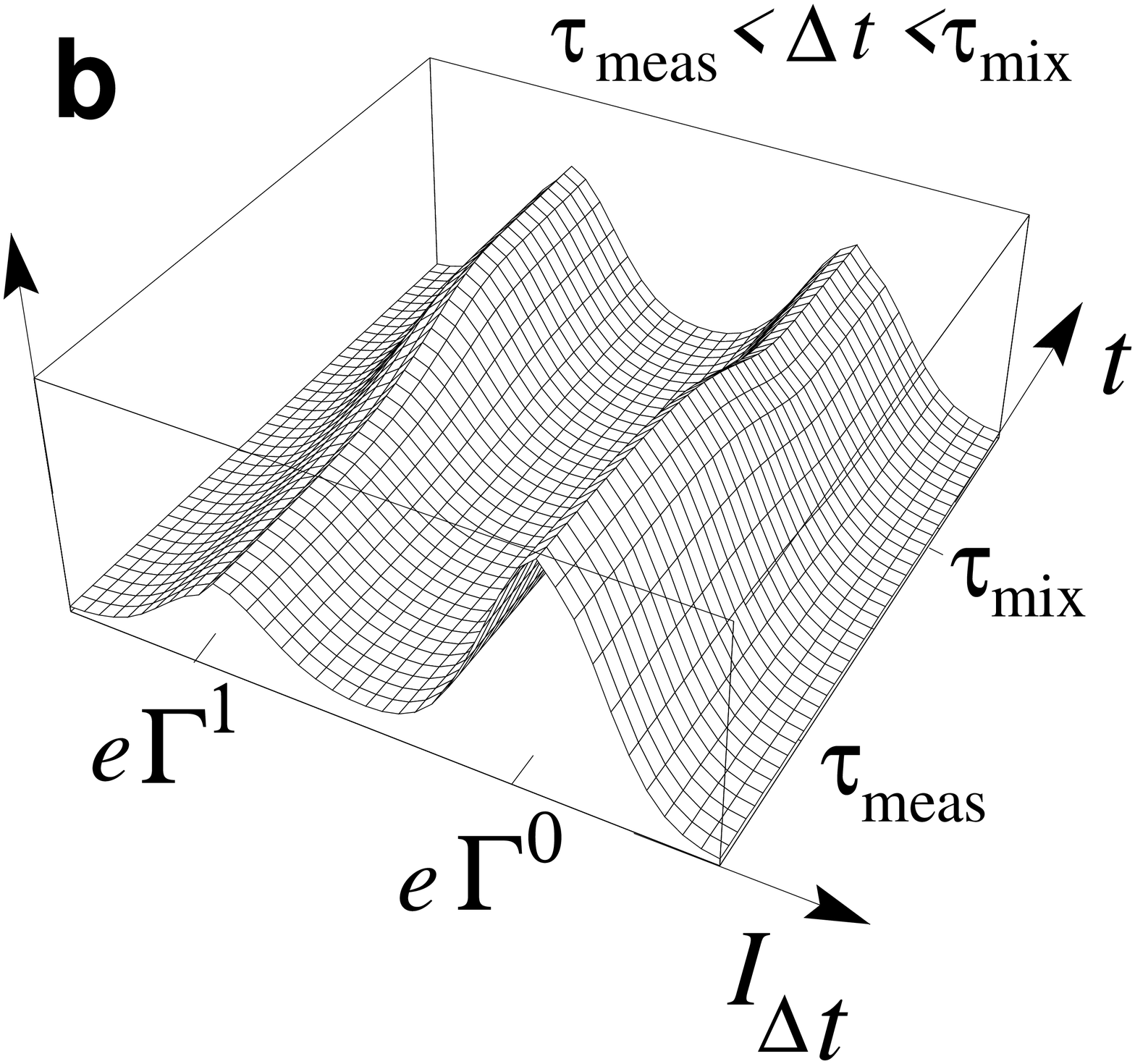,width=0.45\columnwidth}}}
\vskip3mm
\centerline{\hbox{\psfig{figure=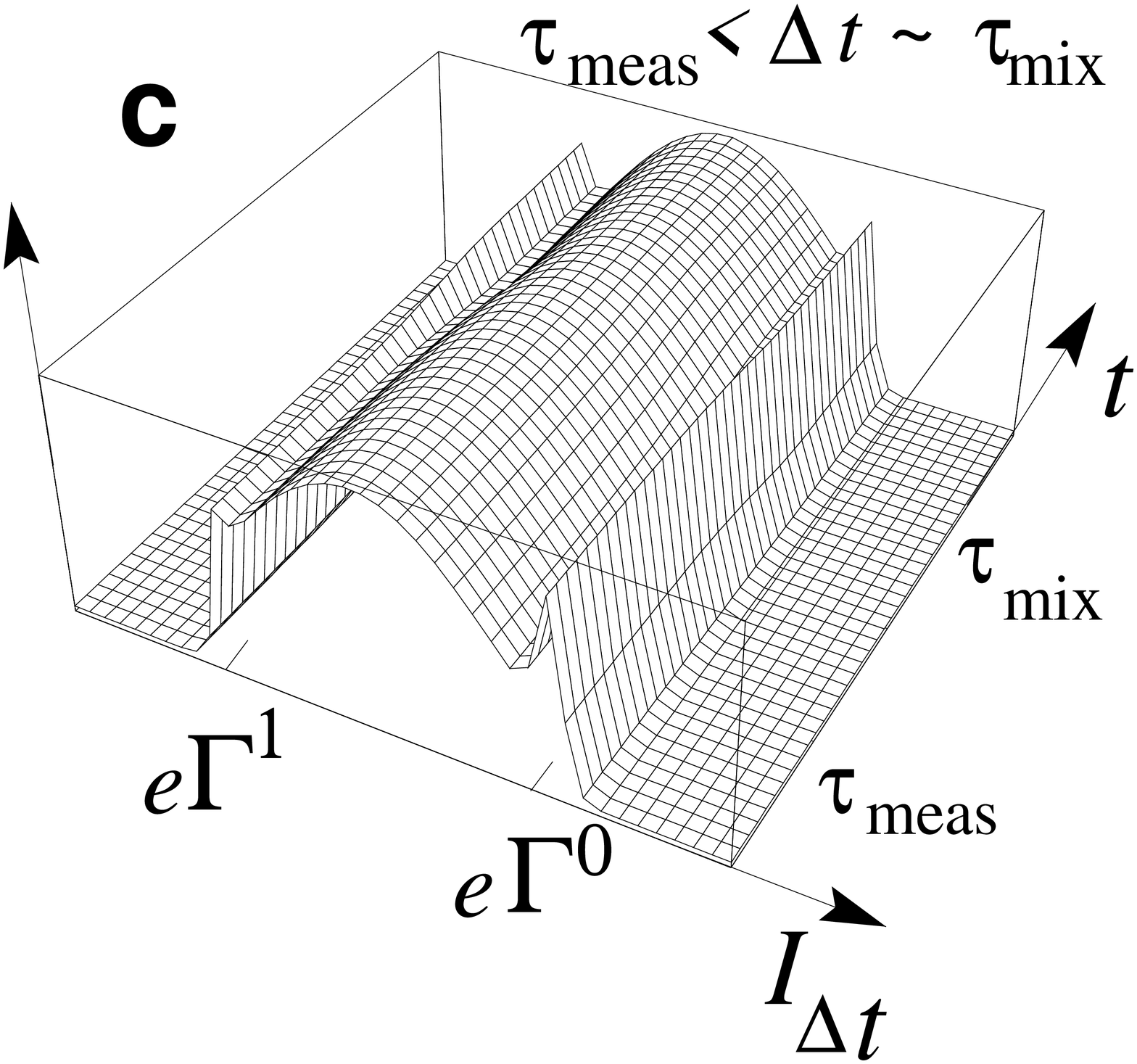,width=0.45\columnwidth}%
\hskip0.05\columnwidth
\psfig{figure=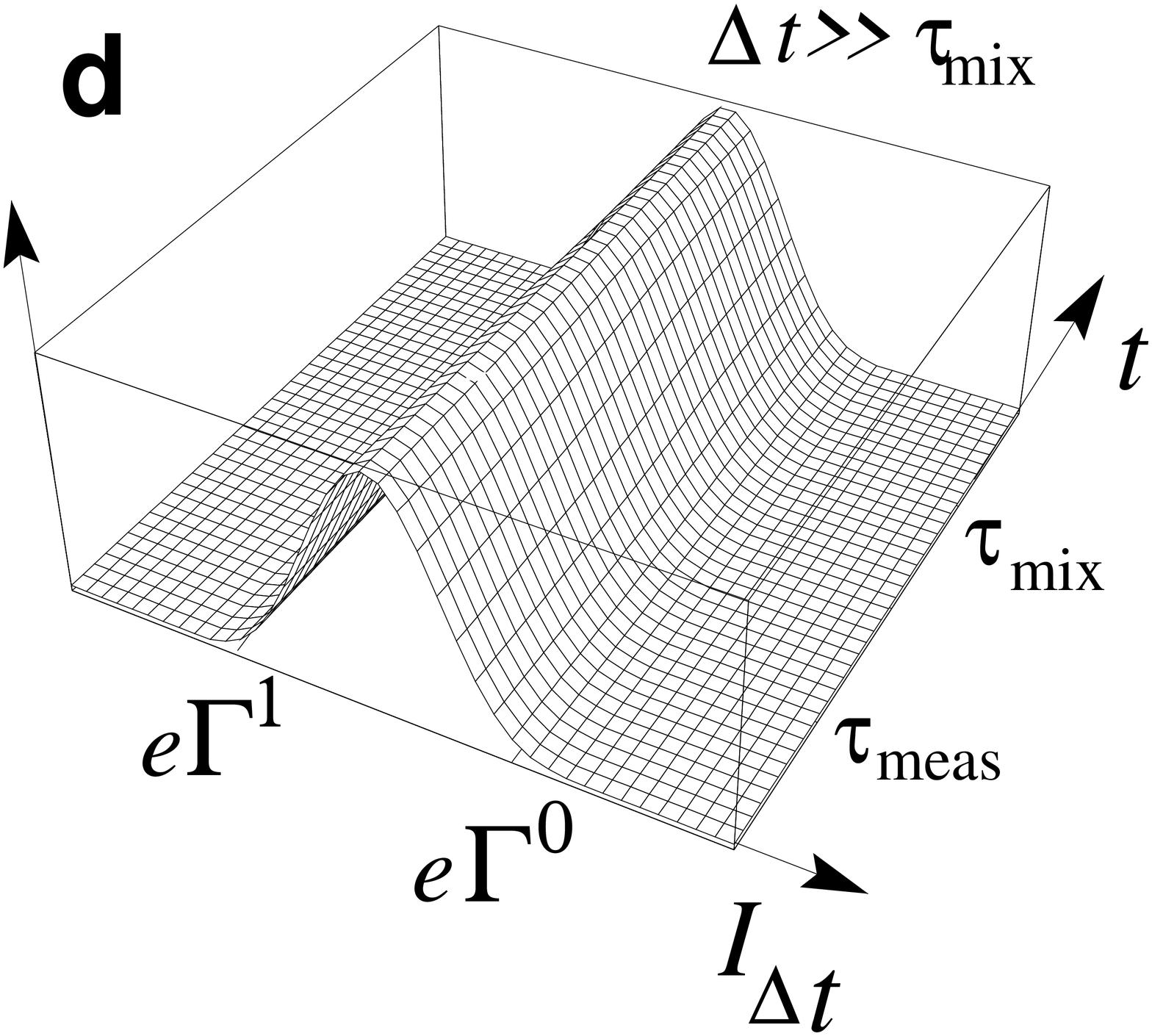,width=0.45\columnwidth}}}
\vskip3mm
\caption[]{\label{Figure:P(I)}
Probability distribution $P(I_{\Delta t},t)$ of the current averaged over 
various time intervals $\Delta t$. The time axis is plotted on a
logarithmic scale.}
\end{figure}
The behavior of
$P(I_{\Delta t},t)$ is displayed in Fig.~\ref{Figure:P(I)} for various
values of  $\Delta t$: 
(a)~If the current is averaged
over very short intervals, $\Delta t\ll \tau_{\rm meas}$
[Fig.~\ref{Figure:P(I)}~(a)], the detector does not have 
enough time to extract the signal from the shot-noise governed background.  
(b)~An effective quantum measurement is achieved if  
$\tau_{\rm meas} < \Delta t < \tau_{\rm mix}$. In this case the
qubit-sensitive signal  can be seen on top of the shot noise. 
Hence, as seen in Fig.~\ref{Figure:P(I)}~(b),  
the measured value of $I_{\Delta t}$ is either close to $e\Gamma^0$ or to
$e\Gamma^1$ (\ref{Eq:Gamma01}). The corresponding probabilities (the
weights of the two peaks) are initially $|a|^2$ and $|b|^2$. They change at 
longer times, $t>\tau_{\rm mix}$, to a 1/2-1/2-distribution due to the
mixing.   
A typical current pattern is a telegraph signal
jumping between $e\Gamma^{0}$ and $e\Gamma^{1}$ on a time scale $\tau_{\rm
mix}$.
(c,d)~If the current is averaged over longer times $\Delta
t\ge\tau_{\rm mix}$,  
the meter-induced mixing erases the information about the qubit's 
state. Several telegraph jumps can occur on this time scale, and for
$\Delta t\gg\tau_{\rm mix}$ one
measures only the time-averaged current between $e\Gamma^{0}$ and 
$e\Gamma^{1}$,  completely insensitive to the initial 
values. This is shown in Fig.~\ref{Figure:P(I)}~(d), while
Fig.~\ref{Figure:P(I)}~(c) 
displays the crossover between (b) and (d).

Notice, that since $P(I_{\Delta t},t\!=\!0)=P(m\!=\!I\Delta t,t\!=\!\Delta t)$,
the zero-time limits of the surfaces in Figs.~\ref{Figure:P(I)}~(a)--(d)
are given by the charge distribution function
plotted in Fig.~\ref{P(m/t,t)}.

We point out that in the Hamiltonian dominated regime 
the current in the SET is sensitive to the 
occupation probabilities of the {\it eigenstates} of $\Hav$, rather
than those in the basis of charge states.
This may appear surprising, since the SET couples to the charge operator of 
the qubit.
More precisely, the current is only sensitive to the {\it expectation
value} of the charge in each eigenstate.  As a consequence, at the
degeneracy point $\eta_{\rm av} = \pi/2$, where the two eigenstates
have the same average charge, both eigenstates lead to the same current in
the SET and no measurement is possible.  The measurement is effective
only when charges in the eigenstates differ, $\tan\eta_{\rm av} \ll 1$.

\subsection{Detector-dominated regime}
\label{subsec:Detector-dom-regime}

When the back-action dephasing dominates over the 
average Hamiltonian, $\tau_{\varphi0}^{-1} \gg \DEav$, 
a perturbative analysis in the charge basis is appropriate.
In this basis the perturbation is the Josephson term
of the qubit's Hamiltonian $\Hcontr$, i.e., $-(1/2)B_x \sigma_x$.  
Starting from the zeroth order, $B_x = \EJ = 0$, we obtain equations
similar to Eqs.~(\ref{Eq:Diagonal_Zeroth_Order_ME},
\ref{Eq:Off_Diagonal_Zeroth_Order_ME}) with the replacements
$\DEav \rightarrow B_{z,\rm av}$ and $\ECpar \rightarrow \ECint$,
where $B_{z,\rm av} \equiv B_z - 2\Nav\ECint$ is the charging
energy of the ``average'' Hamiltonian.  
The
analysis of the diagonal and the off-diagonal modes is performed
similarly. We get  for the dephasing rate $\tau_\varphi^{-1} =
\tau_{\varphi0}^{-1}$.  For the measurement time we reproduce
Eq.~(\ref{Eq:Gamma_Measurement}).  The dynamics of the two
long-living diagonal modes can again be reduced to
Eqs.~(\ref{Eq:P(k,t)NOMIXING}), (\ref{Eq:Diagonal_Approximation}),
(\ref{Eq:REDUCED_MASTER_EQUATION}) and (\ref{Eq:M_RED}) with
\begin{equation}
\label{GMIX_ZENO}
\tau_{\rm mix}^{-1} \approx \EJ^2  \tau_{\varphi0}
\ .
\end{equation}
This result is standard for the Zeno regime, i.e., the regime  when coherent
oscillations are overdamped by dephasing
(cf.~Eq.~\ref{Eq:Tau_Rel_Zeno}). 

The condition for the Zeno regime given above requires a rather
strong dephasing, such that  $\tau_{\varphi0}^{-1}$ exceeds both
components of the qubit's Hamiltonian  (\ref{Eq:Magnetic_Hamiltonian}),
$\tau_{\varphi0}^{-1}\gg\EJ, B_{z,\rm av}$. 
The second part of this condition can be satisfied by tuning the qubit
close to the degeneracy point.  In contrast,
in  the Hamiltonian-dominated regime it is desirable for a good
measurement to switch the qubit away from the degeneracy point. 

The long-time behavior of the charge and current distributions,
$P(m,t)$ and  $P(I_{\Delta t},t)$, is again given by
Eqs.~(\ref{P(m,t)_FINAL}, \ref{Eq:PIt_Final}) but with proper
redefinitions of $\Gamma^0$, $\Gamma^1$, $\tau_{\rm mix}^{-1}$ etc.\
which now refer to the detector-dominated regime. In particular, the
measurement provides information about the initial occupations of the
qubit charge states rather than the eigenstates.

To summarize we present a table with the main results for the two regimes: 
{
\begin{center}
\begin{tabular}{|c||c|c|}
\hline
regime          & Hamiltonian & fluctuation\\
                & dominated   & dominated \\    
                & (coherent)    & (Zeno)\\[2mm]
leading term    & {$\Hav$}      & {$\delta N(t) \dHint$}\\[2mm]
\hline
largest energy  & ${\LARGE\mathstrut}{\DEav}\gg{\tau_{\varphi0}^{-1}}$ &
                ${\tau^{-1}_{\varphi0}}\gg{\DEav}
                $\\[2mm]
preferred basis & eigenbasis            & charge basis\\[2mm]
perturbation & {$E_{\rm int}\sin\eta_{\rm av}\,\rho_x$} 
& {$E_{\rm J}\,\sigma_x$}\\
[2mm]
\hline
${\LARGE\mathstrut}\tau_\varphi^{-1}$ & $\tau_{\varphi0}^{-1}\cos^2\eta_{\rm av} 
+ 
\frac{1}{2}\tau_{\rm mix}^{-1}$ 
& $\tau_{\varphi0}^{-1}$\\[1mm]
\hline
${\LARGE\mathstrut}\tau^{-1}_{\rm meas}$ &
$\displaystyle \propto E_{\rm int}^2 \cos^2\eta_{\rm av}$ &
$\displaystyle \propto E_{\rm int}^2 $\\[2mm]
\hline
${\Huge\mathstrut}\tau^{-1}_{\rm mix}$        & {$\displaystyle 
                          \frac{4\Gamma (E_{\rm int}\sin\eta_{\rm av})^2}
                          {(\Gamma_{\rm L}+\Gamma_{\rm R})^2+\DEav^2}$} &
        {$\displaystyle E_{\rm J}^2 \, \tau_{\varphi0}$}\\[5mm]
\hline
\end{tabular}
\end{center}
}

\subsection{Flux measurements}

\begin{figure}  
\centerline{\hbox{\psfig{figure=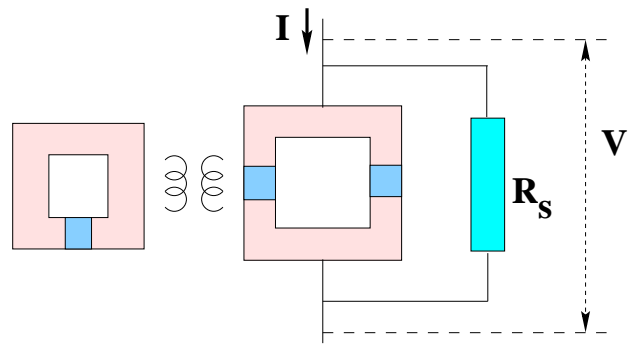,width=0.95\columnwidth}}}
\vspace{5mm}
\caption[]{\label{Figure:SQUID_MEASUREMENT}
Measurement setup for a flux qubit. The qubit (the rf-SQUID
on the left side) is inductively coupled to the 
meter (the shunted dc-SQUID on the right side).}
\end{figure}

To measure the quantum state of a flux qubit (for simplicity we
consider here a single junction rf-SQUID), one needs a dissipative
device sensitive to the magnetic flux. Natural 
candidates are dc-SQUIDs, i.e., superconducting loops interrupted by 
two Josephson junctions. Coupling the qubit and the SQUID 
inductively, as shown in Fig.~\ref{Figure:SQUID_MEASUREMENT}, one 
transfers part of the qubit's flux into the loop of the 
SQUID and, thus, makes the SQUID sensitive to the state of the 
qubit.
    
SQUIDs have been used as ultra-sensitive magnetometers
for many years, and extensive literature covers 
the physics of the SQUIDs in much detail 
\citeaffixed{Tinkham,Likharev_Josephson_Book}{see e.g.}.
Here we mention only some of the recent activities and ideas 
which arose in connection with flux qubits.
Two main strategies were proposed. The first one is to use
underdamped dc-SQUIDs in the hysteretic regime \cite{Mooij}.
This regime is realized when the SQUID is unshunted or 
the shunt resistance is large,
$R_{\rm s} \gg (E_{\rm J}/E_{\rm C})\,R_{\rm K}$, where 
$E_{\rm J}$ and $E_{\rm C}$ are the characteristic Josephson
and charging energies of the SQUID. The higher the shunt
resistance  $R_{\rm s}$,
the lower is the noise of the SQUID in the superconducting 
regime (when no measurement is performed). Therefore, 
in spite of the permanent presence of the SQUID, the 
coherent dynamics of the qubit suffers only weak dephasing. To read
out the state of the  
qubit, the current in the SQUID is ramped, and the value of current
 is recorded where the SQUID switches to the dissipative regime.
As this switching current depends on the flux through the SQUID, 
information about the state of the qubit is obtained. Unfortunately,
the current  switching is a random process, fluctuating even when the
external flux in the SQUID 
is fixed. For currently available system parameters this spread 
is larger than the difference in switching currents corresponding 
to the two states of the qubit. Therefore only statistical (weak repeated) 
measurements are possible in this regime. 

The second strategy is to use overdamped SQUIDs, with $R_{\rm s} < (E_{\rm
J}/E_{\rm C})\,R_{\rm K}$.  When the bias current exceeds the critical value,
the voltage which develops across the shunt resistor depends on the external
flux in the SQUID.  Thus by measuring this voltage one learns about the state
of the qubit.  In this regime the principle of the measurement is identical to
the one presented above for the SET.  Recently \citeasnoun{Averin_SQUID}
analyzed continuous (stationary) measurements in this regime and obtained the
input and output noise characteristics
(cf.~Section~\ref{subsec:Diss_Efficiency}) which determine the relevant time
scales $\tau_\varphi$, $\tau_{\rm meas}$, $\tau_{\rm mix}$.  The main
disadvantage of this strategy is that the SQUID induces dephasing during the
periods of coherent manipulations when no measurement is performed.  The
question still remains to be settled whether a reasonable
compromise between the underdamped and the overdamped limits can be found.

\subsection{Efficiency of the measuring device}
\label{subsec:Diss_Efficiency}

Recently, several devices performing quantum measurements 
have been analyzed. Apart from SETs in the sequential tunneling regime
\cite{Our_PRB,Our_Current_PRL,Korotkov_Selective,Devoret_Schoelkopf_Nature}
it includes SETs in the co-tunneling regime 
\cite{Averin_SET_Cotunneling,Maassen_SET_Cotunneling}, 
superconducting SETs (SSETs) and  dc-SQUIDS \cite{Averin_SQUID}, as
well as quantum point contacts (QPC)  \cite{Averin_Korotkov}.
All these devices are based on the same basic idea: they are dissipative 
systems whose response (conductance, resistance) depends on the state of 
a qubit coupled to them. 

The efficiency of a quantum detector has several aspects.
From a practical point of view the most important is the ability to perform a 
strong, single-shot measurement which requires that the mixing is slower than 
the read-out, $\tau_{\rm mix}\gg\tau_{\rm meas}$. Another desired
property is a low back-action noise of the meter in the off-state,
which can be characterized by a corresponding  dephasing rate.

A further important figure of merit 
is the ratio of the dephasing and measurement times. Quantum 
mechanics demands that a quantum measurement should completely dephase
a quantum state, i.e.\ $\tau_\varphi \le \tau_{\rm meas}$. 
For the example of a SET coupled to a charge qubit the
dephasing time (\ref{Eq:Gamma_phi}) is smaller (or even much smaller)
than the measurement time (\ref{Eq:Gamma_Measurement}). 
 This means that the information becomes available later than it would be
possible in principle.
In this sense the efficiency of the SET in the sequential tunneling regime 
is less then $100\%$~\cite{Our_PRB,Korotkov}. 
The reason for the delay is an entanglement of the qubit
with microscopic degrees of freedom in the SET. To illustrate this
point consider a situation where the initial state of the
system $(a|0\rangle + b|1\rangle)\,|\chi\rangle\,|m=0\rangle$ evolves
into $a|0\rangle\,|\chi_0\rangle\,|m_0\rangle
+b|1\rangle\,|\chi_1\rangle\,|m_1\rangle$, where $|\chi\rangle$ stands
for the quantum state of the uncontrolled environment. One can imagine
a situation  when $m_0 = m_1$, but $|\chi_0\rangle$ and
$|\chi_1\rangle$ are orthogonal. Then the dephasing has
occurred but no measurement has been  performed.
It is interesting to note that the ratio $\tau_\varphi/\tau_{\rm meas}$ 
grows if the SET is biased in an asymmetric way, creating a strong
asymmetry in the tunneling rates, e.g.\ $\Gamma_{\rm L} \ll \Gamma_{\rm R}$. 
In other measurement devices the ratio $\tau_\varphi/\tau_{\rm meas}$
may be close to 1. This includes quantum point contacts 
\cite{Korotkov,Averin_Korotkov} as well as SETs in the co-tunneling regime
\cite{Averin_SET_Cotunneling,Maassen_SET_Cotunneling}. The common
feature of these three examples is that the device consists of one
junction or effectively reduces to it. It should be
kept in mind, however, that a large ratio of  dephasing and measurement
times is not the only figure of merit. For instance, in a SET in the
cotunneling regime the current is low and more difficult to detect as
compared to a SET in the sequential tunneling regime.

The ratio of dephasing and measurement times $\tau_\varphi/\tau_{\rm
meas}$ has been analyzed also in the framework of the theory  
of linear amplifiers \cite{Averin_SQUID,Devoret_Schoelkopf_Nature}.
It can be expressed 
in terms of the noise characteristics of the amplifier
which, in turn, determine the sensitivity 
of the device. In this framework, one considers a detector with the output
signal $I$ and input signal $\phi$, which is coupled to
an observable $Q$ of the detector via a term $\phi Q$. 
The input signal causes a variation of the output,
which can be characterized by the linear response 
coefficient $\lambda\equiv d\langle I \rangle/d\phi$.
Note that usually one operates in a dissipative, 
nonequilibrium regime. 
When used as a quantum detector coupled to a qubit the input variable is
$\phi \propto \sigma_z$, and the coupling is $c \sigma_z Q$ 
(with $c$ being a coupling  constant). 
We first consider a situation where 
tunneling between the qubit's basis states 
is suppressed, i.e., ${\cal H}_{\rm ctrl}=-B_z\sigma_z/2$. In this case the
fluctuations of $Q$ dephase the qubit with rate
\begin{equation}
\tau_\varphi^{-1}=\frac{c^2}{\hbar^2} S_Q
\;.
\label{Eq:eff_deph_rate}
\end{equation}
The symbols $S_Q$ and $S_I$ (introduced below) stand for the
noise power of the corresponding observable if the amplifier 
is decoupled from the qubit:
$S_Q=2\;\langle Q^2_\omega\rangle$ and
$S_I=2\;\langle I^2_\omega\rangle$. A white spectrum is assumed at 
the relevant frequencies.
The two basis states of the qubit produce output signals 
$I^{0/1}=\bar I\pm\Delta I/2$, differing by
\begin{equation}
\label{Eq:DI}
\Delta I=2c\lambda
\;.
\end{equation}
They can be distinguished after the measurement time
\begin{equation}
\tau_{\rm meas}^{-1}=\frac{(\Delta I)^2}{4S_I}
\;.
\label{Eq:eff_meas_rate}
\end{equation}
Hence, the two times are related by
\begin{equation}
\label{eq:G/G}
{\tau_{\rm meas}\over\tau_\varphi}
= {S_Q S_I \over \hbar^2 \lambda^2} = {S_Q S_\phi \over \hbar^2}
\ .
\end{equation}
 In the last form of 
Eq.~(\ref{eq:G/G}) $S_\phi \equiv S_I/\lambda^2$ is the output 
noise in terms of the input, i.e., the noise  which should be
applied to the input to produce the noise $S_I$ at the output.

If the tunneling is turned on, ${\cal H}_{\rm ctrl}=-\frac{1}{2}\Delta E 
(\cos\eta\;\sigma_x + \sin\eta\;\sigma_z)$, both the measurement rate 
(\ref{Eq:eff_meas_rate}) and the rate of pure dephasing (\ref{Eq:eff_deph_rate}) 
acquire an additional factor $\cos^2\eta$, while their ratio (\ref{eq:G/G}) 
persists. Apart from that, the finite tunneling introduces mixing, with rate
\begin{equation}
\tau_{\rm mix}^{-1}= \frac{c^2}{\hbar^2} S_Q \sin^2\eta
\;.
\end{equation}

For the particular case of the SET, which motivates the notations, 
$I$ is the transport current,
$Q=eN$ the charge of its central island, and $\phi$ the  
gate potential externally applied to the middle 
island of the SET ($\phi$ consists of $V_{N}$ and a 
contribution of the qubit, see Eq.~\ref{Effective_Voltages}).
The coupling constant is $c=\Eint/e$ and the linear response is given by 
$\lambda=\Delta I/(2c)=e^2\delta\Gamma/(2\Eint)$ with $\delta\Gamma$ defined by 
Eq.~(\ref{Eq:Gamma0/1}).
In this case, Eqs.~(\ref{Eq:eff_deph_rate}, \ref{Eq:eff_meas_rate}) 
are consistent with what has been described before in 
Eqs.~(\ref{Eq:Gamma_phi}, \ref{eq:G_mes_linear}).

The quantity $\sqrt{S_Q S_\phi}$
is proportional to the ``noise energy'', discussed by
\citeasnoun{Devoret_Schoelkopf_Nature}, 
measured in units of the energy quanta at the given frequency.
For the noise spectra $S_Q$ and $S_\phi$ one can obtain an inequality, similar 
to the Heisenberg uncertainty principle \cite{Braginsky}: 
$S_Q S_\phi\ge \hbar^2$.
Indeed, by virtue of Eq.~(\ref{eq:G/G}) this relation coincides with the
constraint
\begin{equation}
\label{Eq:tphi<tmeas}
\tau_\varphi\le\tau_{\rm meas}
\;.
\end{equation}
Moreover, one can show \cite{Braginsky,Averin_SQUID} that a stronger inequality 
holds. Namely, the quantity
\footnote{This quantity is denoted as ``energy sensitivity'' by 
\citeasnoun{Averin_SQUID}, but \citeasnoun{Devoret_Schoelkopf_Nature}
use this term for a different quantity.}
\begin{equation}
\label{Averin's_epsilon}
\epsilon \equiv {1\over 2}\left(S_Q S_\phi 
- {\mathop{\rm Re}}^2 S_{\phi Q} \right)^{1/2}
\ ,
\end{equation}
is limited by $\epsilon \ge \hbar/2$. Here
$S_{\phi Q}\equiv S_{IQ}/\lambda$, where 
$S_{IQ} \equiv 2 \int dt' \langle I(t)Q(t')\rangle$ characterizes
cross-correlations between the input and the output. Thus one has
\begin{equation}
\label{eq:G/G_Averin}
{\tau_{\rm meas}\over\tau_\varphi} =  
\left(\frac{\epsilon}{\hbar/2}\right)^2 + {({\rm Re}\, S_{\phi Q})^2\over 
\hbar^2}
\ .   
\end{equation}
The optimization of this ratio requires that the detector reaches the
quantum limit of  sensitivity, $\epsilon = \hbar/2$, and
the cross-correlations vanish, ${\rm Re}\, S_{\phi Q}=0$.
\citeasnoun{Averin_SQUID} made the observation 
that the quantum limit is (nearly)
reached in several measurement devices: For a quantum point contact,
an overdamped dc-SQUID, a resistively shunted superconducting SET, or
a normal SET in the  
co-tunneling regime \cite{Averin_SET_Cotunneling} the following two 
relations hold under certain conditions: 
\begin{equation}
\label{eq:Averin's_first}
S_I S_Q \approx |S_{IQ}|^2
\ 
\end{equation}
and
\begin{equation}
\label{eq:Averin's_second}
\lambda \approx {\rm Im}\, S_{IQ}
\ .
\end{equation}
They immediately imply that $\epsilon = \hbar/2$.
Note, that the relations 
(\ref{eq:Averin's_first},\ref{eq:Averin's_second}) 
are requirements for a quantum limited  measurement device. 
They are not valid in general. 
E.g., the near equality in (\ref{eq:Averin's_first}) should 
in general be replaced by the $\ge$ sign.
Even for the detectors where they were found 
they break down (making $\epsilon > \hbar/2$), e.g., at finite 
temperatures or at bias voltages close to the Coulomb blockade threshold 
\cite{Averin_SET_Cotunneling}. 

As for Eq.~(\ref{eq:Averin's_second}), by definition the response coefficient is 
$\lambda = -i \int dt' \theta(t-t') \langle [I(t),Q(t')]\rangle = 
2{\rm Im} \int dt' \theta(t-t') \langle I(t)Q(t')\rangle$.
Hence, Eq.~(\ref{eq:Averin's_second}) implies a vanishing reciprocal  
response coefficient $\lambda' = 2{\rm Im} \int dt' \theta(t-t') \langle
Q(t)I(t')\rangle$.  While in equilibrium 
Onsager's relations imply $\lambda = -\lambda'$,
such an asymmetry can arise in a nonequilibrium stationary state.
It is a characteristic feature of linear measuring devices \cite{Braginsky}.

As for the cross-correlations, $\mathop{\rm Re}S_{\phi Q}$, they vanish, e.g.,
in a dc-SQUID with two identical Josephson junctions, a QPC, symmetrically
coupled to a quantum dot, or a SSET \cite{Averin_SQUID}.

\subsection{Statistics of the current and the noise spectrum}
\label{subsec:Statistics}

In previous subsections we have discussed the statistics of the charge
$m$, which  passed through the detector, and the corresponding current
$I$ after a measurement has been started. In this subsection 
we complement this discussion by investigating the noise properties of
the charge and the current in the stationary regime, i.e., a long time
after the 
qubit-detector coupling  was turned on. The noise spectrum reflects the 
intrinsic properties of the system of qubit and  meter and their
coupling and depends on the corresponding time scales.
We obtain additional evidence for the telegraph behavior of the current.

We study first the charge-charge correlator, which is derived from the joint 
probability distribution of charges at different times 
(\ref{P(m1,m2)_MARKOVIAN}), and obtain
\begin{eqnarray}
\label{mm_CORRELATOR}
\langle m(t_{1})&&\,m(t_{2}) \rangle  
\nonumber \\
= - && {\rm Tr'} \hat\rho_{0}
\partial_{k} 
\left[ U(k,\Delta t) \,
\partial_{k} U\left(k,\bar t-{\Delta t \over 2}\right)
\right]_{k=0}
\ ,
\end{eqnarray}
where $\Delta t \equiv |t_{1}-t_{2}|$ and  $\bar{t} \equiv (t_{1}+t_{2})/2$.
We employ again the two-mode approximation and use
Eqs.~(\ref{Eq:M_RED},\ref{U_RED}) to arrive,
after taking time derivatives, at
\begin{eqnarray}
\langle I(t_{1})\,I(t_{2}) \rangle
 = && e^2 \bar\Gamma^{2} + e^2\, f\,\bar\Gamma \delta(\Delta t)  +
  e^2\, {\delta\Gamma^{2}\over 4} e^{-\Delta t/\tau_{\rm mix}}
\nonumber \\
&& + (\dots)\,e^{-\bar{t}/\tau_{\rm mix}}
\ . 
\end{eqnarray}
Here $\Gamma^{0/1}\equiv\bar\Gamma\pm\delta\Gamma/2$ are the tunneling
rates (\ref{Eq:Gamma0/1}) for  the two qubit states.
The Fano factor $f$ is defined after Eq.~(\ref{Eq:Ptilde}).
In the stationary limit,
$\bar{t} \gg \tau_{\rm mix}$,  we thus obtain for the noise spectrum
\begin{equation}
\label{NOISE}
S_{I}(\omega)=  2e^{2} f \bar \Gamma +
{e^{2}\delta\Gamma^{2}\tau_{\rm mix}\over 1+\omega^2 
\tau_{\rm mix}^{2}}
\ .
\end{equation}
The first, $\omega$-independent term corresponds to the shot noise,
while the second term originates from the telegraph noise. The
ratio of the telegraph and shot noise amplitudes at
$\omega=0$ is
\begin{equation}
\label{NOISE_RATIO}
{S_{\rm telegraph} \over S_{\rm shot}} =  
{\delta\Gamma^2 \tau_{\rm mix}\over 2 f \bar\Gamma }
\approx 4 {\tau_{\rm mix}\over \tau_{\rm meas}}
\ .
\end{equation}
Note that the telegraph noise becomes noticeable on top of the
shot noise in the parameter regime of an effective quantum measurement
($\tau_{\rm mix}\gg \tau_{\rm meas}$).

While the shot-noise contribution reflects intrinsic properties of the
detector (SET), the telegraph noise characterizes the qubit. This
structure of the noise of the output signal is quite general  
\cite{Averin_Korotkov,Averin_SQUID,Korotkov_Osc}. 
For the qubit coupled to the meter, the following relation for the noise of the 
output signal $I$ was derived (using linear response and certain other 
approximations):
\begin{equation}
\label{Averin_Korotkov_Noise}
S_I(\omega)=S_I^0(\omega)+c^2 \lambda^2 S_{\sigma_z}(\omega)
\;, 
\end{equation}
Here the notations introduced in Section~\ref{subsec:Diss_Efficiency} are used, 
with the only difference that $S_I$ now denotes the noise in the {\it presence} 
of the qubit. The first term, $S_I^0$, in the RHS of
Eq.~(\ref{Averin_Korotkov_Noise}) represents the noise of the meter decoupled
from the qubit ($S_{\rm shot}$ for  a SET).  The second term arises
from the qubit and is governed by the dynamics of the its density
matrix. In the preferred basis (cf.
Section~\ref{subsec:ME}--\ref{subsec:Detector-dom-regime}) 
the diagonal elements decay to their stationary values on the  mixing
time scale, while the off-diagonal elements precess with frequency
$\Delta E/\hbar$ and decay within the dephasing time.  Depending on
the qubit's Hamiltonian, both or one of these  
processes contribute to the dynamics of $\sigma_z(t)$.  Hence, the
qubit's contribution, $S_{\sigma_z}(\omega) \equiv 2 \int dt \langle
\sigma_z(0) \sigma_z(t)\rangle \exp(-i\omega t)$, has a ``coherent''
peak at the qubit's eigenfrequency, $\Delta E/\hbar$, of width
$\tau_\varphi^{-1}$, and a ``telegraph'' peak at zero frequency with width
$\tau_{\rm mix}^{-1}$.

To derive the noise spectrum of the qubit under the influence of the
detector, one  can use a Bloch-type master equation for the joint density
matrix [e.g., Eq.(\ref{Eq:MasterEq}) at $k=0$ for a SET]. Under
certain conditions, namely, if the noise spectrum $S_Q(\omega)$ is
white in an interval of frequencies at least up to the qubit's  
level spacing $\Delta E$ (in the SET it is white only up to $\Gamma_{\rm 
R}+\Gamma_{\rm L}$) a simpler set of equations is sufficient
\cite{Averin_SQUID}: 
\begin{equation}
\label{Eq:Averin_Korotkov_ME}
\begin{array}{ccl}
\dot \rho_{00} &=& B_x \mathop{\rm Im} \rho_{01}
\;,\\[2mm]
\dot\rho_{01} &=& 
(i B_z - \tau_{\varphi0}^{-1})\rho_{01} - \frac{i}{2} B_x (\rho_{00}-\rho_{11})
\;.
\end{array}
\end{equation}
Here $\rho_{ij}$ is the qubit's density matrix in the eigenbasis of $\sigma_z$,
$\tau_{\varphi0}^{-1} \equiv c^2 S_Q/\hbar^2$ describes the back-action 
(\ref{Eq:eff_deph_rate}), and $B_x$ and $B_z$ form the Hamiltonian
of the qubit (\ref{Eq:Magnetic_Hamiltonian}).  These equations were used, for
instance, for the analysis of a quantum point contact coupled to a double
quantum dot \cite{Gurvitz,Averin_Korotkov}.

At zero bias, $B_z=0$, one can solve Eq.~(\ref{Eq:Averin_Korotkov_ME})
exactly,  to obtain \cite{Averin_Korotkov}
\begin{equation}
\label{Eq:Coherent_Peak}
S_{\sigma_z} = {4 B_x^2 \tau_{\varphi0} \over
(\omega^2 - B_x^2)^2\tau_{\varphi0}^2 +  \omega^2}
\ .
\end{equation} 
In the Hamiltonian-dominated limit,
$\tau_{\varphi0}^{-1} \ll B_x$, at this bias point no telegraph peak 
appears. Still, the qubit contributes to the noise via the 
last term in Eq.~(\ref{Averin_Korotkov_Noise}). This coherent peak at
$B_x$ has the  height
\begin{equation}
\label{Eq:dSqb_max}
c^2\lambda^2 S_{\sigma_z}^{\rm max} = 4c^2\lambda^2 \tau_{\varphi0} 
= (\Delta I)^2 \tau_{\varphi0}
\;.
\end{equation}
Here $\Delta I$ is the difference in the 
output current (\ref{Eq:DI}) for the two eigenstates of $\sigma_z$
(charge states). Recalling that it is related to the measurement rate 
(\ref{Eq:eff_meas_rate}) in the detector-dominated regime, one concludes that 
the signal-to-noise ratio in 
Eq.~(\ref{Averin_Korotkov_Noise}) is limited, by virtue of 
Eq.~(\ref{Eq:tphi<tmeas}), as
\begin{equation}
\label{Eq:Peak_to_Background}
\frac{c^2\lambda^2 S_{\sigma_z}^{\rm max}}{S_I^0} = 
4\;\frac{\tau_{\varphi0}}{\tau_{\rm meas}} \le 4
\ .
\end{equation}
Thus, the requirements for the observation of coherent oscillations 
in the noise spectrum of the measuring device is directly expressed by
the ratio of dephasing and measurement time.

In the opposite, detector-dominated regime 
$B_x \ll \tau_{\varphi0}^{-1}$, the noise spectrum
(\ref{Eq:Coherent_Peak}) exhibits a telegraph-noise Lorentzian 
(\ref{NOISE}) at low frequencies, 
$4\tau_{\rm mix}/(\omega^2 \tau_{\rm mix}^{2}+1)$, where
$\tau_{\rm mix}^{-1} = B_x^2\tau_{\varphi0}$.

At a general bias point $B_z \ne 0$ in the Hamiltonian-dominated regime the
dynamics of $\sigma_z$ exhibits features at both frequencies, 
$\omega =0$ and $\omega =\Delta E$,
and its noise spectrum has two peaks.  This was shown numerically by
\citeasnoun{Averin_Korotkov} and it is consistent with the result
(\ref{NOISE}). 
A higher-order analysis of the master equation (\ref{Eq:MasterEq}), which 
accounts for the corrections to the oscillating modes 
(\ref{Eq:Off_Diagonal_Zeroth_Order_ME}) due to the mixing, 
allows us to obtain the 
coherent peak which is, however, suppressed due to the strong dephasing 
(cf.~Eq.~\ref{Eq:Peak_to_Background}).

\subsection{Conditional master equation}

We finally would like to comment on recent studies
\cite{Korotkov,Korotkov_Osc,Goan} where the so-called selective or conditional
approach, popular in quantum optics, was employed.  Its purpose is to develop a
framework for the analysis of statistical properties of qubit-related quantities
{\it conditioned} on the dynamics of the output signal, $I(t)$, at earlier
times.  In other words, the goal is to predict the outcome of a measurement of
the qubit's state at time $t$ given the results of the current read-out at
$t'<t$.  A related problem is to produce typical, fluctuating current-time
patterns $I(t)$ which one can encounter in a given experiment.

These problems can be addressed using the proper master equation for
the coupled system (e.g., Eq.~\ref{Eq:MasterEq}).
Monitoring the current amounts to repeated measurements of the 
charge $m$, at sufficiently short time-intervals $\Delta t$.  
Since a closed master equation can be formulated which 
involves only the $m$-diagonal entries of the density matrix
$\hat\rho(m)$, the effect of a measurement with the result $m_0$
amounts to choosing the 
corresponding $\hat\rho(m_0)\delta_{mm_0}$ as the new density
matrix (properly rescaled to ensure the normalization).
Thus a simulation of the evolution proceeds as follows:
From the the master equation with initial matrix $\hat\rho\delta_{m0}$
at $t=0$ one   obtains $\rho(\Delta m,\Delta t)$ at the time of the
first read-out. Then, to simulate an experiment, one selects a
``measured'' value of $\Delta m$ 
with the corresponding probability $P(\Delta m, \Delta t)=\mathop{\rm tr}
\hat\rho(\Delta m,\Delta t)$ (cf.
Section~\ref{subsec:Ham_Dom_Regime}) and  
uses the corresponding density matrix as initial value
for the further evolution during the next time interval.  Repeating this step
many times one produces a typical dependence of $m(t)$ and a
density matrix $\hat\rho_{\rm cond}(t)$ which can be used
to study the conditional statistics of further  measurements.

This procedure can be simplified if sufficiently frequent read-outs are
performed. An expansion in $\Delta t$ allows one to present the step-by-step 
evolution of the density matrix $\hat\rho_{\rm cond}$ as a continuous process
\citeaffixed{Korotkov_Selective}{cf.}:
\begin{eqnarray}
\frac{d}{dt}\hat\rho&=&-i[{\cal H}_{\rm ctrl},\hat\rho]
-\frac{1}{4} \gamma_\varphi^0 [\sigma_z,[\sigma_z,\hat\rho]]
\nonumber\\
&&+\frac{\Delta I}{2S_I^0} [I(t)-\bar I]
\left[ \{\sigma_z,\hat\rho\}-
2\mathop{\rm tr}(\sigma_z\hat\rho)\hat\rho\right]
\;,
\label{Eq:ME_selective}
\end{eqnarray}
where $\Delta I$ is defined in Eq.~(\ref{Eq:DI}).
The measured value of the output signal, $I(t)=\Delta m/\Delta t$, as 
described above, should be chosen randomly with the distribution $P(\Delta 
m,\Delta 
t)$. Using the properties of $P(m,\Delta t)$ at
short $\Delta t$, one can further simplify this procedure and choose
$I(t)$ as 
\begin{equation}
I(t)=\bar I+\frac{\Delta I}{2} \mathop{\rm tr}(\sigma_z\hat\rho)
+\delta I(t)
\;,
\label{Eq:I(t)_selective}
\end{equation}
where $\delta I(t)$ is a random quantity with the noise properties
identical to those of the current in the detector decoupled from the qubit.

We can mention that, similar to  Eq.~(\ref{Eq:Averin_Korotkov_ME}), in the 
derivation of Eq.~(\ref{Eq:ME_selective}) we assumed the back-action noise 
of the detector to be white. Furthermore, $\gamma_\varphi^0$ is the
dephasing rate due to the  
environment and accounts for non-ideality of the detector (see 
Section~\ref{subsec:Diss_Efficiency}). After averaging over the detector 
dynamics, $I(t)$, Eq.~(\ref{Eq:ME_selective}) is recovered.

The equations (\ref{Eq:ME_selective}, \ref{Eq:I(t)_selective}) form a
Langevin-type evolution equation which can be used to produce typical
outcomes of the measurement and to study statistics of the qubit conditioned on
these outcomes.

\section{Conclusions}
\label{sec:Conclusions}

Josephson junction systems in a suitable parameter range can be
manipulated in a quantum coherent fashion. They are promising
physical realizations (the hardware) of future devices to
be used for quantum state engineering. We discussed their modes of
operation in different designs (in the charge and the flux dominated 
regimes), the constraints on the parameters, various dephasing effects, 
and also the physical realization of the quantum mechanical
measurement process. We pointed out the advantages of these
nano-electronic devices as compared to other physical realizations.

We add a few remarks and comparisons. First, in order to demonstrate
that the constraints on the circuit parameters, 
which were derived in previous sections, can be met by available
technologies, we summarize them here and suggest a suitable set. 
\\
(i) Necessary conditions for a Josephson charge qubit are:
$\Delta > \EC \gg E_{\rm J}\,,\,k_{\rm B}T$.  The superconducting
energy gap $\Delta$  has to be  chosen large to suppress quasiparticle
tunneling.  The 
temperature has to be low to assure the initial thermalization,
$k_{\rm B}T \ll \EC,\,\hbar \omega_{LC}$,  and to reduce 
dephasing effects. A sufficient choice is $k_{\rm B}T \sim
E_{\rm J}/2$, since further cooling does not reduce the dephasing
(relaxation) rate in a qualitative way. (Of course, it does so far
from  the degeneracy point, i.e., for $\eta = 0$, or if we switch
off the Hamiltonian, $\Hcontr =0$. However during manipulations $E_{\rm
J}$ is the typical energy difference and sets the time scale for both
the manipulation times and the dephasing.) 

As an explicit example we suggest the following parameters
(the circuit parameters of \citeasnoun{Nakamura_Nature} are in this
regime) and estimate the  corresponding time scales: 
We choose junctions with capacitance $C_{\rm J}= 10^{-15}$~F,
corresponding to a charging energy (in temperature units) 
$\EC \sim 1$~K, 
and a smaller gate capacitance $\Cg = 0.5\cdot 10^{-17}$~F to
reduce the  coupling to the environment. Thus at the working
temperature of $T=50$~mK  the initial thermalization is assured. The
superconducting gap has to be  slightly higher,  $\Delta > \EC$. 
Thus aluminum is a suitable material. We
further choose $E_{\rm J}=100$~mK, i.e., the time scale of one-qubit
operations is $\tau_{\rm op}^{(1)}= \hbar / E_{\rm J}\sim 10^{-10}$~s.\\
(ii) A realistic value of the resistor in the gate voltages circuit is
$R\sim 50~\Omega$. Its voltage fluctuations limit the dephasing time
(\ref{Eq:dephasing}) to values of order 
$\tau_\varphi \sim 10^{-4}$~s, thus allowing for  
$\tau_\varphi / \tau_{\rm op}^{(1)} \sim 10^6$ 
coherent manipulations of a single qubit
\footnote{This may be overly optimistic, and indicating that
  other sources of dephasing need to be considered as well. For instance, at
  these slow time scales the background charge fluctuations may dominate. 
We also note that in the experiment of
\citeasnoun{Nakamura_Nature} a stray capacitance in the probe circuit,
larger than $\Cg$, renders the dephasing time shorter.}.
\\ 
(iii) To assure sufficiently fast 2-bit operations we choose for the design of 
Fig.~\ref{Fig:ManyBits-NewDesign} $L\sim 10$~nH and 
$C_{\rm L} \approx \CJ$. Then the 2-bit operations are 
about $10^2$ times slower than the 1-bit operations and, accordingly,
their maximum number is reduced.\\
(iv) The quantum measurement process introduces additional constraints on
the parameters, which can be met in realistic devices as demonstrated
by the following concise example. The parameters of the qubit are
those mentioned before. For
the junction and gate capacitances of the normal tunnel junctions of the
SET we chose $\CN=1.5\cdot 10^{-17}$~F and $\CgSET =  0.5\cdot 10^{-17}$~F,
respectively, and for the coupling capacitance between  SET
and qubit: $C_{\rm int} = 0.5\cdot 10^{-17}$~F.  We thus obtain:
$\ECSET \approx 25$~K, $\ECqb \approx 1$~K, $\ECint \approx 0.25$~K
(for precise definitions see Appendix~\ref{app:SET}).
We further take $\ngqb =0.25$, $\NgSET  = 0.2$ and $\mu_{\rm L} =
-\mu_{\rm R} = eV_{\rm tr}/2 = 24$~K and $\alpha_{\rm L} = 
\alpha_{\rm  R} = 0.03$. 
This gives  $B_z \approx 2$~K and $\Gamma_{\rm L} = 1.8$~K and
$\Gamma_{\rm R}=7.8$~K (note that due to the 
gate voltage the applied transport voltage is split  asymmetrically).
Thus the measurement time in this regime is $\tau_{\rm meas}
\approx 1.5 \cdot 10^{-8}$~s. Since for this choice of parameters  
$\tau_{\varphi0}^{-1}\approx 4.0 \cdot 10^{-3}$~K~$ \ll B_z \approx \DEav$, the 
``Hamiltonian dominated'' regime is realized. 
Assuming $E_{\rm J} = 0.1$~K we obtain $\tau_{\rm mix}
\approx 0.7 \cdot 10^{-6}$~s. Thus $\tau_{\rm mix}/\tau_{\rm meas}
\approx 45$ and  the separation of peaks should be observable early enough
before the mixing dominates.   
Indeed, the numerical simulation of the system
(\ref{Eq:MasterEq}) for these parameters shows almost
ideal separation of peaks (see Fig.~\ref{PLOT0.1}).   On the other
hand, for $E_{\rm J} = 0.25$~K we obtain $\tau_{\rm mix}/\tau_{\rm meas}
\approx 7$.  This is a marginal situation.  The numerical simulation in this
case (see Fig.~\ref{PLOT0.3}) shows that the peaks first start to
separate, but, soon, the valley between the peaks fills in due to
the mixing transitions.\\

The requirements on the parameters of a flux
qubit circuit can be summarized in a similar way. First, the
parameters should be chosen to  
allow the reduction of the  double-well potential to the two  
ground states forming a two-state quantum system. It is also
desirable that these two basis states have macroscopically different
flux or phase configurations. 
The double well is formed by joining either several Josephson junctions
\cite{Mooij} or a Josephson junction and an inductive term of similar 
strength ($\EJ\sim\Phi_0^2/4\pi^2L$) in an rf-SQUID \cite{Friedman_Cats}).
Since the level spacing within each well is of order
$E_0\sim\sqrt{\EJ\EC}$ and  the barrier height of order $\EJ$, all
these requirements can be satisfied by 
`classical' Josephson junctions with $\EJ\gg\EC$.
Furthermore, the asymmetry of the double well $B_z$, which is controlled by
external fluxes,  and the tunnel splitting $B_x$ should be smaller
than $E_0$ in order to suppress leakage to higher states. Finally, the
temperature should be low enough to allow  
the initialization of the qubit's state and to ensure slow dephasing. In 
summary, the following conditions have to be satisfied:
$k_{\rm B}T\le B_x\ll\sqrt{\EJ\EC}\ll \EJ$. As pointed out
above, it is sufficient  to choose $k_{\rm B}T\sim B_x/2$.
These requirements can be satisfied, e.g., by the parameters of the rf-SQUID 
used by \citeasnoun{Friedman_Cats} ($\EJ\approx 70$~K, $\EC\approx 1$~K, 
$B_x\approx 0.1$~K, and $T\approx 40$~mK) or by similar values discussed by 
\citeasnoun{Mooij} ($\EJ\sim 10$~K, $\EC\sim 0.1$~K, $B_x\sim 50$~mK,
and $T\sim 30$~mK).\\

At present, the most advanced quantum manipulations of a solid state
system, i.e., the coherent oscillations observed by
\citeasnoun{Nakamura_Nature}, have been demonstrated for a Josephson charge
qubit. But flux systems might soon catch up. In the long run, it 
is not clear whether charge or flux systems will 
bring faster progress and further reaching demonstrations of complex quantum
physics.  In fact, a combination of both appears feasible as well.
Therefore we compare shortly the properties of the
simplest charge and flux qubits.\\

A very important quantity is the phase coherence time $\tau_\varphi$,
which has to be compared to the typical operation time scale of the qubit's
dynamics $\tau_{\rm op}$. While effects of various dissipative mechanisms have 
been estimated theoretically, further experimental work is needed to
understand  
dephasing in charge and flux qubits. One potentially dangerous source of 
dephasing, the coupling to the external circuit, can be described by an Ohmic 
oscillator bath. If this contribution to the dephasing dominates, the 
above-mentioned ratio of times 
is determined by the dimensionless parameter
$\tau_{\rm op}/\tau_\varphi \sim \alpha$ (see Section~\ref{sec:Dephasing}).  
For ``unscreened'' charge qubits it is of order of
$\alpha \approx 10^{-2}$, but it can be substantially reduced by the
`screening  ratio' of capacitances (see Section~\ref{sec:Dephasing}).
Putting in numbers corresponding to those of 
\possessivecite{Nakamura_5nsec} experiment (without probe junctions),
we estimate $\tau_\varphi$ to reach several tens of
microseconds. Coherent oscillations for about
5~ns have been observed already in this system, 
in spite of the presence of a non-ideal detector
limiting the phase coherent evolution by quasiparticle tunneling in
the probe junction and also providing a strong coupling to the
external circuit. The phase coherence time should be compared to the
qubit operation time scale of $\tau_{\rm op}\approx10$ to  
100~ps.

For flux qubits considered by \citeasnoun{Mooij} the current circuit
may produce $\alpha \sim 10^{-5}$ and the relaxation times of order
$10~\mu$s, while other dephasing mechanisms
studied would destroy coherence after times of order of hundreds of
microseconds or longer. At the same time, the qubit  
level spacing sets the fastest operation
time to $\tau_{\rm op}\approx0.1$ to 1~ns. 

A major source of errors and dephasing for all charge-degrees of
freedom are the fluctuations of the `offset charges'.  They arise due
to charge transfers in the substrate, e.g., between impurity sites,
and are detrimental for many of the potential applications of
single-charge systems.  Fortunately, they occur typically on long time
scales, and may not take place during a `computation', i.e., a series
of coherent manipulations. 
Similarly for flux qubits nuclear spins provide random magnetic 
fields. Also these fields change only on a long spin-relaxation time
scale and cause  no dephasing in shorter `computations'.

Another important point is the efficiency of quantum detectors used to
read out the charge or flux state.  Estimates show that the newly
developed rf-SETs \cite{Schoelkopf} should make single-shot charge
measurements possible in principle 
($\tau_{\rm mix}\gg\tau_{\rm meas}$).  On the other hand, the flux
read-out with a  SQUID is far from this goal and averaging over
a large number of measurements is needed \cite{Delft_Cats}.

As for the experimental achievements, in both charge and flux systems
superpositions of basis (charge or flux) states have been seen, and the
validity of the two-state model has been confirmed spectroscopically
\cite{Bouchiat_PhD,Nakamura,Friedman_Cats,Delft_Cats}.  As for the coherent
oscillations between the basis states, so far they were seen only in the charge
design \cite{Nakamura_Nature}.
In Appendix \ref{app:spins_dots} we further compare 
Josephson devices to some  
other systems suggested as physical realizations of qubits.\\

To conclude, the fabrication and controlled coherent manipulations of Josephson
junction qubits are possible using present-day technologies.  In these systems
fundamental properties of macroscopic quantum-mechanical systems can be
explored.  First experiments on elementary systems have been performed
successfully.  More elaborate designs as well as further progress of
nano-technology will provide longer coherence times and allow sequences of
coherent manipulations as well as scaling to larger numbers of qubits.  The
application of Josephson junction systems as elements of a quantum computer,
i.e., with a very large number of manipulations and large number of
qubits, will 
remain a challenging issue.  On the other hand, many aspects of quantum
information processing can initially be tested on simple circuits as proposed
here.  We also expect further spin-offs, once the techniques of quantum state
engineering get further developed.

We have further shown that a dissipative quantum system 
coupled to a qubit may serve as a quantum measuring device in an
accessible range of parameters.  Explicitly we studied a
single-electron transistor coupled capacitively to a charge qubit.
We have described the process of measurement by deriving the time
evolution of the reduced density matrix of the coupled system.  We
found that the dephasing time is 
shorter than the measurement time, and we have estimated the mixing
time, i.\,e. 
the time scale on which the transitions induced by the measurement occur.
Similar scenarios are discussed for flux qubits measured by a SQUID coupled to 
it inductively.


\acknowledgments

We thank T.~Beth, C.~Bruder, R.~Fazio, Z.~Hermon, J.~K\"onig, A.~Korotkov,
Y.~Levinson, J.E.~Mooij, Y.~Nakamura, H.~Schoeller, and F.~Wilhelm for
stimulating discussions.  This work has been supported by the
`Schwer\-punkt\-programm Quanten-Informations\-verarbeitung' and the 
SFB 195 of the German Science Foundation (DFG).

\appendix

\section{An ideal model}
\label{app:IdealModel}

\subsection*{The model Hamiltonian}

Quantum state engineering requires coherent manipulations of suitable quantum
systems.  The needed quantum manipulations can be performed if we have
sufficient control over the fields and interaction terms in the
Hamiltonian.  As an introduction and in order to clarify the goal, we
present here an ideal model Hamiltonian and show how the necessary
unitary transformations can be performed.  
We can add that the Josephson junctions
devices discussed in this review come rather close to this ideal model.

As has been stressed by \citename{DVD-Curacao} 
(\citeyear{DVD-Curacao,DVD-Fortschritte}) any physical
system which is considered as a candidate for quantum 
computation, and similar for alternative applications of quantum 
state engineering, should satisfy the following criteria:
(i) First of all, one needs well-defined two-state quantum systems (or
quantum systems with a small number of states).  This implies that
higher states, present in most real
systems, must not be excited during manipulations. 
(ii) One should be able 
to prepare the initial state of the qubits with sufficient accuracy.   
(iii) A long phase coherence time is needed, sufficient to allow for a
large number (e.g., $\ge 10^4$) of coherent manipulations.
(iv) Sufficient control over the qubit's Hamiltonian is
required to perform the necessary unitary transformations, i.e.,
single-qubit and two-qubit logic operations (gates).   
For this purpose
one should be able to control the fields at the sites of each qubit
separately, and to couple qubits together in a controlled way,
ideally with the possibility to switch the inter-qubit interactions on
and off. 
In  physics terms the two types of operations allow creating arbitrary
superpositions and non-trivial coupled states,
such as entangled states, respectively.
(v) Finally, a quantum measurement is needed to read out the quantum
information, either at the final stage or during the computation,
e.g., for the purposes of error-correction.

We consider two-state quantum systems (e.g./ spins), or systems which
under certain condition effectively reduce to two-state systems 
(charge in a box or flux in a SQUID near degeneracy points). 
Any single two-state quantum system can be represented as a spin-1/2
particle, and its Hamiltonian be written as  
${\cal H}(t)=-\frac{1}{2} \bbox{B}(t) \hat{\bbox{\sigma}}$.
Here $\sigma_{x,y,z}$ are Pauli matrices in the space of states 
$\ket{\uparrow}=\left(1\atop 0\right)$ and $\ket{\downarrow}=\left(0\atop 
1\right)$, which form the basis states of a physical quantity (spin,
charge, flux, \ldots) which  is to be manipulated. The ``spin'' is coupled to an 
effective ``magnetic field'' $\bbox{B}$. In an alternative notation used for 
two-state quantum systems the components $B_z$ and $B_{x,y}$ 
correspond to an external bias and a tunneling amplitude, respectively.
Full control of the quantum  dynamics of the spin is possible if the 
magnetic field $\bbox{B}(t)$ can be switched arbitrarily. In
fact, arbitrary single-qubit operations can be performed already if two
of the field components can be controlled, e.g., 
\begin{equation}
\label{Eq:idealH-1qubit}
        {\cal H}_{\rm ctrl}(t)= -\frac{1}{2}  B_z(t) \hat\sigma_z - {1\over
        2} B_x(t) \hat\sigma_x \ . 
\end{equation}
If all three components of the magnetic field can be controlled
the topological (Berry) phase of the system can be manipulated as well
\cite{Fazio_Berry}.   

In order to manipulate a many-qubit system, e.g., to perform quantum
computing, the magnetic field at the site of each spin has to be controlled 
separately. In addition,
one needs two-qubit (unitary) operations which
requires controlling the coupling energies between the qubits. For 
instance, a system with the following model Hamiltonian would be suitable:
\begin{equation}
\label{Eq:idealH}
\Hcontr(t) = - \frac{1}{2}\sum\limits_{i=1}^N
                \bbox{B}^i(t) \hat{\bbox{\sigma}}^i +\sum\limits_{i\ne
                j}
                J^{ij}_{ab}(t) \hat\sigma_a^i \hat\sigma_b^j
                \;,
\end{equation}
where a summation over spin indices $a,b=x,y,z$ is implied.
In (\ref{Eq:idealH}) a general form of coupling is presented, but simpler 
forms, such as pure Ising $zz$-coupling, $XY$-, or Heisenberg coupling are 
sufficient.  

The measurement device, when turned on, and residual interactions with the
environment are accounted for by extra terms ${\cal H}_{\rm meas}(t)$
and ${\cal H}_{\rm res}$, respectively:  
\begin{equation} 
{\cal H} = \Hcontr(t) + {\cal H}_{\rm meas}(t) + {\cal H}_{\rm res} \,.
\end{equation} 
During the manipulations
the meter should be kept in the off-state, ${\cal H}_{\rm meas}=0$.  The
residual interaction ${\cal H}_{\rm res}$ leads to dephasing and relaxation
processes.  It has to be weak in order to allow for a series of coherent
manipulations.

A typical experiment involves preparation of an initial quantum state, 
switching the `fields' $\bbox{B}(t)$ and the coupling energies
$J^{ij}_{ab}(t)$ to effect a specified unitary evolution of the wave function,
and the measurement  of the final state.

\subsection*{Preparation of the initial state}

The initial state can be prepared by keeping the system at low temperatures 
so that it relaxes to the ground state. This is achieved by
turning on a large value of $B_z\gg k_{\rm B}T$
for a sufficiently long time while
$B_x(t) = B_y(t) = 0$.  Then the residual interaction,
${\cal H}_{\rm res}$ relaxes each qubit to its ground state,
$\ket{\uparrow}$.  Switching $B_z(t)$ back to zero leaves the system 
in a well-defined pure quantum state.  If ${\cal H}=0$, there is no
further time evolution.

\subsection*{Single-qubit operations}

A single-bit operation on a given qubit can be performed, e.g., by turning
on $B_x(t)$ for a time span $\tau$.  As a result
of this operation the quantum state evolves according to the unitary
transformation
\begin{equation}
\label{Eq:x-rotation}
U_x\left(\alpha\right)= \exp\left(\frac{i B_{x} \tau
 \hat\sigma_x}{2\hbar}\right) =
\left(
\begin{array}{cc} \cos\frac{\alpha}{2} & i\sin\frac{\alpha}{2} \\
 i\sin\frac{\alpha}{2}&\cos\frac{\alpha}{2} 
\end{array}
\right)
\, ,
\end{equation}
where $\alpha=B_{x} \tau/\hbar$.  Depending on the time span $\tau$,
an $\alpha=\pi$- or an $\alpha=\pi/2$-rotation is performed, producing a spin
flip (NOT-operation) or an equal-weight superposition of spin states. 
Switching on
$B_z(t)$ for some time produces another  needed single-bit
operation: a phase shift between $\ket{\uparrow}$ and
$\ket{\downarrow}$:
\begin{equation}
\label{Eq:z-rotation}
U_z\left(\beta\right)=   \exp\left(\frac{i B_z \tau
 \hat\sigma_z}{2\hbar}\right) =
 \left(\begin{array}{cc}e^{i\beta/2}&0\\0&e^{-i\beta/2}
 \end{array}\right) \, ,
\end{equation}
where $\beta=B_z \tau/\hbar$. With a sequence of these $x$- and $z$-rotations 
any unitary transformation of the qubit state (single-qubit operation) can  be 
performed. There is no need to turn on $B_y$.

\subsection*{Two-qubit operations}

A two-bit operation on qubits $i$ and $j$ is induced by turning on
the corresponding coupling $J^{ij}(t)$. For instance, for the $XY$-coupling, 
$J^{ij}_{ab}\hat\sigma_a \hat\sigma_b= J^{ij} (\hat\sigma_x \hat\sigma_x+ 
\hat\sigma_y \hat\sigma_y)$, in the basis $\ket{\uparrow_i
\uparrow_j}$,  $\ket{\uparrow_i \downarrow_j}$, $\ket{\downarrow_i
\uparrow_j}$, $\ket{\downarrow_i \downarrow_j}$ the result is
described by the unitary operator
\begin{equation}
\label{2bit_Operation}
U_{\rm 2b}^{ij}\left(\gamma\right)=
\left( 
\begin{array}{cccc}
1 & 0 & 0 & 0 \\  0 & \cos\gamma  & i\sin\gamma & 0 \\  0 &
i\sin\gamma & \cos\gamma  & 0 \\ 0 & 0 & 0 & 1
\end{array} 
\right)
\,,
\end{equation}
with $\gamma\equiv 2 J^{ij}\tau/\hbar$. For $\gamma=\pi/2$ the 
operation leads to a swap of the states $\ket{\uparrow_i\downarrow_j}$ and 
$\ket{\downarrow_i\uparrow_j}$ (and multiplication by $i$), while for 
$\gamma=\pi/4$ it transforms the state $\ket{\uparrow_i \downarrow_j}$ into  an 
entangled state $\frac{1}{\sqrt{2}} \left(\ket{\uparrow_i \downarrow_j} + i
\ket{\downarrow_i \uparrow_j} \right)$.  

We note that apart from the sudden switching of $B_{z,x}^i(t), J^{ij}(t)$,
discussed above for illustration, one can also use other techniques to
implement single-bit or two-bit operations.  For instance, one can induce Rabi
oscillations between different states of a qubit or a qubit pair by ac resonance
signals; or perform adiabatic manipulations of the qubits' Hamiltonian to
exchange different eigenstates (with occupations remaining unchanged).  We
discuss some of these methods for particular physical systems in
Sections~\ref{sec:JosChargeBit} and \ref{sec:FluxQubits}.

\section{Quantum logic gates and quantum algorithms}
\label{app:gates}

In Appendix~\ref{app:IdealModel} and
Sections~\ref{sec:JosChargeBit},~\ref{sec:FluxQubits} we showed how elementary
quantum logic gates can be realized by simple manipulations of concrete physical
systems.  Details such as the application of a magnetic field pulse or the type
of the two-qubit coupling depend on the specific model.  On the other hand,
quantum information theory discusses quantum computation in
realization-independent terms.  For instance, it is customary to build quantum
algorithms out of specific, `standard' single- and two-qubit gates, some of
which will be discussed below.  Hence, one needs to know how to express these
standard gates in terms of the elementary operations specific for a given
physical model.  Furthermore, one may be interested in an optimized
implementation, in terms of time, complexity of manipulations, or amount of
additional dissipation.  Here we give several examples of standard gates and
quantum algorithms and cover optimization issues later.

\subsection*{Single- and two-qubit gates}

The quantum generalization of the NOT gate,
\begin{equation}
{\rm NOT}=\left(\begin{array}{cc} 0&1\\1&0
\end{array}\right)
\,,
\end{equation}
permutes the basis vectors $\ket{0}\to\ket{1}$ and
$\ket{1}\to\ket{0}$.  It can be performed as the  $x$-rotation
(\ref{Eq:x-rotation}) with the time span corresponding to $\alpha=\pi$
(up to an unimportant overall phase factor):
$U_x(\alpha=\pi)=i\cdot\mbox{NOT}$.  Unlike to classical computation,
in quantum logic there exists a logic gate,  called $\sqrt{\rm NOT}$,
which when applied twice produces the NOT-gate:
\begin{equation}
\sqrt{\rm NOT}=\frac{1}{2}
\left(\begin{array}{cc}
\phantom{-}1+i&-1+i\\-1+i&\phantom{-}1+i
\end{array}\right)
\,.
\end{equation}
This gate~\footnote{$\sqrt{\rm NOT}$ is not uniquely defined: the matrices 
$i^{-n/2} U_x(\alpha=n\pi/2)$ with $n=1,3,5,7$ produce NOT, when squared.}  is 
obtained by an $x$-rotation (\ref{Eq:x-rotation}) with $\alpha=\pi/2$,
more precisely $U_x(\alpha=\pi/2)=i^{1/2}\cdot\sqrt{\rm NOT}$.

Another important, essentially quantum mechanical single-bit operation
is the Hadamard  gate:
\begin{equation}
\label{Eq:Hadamard}
\picH={\sf H}=\frac{1}{\sqrt{2}}
\left(\begin{array}{cc}
1&\phantom{-}1\\1&-1
\end{array}\right)
\,.
\end{equation}
It transforms basis vectors into superpositions:
$\ket{0}\to(\ket{0}+\ket{1})/\sqrt{2}$,
$\ket{1}\to(\ket{0}-\ket{1})/\sqrt{2}$.  This gate is of use to
prepare a specific initial state: when applied
to every qubit of the system in the ground state  $\ket{0\dots 0}$, it
provides an equally-weighted superposition of all basis states:
\begin{equation}
\label{Eq:eq-weighted-sup}
{\sf H}\otimes\dots\otimes{\sf H}\ket{0\dots 0}=
\frac{1}{2^{N/2}} \sum\limits_{d_1,\dots,d_N=0,1}
\ket{d_1\dots d_N}
\,.
\end{equation}
The terms in the sum can be viewed as binary representations of all
integers from  $0$ up to $2^N-1$.  Thus the state
(\ref{Eq:eq-weighted-sup}) is a superposition of all these integers.
When used as input of a  quantum algorithm,
it represents $2^N$ classical inputs.  Due to linearity of quantum
time evolution they are processed simultaneously, and also the output
is a superposition of $2^N$ classical results. This {\em quantum
parallelism} is a fundamental property of quantum computation and is responsible
for the  exponential speed-up of certain quantum algorithms.

Among the two-qubit gates an important one is  the exclusive-OR (XOR)
or  controlled-NOT (CNOT) gate:
\begin{equation}
{\rm CNOT}=\left(\begin{array}{cccc} 1&0&0&0\\ 0&1&0&0\\ 0&0&0&1\\
0&0&1&0
\end{array}\right)
\,.
\end{equation}
When applied to classical (basis) states it flips the second bit only
if the  first bit is $1$. It was shown by \citeasnoun{9authors} that
the CNOT gate together with single-bit operations forms a universal
set, sufficient for any  quantum computation. In other words, any
unitary transformation of a many-qubit  system can be decomposed into
single-bit gates and CNOT-gates.  This explains the  importance of
CNOT in the quantum  information-theoretical literature. However, it
should be pointed out that {\em  almost any} two-qubit gate (an
exception is the classical SWAP-gate), when combined with single-bit
operations, forms a universal set.

Let us also mention another useful two-bit gate, the controlled phase
shift:
\begin{equation}
\label{Eq:Rphi}
\picRphi=R(\phi)=
\left(\begin{array}{cccc}
1&0&0&0\\ 0&1&0&0\\ 0&0&1&0\\ 0&0&0&e^{i\phi}
\end{array}\right)
\,.
\end{equation}
It shifts the phase of the state $\ket{1}$ of the second qubit  when
the  first qubit is in the state $\ket{1}$. (We use an unconventional symbol for 
this operation, which stresses its symmetry with respect to the transposition of 
qubits.)

\subsection*{Quantum Fourier Transformation}

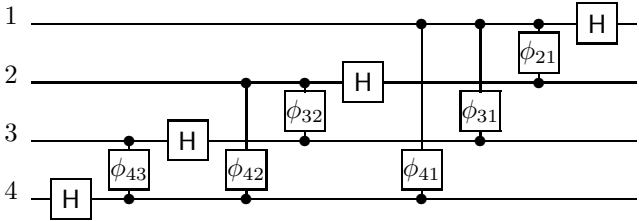
\begin{figure}
\centerline{%
\setlength{\unitlength}{0.006\columnwidth}
\begin{picture}(175,65)
  \put(-2,15){4} \put(-2,30){3} \put(-2,45){2} \put(-2,60){1}
  \put(10,10){\framebox(10,10){\sf H}}
  \put(25,17.5){\framebox(10,10){$\phi_{43}$}}
  \put(40,25){\framebox(10,10){\sf H}}
  \put(55,17.5){\framebox(10,10){$\phi_{42}$}}
  \put(70,32.5){\framebox(10,10){$\phi_{32}$}}
  \put(85,40){\framebox(10,10){\sf H}}
  \put(100,17.5){\framebox(10,10){$\phi_{41}$}}
  \put(115,32.5){\framebox(10,10){$\phi_{31}$}}
  \put(130,47.5){\framebox(10,10){$\phi_{21}$}}
  \put(145,55){\framebox(10,10){\sf H}}
  \put(5,15){\line(1,0){5}}     \put(20,15){\line(1,0){140}} 
  \put(5,30){\line(1,0){35}}    \put(50,30){\line(1,0){110}}
  \put(5,45){\line(1,0){80}}    \put(95,45){\line(1,0){65}}
  \put(5,60){\line(1,0){140}}   \put(155,60){\line(1,0){5}}
  \put(30,30){\circle*{3}}      \put(30,15){\circle*{3}}
  \put(30,30){\line(0,-1){2.5}} \put(30,15){\line(0,1){2.5}}
  \put(60,45){\circle*{3}}      \put(60,15){\circle*{3}}
  \put(60,15){\line(0,1){2.5}}  \put(60,45){\line(0,-1){17.5}}
  \put(75,45){\circle*{3}}      \put(75,30){\circle*{3}}
  \put(75,30){\line(0,1){2.5}}  \put(75,45){\line(0,-1){2.5}}
  \put(105,60){\circle*{3}}     \put(105,15){\circle*{3}}
  \put(105,15){\line(0,1){2.5}} \put(105,60){\line(0,-1){32.5}}
  \put(120,60){\circle*{3}}     \put(120,30){\circle*{3}}
  \put(120,30){\line(0,1){2.5}} \put(120,60){\line(0,-1){17.5}}
  \put(135,60){\circle*{3}}     \put(135,45){\circle*{3}}
  \put(135,45){\line(0,1){2.5}} \put(135,60){\line(0,-1){2.5}}
\end{picture}
}
\caption[]{\label{Fig:FT}%
A realization of the Fourier transformation for 4 qubits, i.e. $2^4$
coefficients. The phase shifts are $\phi_{jk} = \pi /2^{j-k}$.}
\end{figure}

As an example we discuss the quantum algorithm for a discrete Fourier 
transformation.  $N$ qubits allow us to represent
the integers $j = 0,
\dots, 2^N-1$ as basis states  $\ket{0},\dots,\ket{2^N-1}$. Starting from a 
superposition of these states with amplitudes $c_j$ and
applying the combination of controlled phase shifts and Hadamard gates
shown in Fig.~\ref{Fig:FT} one obtains an output:
\begin{equation}
\sum\limits_{j=0}^{2^N-1} c_i\ket{j}\to
\sum\limits_{k=0}^{2^N-1} \tilde c_k\ket{k}
\,,
\end{equation}
where the output amplitudes $\tilde c_k$ and input amplitudes  $c_j$ are
related by the discrete Fourier transform:
\begin{equation}
\tilde c_k=\frac{1}{2^N}\sum\limits_{j=0}^{2^N-1}
\exp\left(\frac{2\pi i k j}{2^L}\right) c_j
\,.
\end{equation}
While in classical computation the time needed for the Fourier
transform grows  exponentially with the number of bits $N$, the
quantum algorithm in  Fig.~\ref{Fig:FT} takes $\propto N^2$ steps. The
quantum Fourier transformation was developed  by \citeasnoun{Shor94}
and later improved by \citeasnoun{Coppersmith} and Deutsch
\citeaffixed{Ekert-Josza}{see}.
Its exponential speed-up compared to classical algorithms
is crucial for the performance of \possessivecite{Shor94} 
algorithm for the factorization of large integers.

\subsection*{Quantum computation and optimization}

The unitary transformations needed for quantum computation or any simple quantum
manipulation should be realized in a particular physical system.  For this
purpose they should be decomposed into elementary unitary gates.  This
decomposition is not unique and should be optimized with respect to various
parameters, for instance, the time, the number of steps, or the complexity of
the manipulations involved.  In some physical systems the manipulations involve
additional dissipation (as compared to the idle state) which should be optimized
as well.  In this subsection we discuss how certain unitary logic gates can be
realized in a spin system with model Hamiltonian (\ref{Eq:idealH}).  The
results are also useful for many other physical realizations of qubits (see
Sections~\ref{sec:JosChargeBit}, \ref{sec:FluxQubits}) since many of them have
similar Hamiltonians with similar control parameters.

First we consider the Hadamard gate (\ref{Eq:Hadamard}).  It can be performed,
up to an overall phase factor, as a sequence of elementary operations
(\ref{Eq:x-rotation}) and (\ref{Eq:z-rotation}):  ${\sf H}\propto
U_x(\alpha=\pi/4)U_z(\beta=\pi/4)U_x(\alpha=\pi/4)$.  However, it can also be
performed faster by simultaneous switching of $B_x$ and $B_z$:
\begin{equation}
{\sf H}\propto
\exp\left(-i\frac{\pi}{2}\frac{\sigma_x+\sigma_z}{\sqrt{2}}\right).
\end{equation}

The CNOT gate in the model system (\ref{Eq:idealH}) can be implemented
by a  combination of 2 two-qubit gates $U_{\rm 2b}$ (\ref{2bit_Operation}) and 
several
single-qubit gates \citeaffixed{Imamoglu}{see also}:
\begin{eqnarray}
\label{Eq:CNOTideal}
\mbox{CNOT}
&\propto& U_x^2 \left( \frac{\pi}{2}\right) U_z^2
\left(-\frac{\pi}{2}\right) U_x^2 \left(-\pi\right) U_{\rm
2b}\left(-\frac{\pi}{2}\right)
\times
\nonumber\\
&&U_x^1 \left(-\frac{\pi}{2}\right) U_{\rm
2b}\left(\frac{\pi}{2}\right) U_z^1 \left(-\frac{\pi}{2}\right) U_z^2
\left(-\frac{\pi}{2}\right)
\,.
\end{eqnarray}
Similarly, the controlled phase shift gate (\ref{Eq:Rphi}) is produced
by the  sequence
\begin{eqnarray}
\label{Eq:Rphiideal}
R(\phi) &\propto& U^2_x\left(-\frac{\pi}{2}\right) U_{\rm
2b}\left(-\frac{\pi}{2}\right) U^1_x\left(-\frac{\phi}{2}\right)
U_{\rm 2b}\left(\frac{\pi}{2}\right)
\times
\nonumber\\
&& U^2_x\left(\frac{\pi}{2}\right) U^1_z\left(-\frac{\phi}{2}\right)
U^2_z\left(-\frac{\phi}{2}\right)
\,.
\end{eqnarray}

One can see from these examples that it takes quite a number of
elementary gates to perform the CNOT or $R(\phi)$ and further
optimization is desired.  In many realizations two-qubit elementary
gates are more costly (complicated or longer) than single-qubit gates.
With this taken into account one can ask whether
Eqs.~(\ref{Eq:CNOTideal}) and (\ref{Eq:Rphiideal}) can be reduced to
just one two-bit gate?  An analysis
of this problem \cite{LocalEquiv} shows that
it is not possible if the two-bit elementary gate (\ref{2bit_Operation}), 
produced by the $XY$-coupling Hamiltonian, is used. The result is the same for 
the Heisenberg spin coupling, but the Ising-type coupling, 
$\propto\hat\sigma_z\hat\sigma_z$, allows to achieve $R(\phi)$ [and hence 
$\mbox{CNOT}={\sf H}^2 R(\pi) {\sf H}^2$]
with only one two-bit operation: $R(\phi)\propto
U_z^1(-\phi/2) U_z^2(-\phi/2) \exp(i\phi\sigma_z^1\sigma_z^2/4)$.
The optimization of certain quantum logic
circuits was also discussed in connection with qubits based on spins
in quantum dots~\cite{LossOptim}.

\section{Charging energy of a qubit coupled to a SET}
\label{app:SET}

The charging energy of the system
shown in Fig.~\ref{CIRCUIT} is a quadratic function of the charges $n$ and $N$:
\begin{eqnarray}
\label{CHARGING_ENERGY}
&&{\cal H}_{\rm C} (n,N,V_n,V_N)=4\ECqb n^2 + \ECSET N^2  
\nonumber \\
&&+\ECint N(2n-1) + 2en V_{n} + eN V_{N} \, .
\end{eqnarray}
The form of the mixed term $\propto N(2n-1)$ is chosen for 
later convenience. 
The charging energy scales $\ECqb$, $\ECSET$ and $\ECint$ are
set by the capacitances between all the islands.
Elementary electrostatics shows that they can be written as
\begin{eqnarray}
\label{CHARGING_ENERGY2}
\ECqb &=& e^2(\CgSET + C_{\rm int} + 2\CN)/ 2A \approx e^2/2C_{\rm J}
\ ,\nonumber\\
\ECSET &=& e^2(\Cgqb +C_{\rm int}+C_{\rm J})/2A \approx  e^2/(4\CN)
\ ,\nonumber\\
\ECint &=& e^2C_{\rm int}/ A \approx e^2C_{\rm int}/(2C_{\rm J} \CN)
\ .
\end{eqnarray} 
Here we introduced
\begin{eqnarray}
A &\equiv& (\Cgqb + C_{\rm J})(\CgSET + C_{\rm int} + 2\CN) 
\nonumber \\ 
&+&   C_{\rm int}(\CgSET+2\CN) \approx 2 C_{\rm J} \CN 
\ .
\end{eqnarray}
For simplicity we assumed that the two tunnel junctions of the SET have
equal capacitances $\CN$, and the approximate results refer to the
limit $\Cgqb,C_{\rm int},\CgSET \ll \CN \ll C_{\rm J}$, which we
consider useful.  The effective voltages  $V_{n}$
and $V_{N}$ depend in general on the gate voltages $\Vgqb$, $\VgSET$ and
the transport voltages applied to the SET's electrodes. However, for a
symmetric setup (equal junction capacitances) and symmetrically distributed
transport bias (as shown in Fig.~\ref{CIRCUIT}), $V_{n}$ and
$V_{N}$ are controlled only by the two gate voltages:
\begin{eqnarray}
\label{Effective_Voltages}
&&V_N = \VgSET {\CgSET(\Cgqb + C_{\rm int}+C_{\rm J})\over A} 
              \nonumber \\
        &&\ \ \ \ \ + \Vgqb {C_{\rm int} \Cgqb \over A} + {\ECint\over e}
\ ,\nonumber \\
&&V_n = \VgSET {\CgSET C_{\rm int}\over A} + \Vgqb{(\CgSET +C_{\rm
              int} + 2 \CN)\Cgqb\over A}
\ .
\nonumber \\
\end{eqnarray}

The total charging energy can thus, up to a nonessential constant, be presented
as a sum of the contributions of the qubit (\ref{Eq:1bit_Hamiltonian_Eqb}), the
SET (\ref{SET_HAMILTONIAN}), and the interaction term $\Eint N (2n-1)$ (cf.
Eq.(\ref{INTERACTION_HAMILTONIAN})).  The effect of the SET is to renormalize
the parameters of the qubit Hamiltonian~(\ref{Eq:1bit_Hamiltonian_Eqb}):  $\EC$
and $\ng$ should be substituted by $\ECqb$ and $\ngqb \equiv
-eV_n/4\ECqb$.

\section{Derivation of the master equation}
\label{app:Diagrams}

\begin{figure}  
\vskip3mm
\centerline{\hbox{\psfig{figure=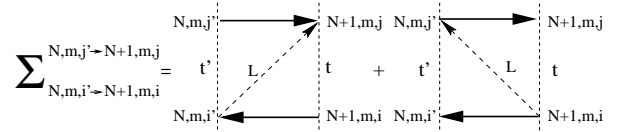,width=0.9\columnwidth}}}
\vskip1mm
\caption[]{\label{Fig:DIAGRAM_GAIN}
Example of a self-energy diagram for an `in'-rate.}
\end{figure}

\begin{figure}
\centerline{\hbox{\psfig{figure=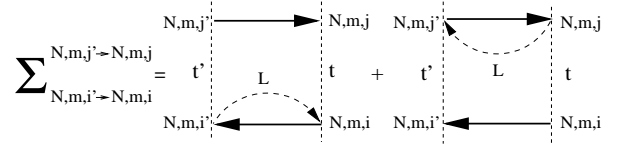,width=0.9\columnwidth}}}
\vskip1mm
\caption[]{\label{Fig:DIAGRAM_LOOSE}
Example of a self-energy diagram for an `out'-rate.}
\end{figure}

We briefly review the rules for the evaluation of diagrams; for more details
including the discussion of higher order diagrams we refer  to
\possessivecite{Schoeller_PRB} paper.  Typical diagrams, which are analyzed
below, are displayed in Figs.~\ref{Fig:DIAGRAM_GAIN}
and~\ref{Fig:DIAGRAM_LOOSE}. The horizontal lines, discussed in detail below,
describe the time evolution of the system governed by the zeroth order
Hamiltonian ${\cal H}_0$.  The directed dashed lines stand for tunneling
processes, in the example considered the tunneling takes place in the left
junction.  According to the rules, the dashed lines contribute the following
factor to the self-energy $\Sigma$:
\begin{equation}
\label{LINE}
        \alpha_{\rm L}\, \left({\pi k_{\rm B}T\over\hbar}\right)^2 \,
        {\exp\left[\pm \frac{i}{\hbar}\mu_{\rm L}(t-t')\right] \over
        \sinh^2\left[{\pi k_{\rm B}T\over\hbar} \left(t-t' \pm
        i\delta\right)\right]} 
\ , 
\end{equation}
where $\alpha_{\rm L} \equiv h / (4\pi^2 e^2 R_{\rm T,L})$ is the dimensionless
tunneling conductance, $\mu_{\rm L}$ is the electro-chemical potential of the
left lead, and $\delta^{-1}$ is the high-frequency cut-off, which is at most of
order of the Fermi energy.  The sign of the infinitesimal term $i\delta$ depends
on the direction of the dashed line in time.  It is negative if the direction of
the line with respect to the Keldysh contour coincides with its direction with
respect to absolute time (from left to right), and positive otherwise.  For
example, the right part of Fig.~\ref{Fig:DIAGRAM_GAIN} should carry a minus 
sign, while the left part carries a plus sign.  Furthermore, the sign in front 
of $i\mu_{\rm L}(t-t')$ is negative (positive), if the line goes forward 
(backward) with respect to absolute time.  Finally, the first order diagrams are 
multiplied by $(-1)$ if the dashed line connects two points on different 
branches of the Keldysh contour.

The horizontal lines describe the time evolution of the system between
tunneling processes. For an isolated central island they turn into
exponential factors $e^{\pm \frac{i}{\hbar}E(t-t')}$, depending
on  the charging  energy of the system.  However, in the present case
the island is coupled to the qubit, and we need to account for the
nontrivial time evolution of the latter.  For instance,  the upper
line in the left part of Fig.~\ref{Fig:DIAGRAM_GAIN} corresponds to
$\langle N,j|e^{-\frac{i}{\hbar}{\cal H}_0(t-t')}|N,j'\rangle$, while
the lower line corresponds to  $\langle N+1,i'|e^{\frac{i}{\hbar}{\cal
H}_0(t-t')}|N+1,i\rangle$.

In the present problem we assume that the tunneling conductance of 
the SET is low compared to the quantum conductance. In this case
lowest order perturbation theory in the single-electron tunneling, describing
sequential tunneling processes, is sufficient.  The diagrams for
$\Sigma$ can be split into two classes, depending on whether they
provide expressions for off-diagonal ($N'\ne N$) or diagonal ($N'=N$)
elements of $\Sigma$ in $N$.  In analogy to the scattering integrals
in the Boltzmann equation these can be labeled `in' and `out'
terms, in the sense that they describe the increase or decrease of a
given element $\hat{\rho}^{i,N,m}_{j,N,m}$ of the density matrix
due to transitions from or to other $N$-states.  Examples for the
`in' and `out' terms are shown in Figs.~\ref{Fig:DIAGRAM_GAIN} and
\ref{Fig:DIAGRAM_LOOSE}, respectively.

We now are ready to evaluate the rates in Figs.~\ref{Fig:DIAGRAM_GAIN}
and \ref{Fig:DIAGRAM_LOOSE}. For example, the `in'
tunneling  process  in the left junction is expressed as
\begin{eqnarray}
\label{LEFT_RATE_MATRIX}
        &&\Sigma^{j',N,m \rightarrow j,N+1,m} _{i',N,m \rightarrow
        i,N+1,m}(\Delta t)=  -\alpha_{\rm L}\left({\pi k_{\rm
        B}T\over\hbar}\right)^2 \nonumber \\ &&\times \left\{
        {\exp\left[-\frac{i}{\hbar} (\tilde E^{N}_{N+1}+W^{N}_{N+1})
        \Delta t \right]   \over \sinh^2\left[ {\pi k_{\rm
        B}T\over\hbar}(\Delta t + i\delta) \right] } \right.  \\ &&
        \left.  + {     \exp\left[ -\frac{i}{\hbar} (\tilde
        E^{N+1}_{N}+W^{N+1}_{N}) \Delta t \right]   \over
        \sinh^2\left[ {\pi k_{\rm B}T\over\hbar}(\Delta t - i\delta)
        \right] } \right\}^{j'j}_{i'i}  \nonumber \ ,
\end{eqnarray}  
where
$\tilde E^{N_1}_{N_2} \equiv
\left[\ECSET(N_1-\NgSET)^2 - \mu_{\rm L}\,N_1\right] - 
\left[\ECSET(N_2-\NgSET)^2-  \mu_{\rm L}\,N_2\right]
$ is the Coulomb energy gain for the tunneling in the left
junction in the absence of the qubit, and the operators
\begin{equation}
W^{N_1}_{N_2} = {\cal H}_{0}^{\rm T}(N_1)\otimes 1 - 1\otimes {\cal
H}_{0}(N_2)\ .
\label{W_N1_N2}
\end{equation}
provide corrections to the energy gain sensitive to the qubit's state.
Here ${\cal H}_{0}(N)$ is the $N$-th block of the Hamiltonian  ${\cal
H}_{\rm ctrl}+{\cal H}_{\rm int}$ (note that ${\cal H}_{\rm ctrl}$
and ${\cal H}_{\rm int}$ are block-diagonal with respect to $N$).  The
indices $j',j$ and $i',i$ relate to the left and right side of the
tensor product in (\ref{W_N1_N2}) correspondingly. 

The form of the master equation (\ref{MASTER_EQUATION}) suggests  the
use of the Laplace transform, after which the last term in
(\ref{MASTER_EQUATION}) becomes $\Sigma(s)\hat{\rho}(s)$.  We
Laplace transform (\ref{LEFT_RATE_MATRIX}) in the regime  $\hbar s,
|W^{N}_{N+1}|,|W^{N+1}_{N}| \ll \tilde E^{N}_{N+1}$, i.e.,  we assume
the density matrix $\hat{\rho}$ to change slowly on the time  scale
given by $\hbar/\tilde E^{N}_{N+1}$. This assumption should be
verified later for self-consistency. The inequalities also mean that
we choose the operation regime of the SET far enough from the Coulomb
threshold. Therefore the tunneling is either energetically allowed for both
states of the qubit or it is blocked for both of them. At low
temperatures  ($k_{\rm B}T \ll \tilde E^{N}_{N+1}$) and for  $\tilde
E^{N}_{N+1}\delta \ll \hbar$ we obtain
\begin{eqnarray}
\label{LEFT_RATE_LAPLACE}
    &&\Sigma^{N,m,j' \rightarrow N+1,m,j}_{N,m,i' \rightarrow
    N+1,m,i}(s) \approx  \nonumber \\  && \Bigg\{
    \frac{\pi}{\hbar}\alpha_{\rm L} \Theta(\tilde E^{N}_{N+1})  \Big[
    2\tilde E^{N}_{N+1} + (W^{N}_{N+1} - W^{N+1}_{N})  \Big] \nonumber
    \\ &&- \alpha_{\rm L} D(\tilde E^{N}_{N+1}) \Big[ 2s +
    \frac{i}{\hbar}(W^{N}_{N+1} + W^{N+1}_{N}) \Big]
    \Bigg\}^{j'j}_{i'i}  \ ,
\end{eqnarray}
where   $D(\tilde E^{N}_{N+1}) \approx 1+\gamma + 
\ln(\tilde E^{N}_{N+1}\delta/\hbar)$
and $\gamma \approx 0.58$ is Euler's constant.  The first term of
(\ref{LEFT_RATE_LAPLACE}) is the standard  Golden rule tunneling rate
corresponding to the so-called  orthodox theory of single-electron
tunneling~\cite{Averin-Likharev}.  The rate depends strongly on the
charging energy difference,  $\tilde E^{N}_{N+1}$, before and after
the process, which in the present problem is modified according to the
quantum state  of the qubit (the $W$-terms). At finite temperatures
the step-function  is replaced by $\Theta(E) \rightarrow
[1-\exp{(-E/k_{\rm B}T)}]^{-1}$.  We denote the full matrix of such
rates $\check \Gamma$. As has already been mentioned, we concentrate 
on the regime when the leading tunneling process in the SET is the
sequential tunneling, involving only two adjacent charge states,  say,
$N=0$ and $N=1$ (to avoid confusion with  the states of the qubit we 
keep using the notation $N$ and $N+1$). Let us, for example,
calculate a submatrix of $\check \Gamma $ which originates 
from the first term on the RHS of (\ref{LEFT_RATE_LAPLACE}) and corresponds to 
the tunneling process $N \rightarrow N+1$, $m \rightarrow m$ in the left 
junction.
This submatrix $\check \Gamma_L$ is a super-operator
which acts on a $2 \times 2$-matrix $\hat \rho$ as
\begin{equation}
\hbar \check\Gamma_L\hat\rho =
2\pi\alpha_L \tilde E_{N+1}^{N} \hat\rho
+\pi\alpha_L\;[\dHint,\hat\rho]_+
\; ,
\end{equation}
where $\dHint \equiv {\cal H}_{0}(N+1) - {\cal H}_{0}(N) = \ECint \sigma_z$. 

The last, logarithmically diverging term of
Eq.~(\ref{LEFT_RATE_LAPLACE}) produces the commutator term in the
RHS of the master equation (\ref{MASTER_EQUATION}).  These terms turn
out to be unimportant in the first order of the  perturbation
theory. Indeed, for the left junction we obtain the  following
contribution to the RHS of (\ref{MASTER_EQUATION}):
\begin{equation}
\label{LEFT_DIVERGENT_TERM}
\alpha_{\rm L} \hat D_{\rm L}
\left(
{d\hat\rho\over dt} -  {i\over \hbar}\left[\hat\rho,{\bar {\cal
H}}_{0}\right] 
\right)
\ ,
\end{equation}   
where ${\bar {\cal H}}_{0}\equiv
\frac{1}{2}({\cal H}_{0}(N)+{\cal H}_{0}(N+1))$
and $\hat D_{\rm L}$ is a matrix in $N$ and $m$ spaces. The
eigenvalues of the  matrix $\hat D_{\rm L}$ are at most of order
$D(\tilde E^{N}_{N+1})$.  Neglecting terms of order $\alpha_{\rm L}
D(\tilde E^{N}_{N+1}) \ECint$ in  Eq.~(\ref{LEFT_DIVERGENT_TERM}), we
can replace ${\bar {\cal H}}_{0}$ by ${\cal H}_0$.  Our analysis shows
that these neglected ``coherent-like'' terms do not change the results
as long as  $\alpha_{\rm L}\,|\ln(\tilde E^{N}_{N+1}\delta/\hbar)|\ll
1$. A similar analysis can be carried out for the right tunnel junction  of
the SET. 

Now we can transfer all the ``coherent-like'' terms into the LHS of
the master  equation,
\begin{equation}
\label{MASTER_EQUATION_CORRECTED}
        \left(1-\alpha_{\rm L} \hat D_{\rm L} -\alpha_{\rm R} \hat
          D_{\rm R}\right)  \left\{ {d\hat{\rho}(t)\over dt} -
          {i\over\hbar}[\hat{\rho}(t), {\cal H}_0]  \right\} =
          \frac{1}{\hbar}\hat{\Gamma} \hat{\rho}(t),
\end{equation}
and multiply Eq.~(\ref{MASTER_EQUATION_CORRECTED}) from the left by
$(1-\alpha_{\rm L}\hat D_{\rm L} - 
\alpha_{\rm R}\hat D_{\rm R})^{-1} 
\approx (1 + \alpha_{\rm L}\hat D_{\rm L} + 
\alpha_{\rm R}\hat D_{\rm R})$ so that the
corrections move back to the RHS.  Since $\check\Gamma$ is itself
linear in  $\alpha_{\rm L}$ and $\alpha_{\rm R}$, the corrections are of
second order in $\alpha$ (more accurately, they are small if
$\alpha\,|\ln(\tilde E^{N}_{N+1}\delta/\hbar)| \ll 1$ for both
junctions).  Thus we drop the ``coherent'' corrections and arrive at
the final form of the master equation:
\begin{equation}
\label{MASTER_EQUATION_FINAL}
        {d\hat{\rho}(t)\over dt} - {i\over\hbar}[\hat{\rho}(t),
        {\cal H}_0]   = \frac{1}{\hbar}\check{\Gamma} \hat{\rho}(t)
        \ .
\end{equation}

We have shown that under the assumption of sufficiently slow dynamics of the 
qubit and SET, $\EJ,B_z(\Vg),\Eint,\Gamma_{\rm L/R}\ll\tilde E^{N}_{N+1},\tilde 
E^{N+1}_{N}$,
the evolution of the system reduces to Markovian dynamics as described by
the master equation (\ref{MASTER_EQUATION_FINAL}).

\section{Quantum Dots and Spins}
\label{app:spins_dots}

Ultra-small quantum dots have also been suggested as possible
realizations of qubits.  A quantum dot is, for instance, a small area
in a two-dimensional 
electron gas (2DEG) where electrons are confined by the surrounding potential
walls created by metal structures on top of the systems and applied gate
voltages.  Usually the size of the dot is larger than the Fermi wave 
length of the 2DEG, $\lambda_{\rm F} \approx 30-80 {\rm nm}$.  Then, many
electrons are confined in the dot.  To construct qubits one needs smaller dots
whose size is or order of $\lambda_{\rm F}$.  Such dots resemble artificial
atoms since they contain only a small number of electrons filling discrete
energy levels.  The simplest qubit would be formed by an artificial `hydrogen
molecule' with one electron confined in a double dot. Of course, a
system with many electron states is sufficient as well, provided the
level spacing is large enough such that by proper 
tuning only one level in each dot plays a role.
In these systems the coherent tunneling of single electron
charges, controlled by external gate voltages, has  
been demonstrated  \cite{dot-molecules}, and many of the ideas outlined in
Section~\ref{sec:JosChargeBit} for Josephson junction devices 
apply as well.  However, there appear several difficulties:  one has to
push the fabrication to the limits (in contrast to the superconducting
box, where we exploit the macroscopic coherence 
of the superconducting state), and the system is likely to be
disordered due to variations in the difficult fabrication process.  These
problems may be overcome,  e.g., by using chemically well-defined
cluster molecules or other advanced growth techniques, but still at this stage
there is not much of an activity to be reported exploiting the charge
degrees of freedom in quantum dots.

Spin degrees of freedom have also been proposed as candidates for
qubits in mesoscopic systems \cite{Loss,Kane}.  The most obvious advantage of
this approach is the fact that the Hilbert space of a spin-1/2 particle
is really two-dimensional \footnote{Strictly speaking, this property is
lost due to interactions in a many-body environment.}.  In contrast,
in most other proposals the two-dimensional Hilbert space of a qubit is a
subspace of a much larger physical Hilbert space, to which the system can
leak out. 
The proposal of \citeasnoun{Loss} deals with spin states of electrons
in quantum dots. The dot has to be small enough and the temperature
low enough that the orbital degrees of freedom of the electron are
fixed,  and the only relevant
degree of freedom is the electron's spin.  The spin degree of freedom is quite
decoupled from the voltage fluctuations of the gates.  Therefore the dephasing
time of the qubit may be quite long (of order of microseconds).

To manipulate individual qubits one has to apply individual (local) magnetic
fields to each spin.  An alternative method is to apply a homogeneous
magnetic field and shift 
electrostatically the position of the spin in the dot to areas with different 
$g$-factors \cite{Loss_DiVincenco_Turkey}. In this way a resonance
condition can be controlled for individual spins.  The two-bit
coupling is achieved by controlling the tunneling between 
neighboring quantum dots.  This may be done by changing the gate
voltage configuration, 
which changes the potential barrier between the dots.  In this way the
overlap of the wave functions of the electrons in the dots is
controlled and so is the  exchange splitting
between the  singlet and the triplet states of the two spins
produced by  the Coulomb interaction. This may be
represented as an interaction Hamiltonian of the Heisenberg type, $H_{\rm
int} = J(t)\bbox{S}_1\bbox{S}_2$. Since the overlap integral
 depends exponentially on the barriers and can be made
very small, also the interaction strength $J(t)$ can be tuned by gate
voltages to exponentially low values, 
completely decoupling the qubits. If turned on, the Heisenberg
interaction allows one to perform the `square root of swap' operation
(see Eq.~\ref{2bit_Operation}), which,
together with the one-bit gates, constitute a universal set.

Another setup which exploits the spin states was proposed by \citeasnoun{Kane}.
in this case the qubit is the nuclear spin $1/2$ of the phosphorus
atom placed as a 
donor in the silicon host.  The host nuclei have zero spin.  The nuclear spins
are addressed by gating the shallow bound states of electrons around the
positively charged donor ions.  If a high enough magnetic field is applied, the
electronic spins are completely polarized.  The hyperfine interaction makes the
nuclear spins feel an additional effective magnetic field proportional to the
density of the electronic wave function at the nucleus.  The wave
functions of the shallow states extend thousands of \AA\ away from the donors'
nuclei.  Therefore one can manipulate these states by gate voltages, changing
their wave functions' shape and, effectively, the local magnetic field for
individual qubits.  If one allows the shallow states from two donor atoms to
overlap, an effective exchange interaction between the nuclear spins emerges.
The wave function overlap may be again controlled by gates.  Thus 
a controllable  two-bit coupling is achieved.

A direct measurement of the quantum state of an individual spin is a 
difficult task.  Instead, in both proposals of \citeasnoun{Loss} and
\citeasnoun{Kane}, it is 
suggested to translate the spin state into a charge state which could
be measured easier.  Both proposals are interesting from a physics
point of view and have many advantages, allowing  in principle
for a large number of phase coherent manipulations. On the other hand,
both proposals operate at the limit of present day technology
and even the simplest  quantum manipulations still have to be
demonstrated. 


\end{document}